\documentclass[preprint,authoryear,review,11pt]{elsarticle}
\usepackage[dvipsnames]{xcolor}
\usepackage{setspace}
\usepackage{amsmath}  
\usepackage{amsfonts} 
\usepackage{amssymb}  
\usepackage{bm}
\usepackage{dsfont}
\usepackage{tcolorbox}
\usepackage{float}
\usepackage{xurl}
\usepackage{hyperref}
\usepackage[title]{appendix}
\hypersetup{
    colorlinks = true,
    allcolors = [rgb]{0,0.29,0.6}
    }
\usepackage{pdflscape}
\usepackage{enumitem} 
\usepackage{epstopdf}
\usepackage[official]{eurosym}
\usepackage{MnSymbol,wasysym}
\usepackage{caption}
\usepackage{subcaption}
\usepackage{graphicx}
\usepackage{rotating}
\usepackage{booktabs}
\usepackage{tabularx}
\usepackage{threeparttable}
\usepackage{pdflscape}
\usepackage{adjustbox,lipsum}
\usepackage{algorithm,algpseudocode}
\usepackage{changepage}
\usepackage{multirow}
\usepackage{dcolumn}
\usepackage{enumitem}
\usepackage[a4paper, left=1.3in, right=1.3in, top=1in, bottom=1in]{geometry}
\usepackage{graphicx}%
\usepackage{tikz}
\usepackage{multirow}%
\usepackage{amsmath,amssymb,amsfonts}%
\usepackage{amsthm}%
\usepackage{mathrsfs}%
\usepackage[title]{appendix}%
\usepackage{subcaption}%
\usepackage{textcomp}%
\usepackage{manyfoot}%
\usepackage{booktabs}%
\usepackage{algorithm}%
\usepackage{algorithmicx}%
\usepackage{algpseudocode}%
\algrenewcommand\algorithmicrequire{\textbf{Input:}}
\algrenewcommand\algorithmicensure{\textbf{Output:}}
\algnewcommand{\Initialize}[1]{%
  \State \textbf{Initialize:}
  \Statex \hspace*{\algorithmicindent}\parbox[t]{.8\linewidth}{\raggedright #1}
}
\usepackage{listings}%
\usepackage{natbib}

\usepackage[textwidth=1.2cm, textsize=tiny]{todonotes}
\theoremstyle{thmstyleone}%
\newtheorem{theorem}{Theorem}
\newtheorem{proposition}[theorem]{Proposition}%
\newtheorem{corollary}[theorem]{Corollary}

\theoremstyle{thmstyletwo}%
\newtheorem{remark}{Remark}%
\usetikzlibrary{patterns}

\theoremstyle{thmstylethree}%
\newtheorem{definition}{Definition}%
\newtheorem{assumption}{Assumption}

\ifx\proof\undefined
  \newenvironment{proof}[1][Proof]{\par\noindent\textbf{#1.}\ }{\hfill$\square$\par}
\fi

\begin{document}

\begin{frontmatter}



\title{Disaster Risk Financing through Taxation: A Framework for Regional Participation in Collective Risk-Sharing}

\author[fallou]{Fallou Niakh\corref{cor1}} 
\ead{fallou.niakh@ensae.fr} 

\author[arthur]{Arthur Charpentier} 
\ead{charpentier.arthur@uqam.ca} 

\author[fallou]{Caroline Hillairet}
\ead{caroline.hillairet@ensae.fr}

\author[arthur]{Philipp Ratz}
\ead{philippratz@outlook.com}

\cortext[cor1]{Corresponding author}

\affiliation[fallou]{organization={ENSAE IP Paris, CREST UMR 9194}, 
addressline={5 Avenue Henry Le Chatelier}, 
city={Palaiseau},
postcode={91120}, 
country={France}}

\affiliation[arthur]{organization={Mathematics Departement, UQAM},
addressline={avenue du Président Kennedy}, 
city={Montréal},
postcode={H2X 3Y7}, 
country={Canada}}


\begin{abstract}
\noindent
We consider an economy composed of different risk profile regions wishing to be hedged against a disaster risk using multi-region catastrophe insurance. Such catastrophic events inherently have a systemic component; we consider situations where the insurer faces a non-zero probability of insolvency. To protect the regions against the risk of the insurer's default, we introduce a public-private partnership between the government and the insurer. When a disaster generates losses exceeding the total capital of the insurer, the central government intervenes by implementing a taxation system to share the residual claims. In this study, we propose a theoretical framework for regional participation in collective risk-sharing through tax revenues by accounting for their disaster risk profiles and their economic status.
\text{ }  \\
\noindent

\begin{keyword}
Disaster risk \sep Multi-region catastrophe insurance \sep Risk sharing \sep Government intervention \sep Taxation \\
JEL: G22, H23, D81, Q54
\end{keyword}

\end{abstract}
\end{frontmatter}

\section{Introduction}
The effects of climate change are far-reaching and diverse. Whereas some of them, such as temperature increases, are easily detectable, the associated impacts in climate extremes are more challenging to estimate, as they fall outside of observational data~\citep{rummukainen2012changes}. At the same time, developments in exposed areas can increase losses incurred during severe weather events, with some estimates predicting growth in damages larger than that of the US economy~\citep{dinan2017projected}. This poses challenges for the insurance and reinsurance industries, where estimating the future costs of associated losses is critical.\\
{Recent developments in predictive modeling have attempted to address these challenges. For instance, ~\cite{gao2022leveraging} use high-resolution weather data in a spatial point process framework to predict hail damage claims, while ~\cite{shi2024leveraging} propose a deep learning triage model that leverages weather dynamics for real-time insurance claim prioritization. These tools offer promising directions to better anticipate and localize financial losses, but they also highlight the complexity and heterogeneity of risks, which remain difficult to aggregate within traditional actuarial models.}\\
Although modeling these risks is difficult, some real-world effects can already be seen. From 1969 to 1998, 36 U.S. insurers became insolvent due to natural disasters 
\citep{charpentier_natural_2014}. Hurricane Hugo alone was responsible for 20 of these insolvencies between 1989 and 1993 \citep{matthews1999insolvency}. Similarly, the hurricanes that hit Florida in 2004 and 2005 contributed to the insolvency of 27 insurers from 2004 to 2011 
\citep{florida_risk_2011}. Recent events have underscored the growing challenges posed by climate-related disasters. In the fall of 2024, Florida was hit by two devastating cyclones, which caused billions of dollars in damage and highlighted the increasing frequency of extreme weather events in high-risk areas. This has further exacerbated the pressure on insurers, who have already stopped renewing contracts for properties at high risk of wildfires or hurricanes~\citep{CBS_calif,farmers_florid}.
These events also reflect the risk of systemic propagation when correlated losses spread geographically, overwhelming the capacity of individual insurers or reinsurers. Such dynamics resemble a form of geographical contagion, a phenomenon conceptually explored by ~\cite{qiu2023optimal} in the context of default risk interdependence, where one shock increases the likelihood of others within a shared environment.\\
In France, climate change has ignited significant initiatives. The Langreney Report \citep{merad2024adapter} provided a comprehensive analysis of the increasing impacts of climate change on the insurance sector, urging policymakers and insurers to act quickly to mitigate these risks. In response to these concerns, the French Caisse Centrale de Réassurance (CCR) launched the Observatory of Insurability in October 2024. This initiative aims to evaluate and enhance the insurability of climate risks ~\citep{ccr2024}.\\

These historical examples and recent events highlight the increasing vulnerability of the insurance and reinsurance sectors in a changing climate. In parallel, spatial dependence among risks, especially for geographically distributed portfolios, has been widely acknowledged. For example, ~\cite{tadayon2024spatial} propose a spatial copula-based model to capture joint dependence in claim frequency and severity in auto insurance. Such approaches offer useful insights into the construction of more resilient regional risk-sharing mechanisms, especially when applied to natural catastrophe settings.\\
This situation presents a significant challenge for civil society, as insurance takers face the combined problem of not being able to insure their property in the first place and being exposed to an insolvency risk (or to coverage limits) even when an insurance contract is obtained. In this article, we study different mechanisms that can be used to mitigate these effects. These challenges have been well documented in the insurance literature. For instance, \cite{wu2015reexamining} examines the limitations of traditional diversification and risk transfer instruments in smoothing catastrophic risks, highlighting the difficulty of achieving stable coverage under extreme conditions, while \cite{cai2007optimal} explore optimal retention under tail risk constraints, offering insights into how capital should be distributed among claimants when total claims exceed reserves. Building on this, \cite{wu2020equilibrium} develops a market equilibrium framework incorporating disaster-resilient technologies and government interventions, demonstrating the importance of hybrid systems where public mechanisms complement private coverage. In general, two types of insurance schemes can be considered
\begin{enumerate}[noitemsep]
    \item Private market insurance, where only policyholders are at risk of their insurer going bankrupt.
    \item Government programs, where policyholders share risk collectively through tax revenues.
\end{enumerate}
The latter group of programs has some real-life examples, such as the Caribbean Catastrophe Risk Insurance Facility (CCRIF) and the African Risk Capacity (ARC). CCRIF, established in 2007, is a parametric insurance facility that allows Caribbean and Central American governments to access immediate liquidity following hurricanes and earthquakes, thus reducing the fiscal burden of natural disasters \citep{thirawat2017disaster}. Similarly, ARC is a specialized agency of the African Union that provides risk pooling and early financing for extreme weather events, including droughts and floods, enabling sovereign African states to take climate-adaptive action \citep{awondo2019efficiency}. These programs are examples of multi-country public-private partnerships that rely on regional risk pooling and pre-financed instruments to stabilize post-disaster expenditures.

Due to the complexities of localized losses, other mechanisms, such as the Swiss Natural Hazard Insurance, operate under a decentralized but mandatory insurance scheme. In Switzerland, natural hazard insurance is compulsory in most cantons and is administered by cantonal insurance providers (often public) alongside the private sector, offering broad coverage through risk pooling, cross-subsidization, and uniform premiums \citep{gurenko2004catastrophe}. This system blends actuarial principles with social solidarity to ensure insurability even in high-risk areas.

Similar concerns have been raised about the design and sustainability of public disaster insurance programs in the U.S., notably the National Flood Insurance Program. \cite{michel2011redesigning} argue for risk-based pricing combined with targeted subsidies, highlighting the importance of aligning incentives and solvency guarantees—issues we address through an ex-post taxation mechanism in the case of insurer default.
Academic research by \cite{charpentier_natural_2014} suggests that government programs are more efficient than the free market in providing natural catastrophe insurance. Insurance with an unlimited guarantee from the government is shown to be a better option than limited-liability insurance. In simpler terms, government programs allow adverse financial effects to be shared among policyholders, making them less risky and more attractive. Risk-averse policyholders are more willing to pay higher rates for unlimited-guarantee insurance, reducing the probability of insolvency. Whereas such government programs are often based on the principle of solidarity across regions, as in \cite{charpentier_natural_2014},
there is also a strong theoretical basis for their efficiency under interregional risk heterogeneity. \cite{boadway1996efficiency} show that intergovernmental transfers can enhance economic efficiency in federal systems, particularly when regions face asymmetric shocks or market failures. Building on this idea, we investigate how a taxation-based risk-sharing mechanism can allocate disaster costs across regions in a way that is both individually rational and Pareto optimal.
 
Here, we explore how collective insurance and taxation schemes can be rationalized using economic principles of risk sharing grounded in utility theory. In particular, our approach is informed by the economic premium principle of \cite{buhlmann1980economic}, which frames premium setting and loss sharing as cooperative decisions between agents with differing risk preferences. 

In addition to the two types of government-driven and market-driven insurance, some hybrid mechanisms also exist, usually hierarchical risk-sharing mechanisms where the government intervenes in addition to the insurer for higher losses. The selection of a particular mechanism for catastrophe insurance varies from country to country, depending on various factors such as national conditions and the developmental status of the insurance sector. For instance, California's earthquake insurance follows a government-driven mechanism where the California Earthquake Authority (CEA) mandates insurance companies to sell insurance policies and assess disaster losses. The CEA is responsible for bearing the earthquake risk \citep{roth1998earthquake}. In contrast, the catastrophe insurance system in Britain is highly developed and follows a fully market-oriented mechanism where the government only plays a supervisory role. Japan has established a hierarchical risk-sharing structure for earthquake insurance. This involves selling earthquake insurance as an add-on to fire insurance, with claims distributed among entities such as Japan Earthquake Reinsurance Co. (JER), insurance companies, and the government, depending on the total claim size \citep{ishiwatari2012chapter}. The catastrophe insurance model in China is a hybrid approach characterized by ``government guiding the market operation''. This approach is influenced by international experience and national conditions. Due to low catastrophe insurance coverage, the Chinese government provides residents with proportional or full premium subsidies \citep{wang2023risk}. This type of direct public subsidy has also been studied in the context of long-term care insurance. \cite{courbage2024effects} analyze how subsidies targeted at different dependency levels influence participation and efficiency in insurance programs. While their focus is on health-related risks, the underlying logic, that well-calibrated public support can strengthen market-based mechanisms, applies equally to catastrophe insurance. The mechanism that we develop in this article covers all of these mechanisms as special cases of hierarchical risk-sharing and allows us to compare the different approaches to ease decision-making. These initiatives represent common risk diversification measures, bringing together the governments of different regions to collectively manage and mitigate risks on a larger scale. From a theoretical perspective, \cite{farhi2017fiscal} show that in a federation with incomplete markets, optimal fiscal transfers can improve welfare by smoothing regional shocks. Their framework motivates our exploration of a taxation-based disaster risk-sharing scheme, where a central authority reallocates resources across heterogeneous regions facing correlated catastrophic risk.

The paper is then organized as follows. Section \ref{sec:main_model} describes the general model considered in this study, encompassing all three mechanisms described above. We then introduce risk-sharing on the part of the government as a general taxation framework in Section \ref{sect3}. Although our approach admits an analytical solution, the derivations become quickly intractable when many regions are present in the mechanism, which in the age of micro-level data seems a likely scenario. Sections \ref{sect4} and \ref{sec:numerical_algo} present, therefore, an algorithm and a numerical implementation thereof to determine optimal taxation rules in large-scale mechanisms. We illustrate the implications of the mechanism in Section \ref{sec:application} and conclude in Section \ref{sec:conc}.
Details on some technical parts are postponed in the Appendix.

\section{The model}\label{sec:main_model}

To cover the three mechanisms (government-driven, market-driven, and hierarchical risk-sharing) outlined above, we consider three types of agents throughout our study. The first type is the heterogeneous regions within a population that face the risk of a catastrophic event. The heterogeneity across the regions stems from both their level of wealth and their exposure to natural disasters. The second type is a representative private primary insurer that offers traditional insurance policies to help protect these regions from potential losses up to either a coverage limit or a risk of insolvency. The third type is a government that aims to provide unlimited guarantees to the regions. 

\subsection{Micro-foundations}

Consider an economy consisting of $n$ regions labeled from $1$ through $n$ and undergoing a catastrophic event. Each region $i$ has an initial wealth level or endowment, denoted as $w_i \in \mathbb R^+$. This endowment can be considered as part of the gross domestic product (GDP) for each region in a given year. After experiencing a catastrophe, we define the loss incurred by region $i$ as 
\begin{equation}\label{spef}
    X_i=\tau_iw_i\enspace, 
\end{equation} 
where $\tau_i$ is a random variable taking values in $[0,1]$ representing the proportion of wealth that region $i$ loses due to the catastrophic event. This allows us to model catastrophic events that affect certain regions with a given severity but may result in substantially different losses, given the value of the underlying assets. Hence, the framework enables us to split a region's risk into two components: its disaster risk profile ($\tau_i$) and its economic status ($w_i$).

Given that disasters often exhibit a strong spatial correlation of the severity of a catastrophic event, we assume that the random variables $X_1, X_2, ..., X_n$ are defined on a common probability space $(\Omega, \mathcal{F}, \mathbb{P})$. Let $L^{\infty}_{+}(\Omega, \mathcal{F}, \mathbb{P})$ denote the set of bounded, non-negative random variables. We also assume this space is rich enough to contain all the random variables mentioned in this paper. The variables $(X_i)_{i=1, \dots, n}$ are not assumed to be independent nor identically distributed across regions, as the regions may be subject to highly correlated risks, and they may have different levels of risk exposure. These characteristics have been empirically observed in societies exposed to catastrophic risks~\citep{meyer2017ostrich}. The aggregate wealth loss (referred to as the total loss) in the whole economy is then denoted by
\begin{equation}
 S=\sum^{n}_{i=1}X_i\enspace.   
\end{equation}

In our population, some regions are exposed to higher risk or lower risk. Safer regions usually do not favor catastrophe insurance policies, but might still want to cover some extreme risks. Conversely, high-risk regions are more likely to purchase catastrophe insurance, but general insurance companies might be less willing to offer such policies due to the potential disruption to their operational stability.

Here, we explore the joint purchase of catastrophe insurance by multiple regions. Multi-region catastrophe insurance can collectively enable high-risk and low-risk regions to buy insurance, forming a risk pool. Even though these regions are in close proximity, they do not necessarily experience the same major disasters in the same year. Therefore, member regions can share each other's financial burden and mitigate the risks faced by individual regions. For insurance companies, offering multi-region catastrophe insurance can help mitigate the instability of solely insuring high-risk regions. In this study, we model the risk aversion of the different regions using utility functions. As is customary in decision theory, a utility function is assumed to be strictly increasing and strictly concave, reflecting the region's cautious attitude towards uncertainty. Since we examine utility over positive wealth, we restrict our analysis to the utility function $u: [0,+\infty) \mapsto(-\infty,+\infty)$. Typical examples of such functions include those with constant absolute risk aversion, such as the exponential utility family $           u(x)=-\gamma e^{-\frac{x}{\gamma}}$ where $\gamma > 0$ is the absolute risk aversion parameter, and those with constant relative risk aversion, corresponding to the power utility family $u(x)=(b+x)^c$ where $0<c<1$ and $b>0$ is the translation parameter.  The satisfaction of each region is modeled by its utility functions $u_i$ to incorporate the heterogeneous risk aversion feature into the collective risk-sharing problem.

\subsection{Ex-ante risk transfer through insurance}

The traditional mechanism transfers risk from the regions to a single insurer. To cover losses, the insurer, indexed as $0$, provides an initial capital $k_0 \in \mathbb R^+$ and collects a premium from each region $i$ according to the following premium formula (as in \cite{wang2023risk})
\begin{equation}\label{eq:insurancepremium}
    \pi_i:= (1+\theta)\mathbb{E}[X_i] + \eta \sigma [X_i]  \ \text{and} \; k:=\sum^{n}_{i=1}\pi_i \enspace,
\end{equation}
where $\theta \in \mathbb R^+$ is a premium loading factor of the insurance policies, and $\eta \in \mathbb R^+$ is used to calculate the safety premium as a percentage of the standard deviation $\sigma [X_i]$ of $X_i$. Hence, the total capital of the insurance firm is given by 
\begin{equation}
    K := k + k_0\enspace.
\end{equation}
To model the case of insolvency\footnote{ Remark that this framework could also include the case of a total coverage limit,  with a premium formula \eqref{eq:insurancepremium} that should be adapted.}, we suppose that $\mathbb{P}(S > K) > 0$, that is, in particular $K < \sum^{n}_{i=1}w_i$, such that the total capital is insufficient to exclude the possibility of default. Therefore, the contracts are defaultable, meaning the insurer may not fully pay the region's claims. 

In those scenarios where the insurance company defaults on its contracts, the insurer's compensation will be strictly smaller than the experienced losses. On the contrary, in favorable scenarios where the aggregate losses $S$ are smaller than the collected premium, the region will receive some surplus in addition to its experienced losses.\\
In what follows, the random vector $(Y_i)_{i \in [ 1, n ]} $ denotes the payment of the insurer to the regions.  These payments depend on three distinct types of scenarios. 

\begin{enumerate}
    \item \textbf{Favorable scenarios $\Omega_+$}: Favorable scenarios correspond to the states where the total losses of the regions are lower than the total premium collected by the insurer: $$\Omega_+:=\{\omega, \, \text{s.t.} \;  S(\omega) < k\}.$$ In those scenarios, the losses $X_i$ are fully compensated by the insurer, and the surplus $k-S$ is distributed among the regions and the insurer: $Y_i \geq X_i$ for all $i=1, \cdots, n$.
     \item \textbf{Default scenarios $\Omega_-$}: Default scenarios correspond to the states where the insurer defaults on its contracts:  $$\Omega_-:=\{\omega, \, \text{s.t.} \;  S(\omega) > K\}.$$ In this case, the insurance company allocates the total capital $K$ to the regions as partial compensation and  $Y_i \leq  X_i$ for all $i=1, \cdots, n$.
    \item \textbf{Intermediate scenarios $\Omega_=$}: Intermediate scenarios  corresponds to the  states   where  $S$ exceeds the total premium collected but is less than $K$: $$\Omega_=:=\{\omega, \, \text{s.t.} \;   k \leq S(\omega) \leq  K\} .$$ In this situation, the insurer fully compensates the regions, and no surplus sharing occurs: $Y_i = X_i$ for all $i=1, \cdots, n$.
\end{enumerate}
This paper concentrates on optimally determining the ex-post risk sharing through taxation and does not address how to set the level of the capital $K$ nor the amount $Y_i$. This means that the deterministic vector $\pi$ and the random vector $(Y_i)$ are exogenous. The value of the compensation $Y_i$ in favorable and default scenarios is detailed hereafter, based on previous results in \cite{coculescu_fairness_2022}.

\subsubsection{Surplus sharing mechanism}
In the set $\Omega_+$ of favorable scenarios, after the coverage period ends, a surplus-sharing mechanism rewards regions for exposing themselves to the risk of insurer default:
\begin{equation}
    Y_i  \geq X_i  \quad \text{on} \quad \Omega_+.
\end{equation}
More specifically, the positive random vector $(Y_i-X_i)^+$ is the benefit sharing of the regions $i$ for the surplus $k-S$. This surplus can be shared through different mechanisms, among which the standard proportional sharing, in which each region $i$ receives some constant proportion $c_i$ of the insurer's benefit $k-S$. In this case, these constant proportions $c_i$  are specified at the beginning of the contract. This leads to the amount paid by the insurer to region $i$ given by
\begin{equation}
  Y_i= X_i + c_i(k - S)^+ \quad \text{on} \quad \Omega_+
   \label{benef}
\end{equation}
where each $c_i$ is a non-negative constant and $c_0 + \sum^{n}_{i=1} c_i=1$ with $c_0 $ the share of benefit allocated to the insurer. We assume naturally that $c_i(k - S)^+ \leq \pi_i$ to exclude the possibility of free enrichment.

\subsubsection{Capital allocation mechanism}
In the set $\Omega_-$,  the insurer cannot fully cover the losses. The total available capital $K$ is distributed among the regions for partial compensation, leading to $Y_i < X_i$ for the affected regions (and $Y_i=X_i=0$ otherwise) and $\sum^{n}_{i=1} Y_i=K$.  It may be natural to assume that the insurer distributes the available capital $K$  among the regions proportionally to their respective claims. This partial compensation ensures that each region receives a fair share of the available funds. In this case, the allocated capital (partial compensation) $Y_i$ for region $i$ is given by
\begin{equation}
    Y_i=\frac{X_i}{S} K  \quad \text{on} \quad \Omega_-.
    \label{capt}
\end{equation}
Gathering the three scenarios and Equations \eqref{benef} and \eqref{capt}
allows us to write the actual payment $Y_i$ as follows:

    \begin{equation}\label{eq:sharing}
    Y_i= [X_i + c_i(k - S)]\mathbf{1}_{\Omega_+} + X_i\mathbf{1}_{\Omega_=} + \frac{X_i}{S} K \mathbf{1}_{\Omega_-},
    \end{equation}
and the wealth of the insurer after the event is 
    \begin{equation}
       [k_0 + c_0(k - S)]\mathbf{1}_{\Omega_+} + (K-S)\mathbf{1}_{\Omega_=}.
    \end{equation}
\medskip
\begin{remark}
In case of  ex-ante risk sharing different than \eqref{benef} and \eqref{capt}, $Y_i$ can be written similarly for $i= 1, \cdots, n$
$$Y_i= [X_i + B_i]\mathbf{1}_{\Omega_+} + X_i\mathbf{1}_{\Omega_=} + K_i \mathbf{1}_{\Omega_-}$$
where the non-negative random variables $B_i$ and $K_i$ satisfy $\sum^{n}_{i=1} K_i=K$ and $\sum^{n}_{i=1} B_i \leq (k-S).$ 
Adopting other forms of $B_i$ and $K_i$ than the ones in \eqref{benef} and \eqref{capt} will not change the ex-post risk sharing.\\
\end{remark}

A similar setup for the design of fair insurance contracts in the presence of default risk of the insurance company has been studied in \cite{coculescu_fairness_2022}. Here, too, these three scenarios are considered, and fairness considerations are addressed not only at the initial stage of premium collection but also at the payout stage, even in scenarios where the insurer might default. Indeed, when a contract is exposed to the default risk of the insurance company, \cite{coculescu_fairness_2022} highlights that ex-ante equilibrium considerations require a certain participation in the benefit of the company to be specified in the contracts. Using cooperative game theory,  \cite{coculescu_fairness_2022} determines the fair contracts' design problem of a cost vector $(\pi_1, ..., \pi_n)$ and a payoff vector $(Y_0, \dots, Y_n)$, that is it investigates the optimal ex-ante risk transfer between the insurer and the insured agents. Despite the model setup commonalities for the insurer between \cite{coculescu_fairness_2022} and the present paper, we are interested here in optimal ex-post default risk sharing between the insured regions. Therefore, in our setting, the benefits of participation are considered as exogenous (even if the optimal fair benefit participation of \cite{coculescu_fairness_2022} could be considered as a special case), and we concentrate on determining the optimal ex-post taxation rule for sharing the residual losses between the regions.

\subsection{Ex-post risk sharing through Government intervention}\label{sec:gov_setup}
Once the ex-ante risk transfer mechanism between the insurer and the regions is fixed,  this section focuses on the government intervention strategy to guarantee regions full coverage in an optimal way, which is the main contribution of the paper. The government is a central coordinator to encourage high-risk and low-risk regions to participate in joint disaster risk mitigation. 

\begin{figure}[h!]
    \begin{center}
        \includegraphics[width=.8\linewidth]{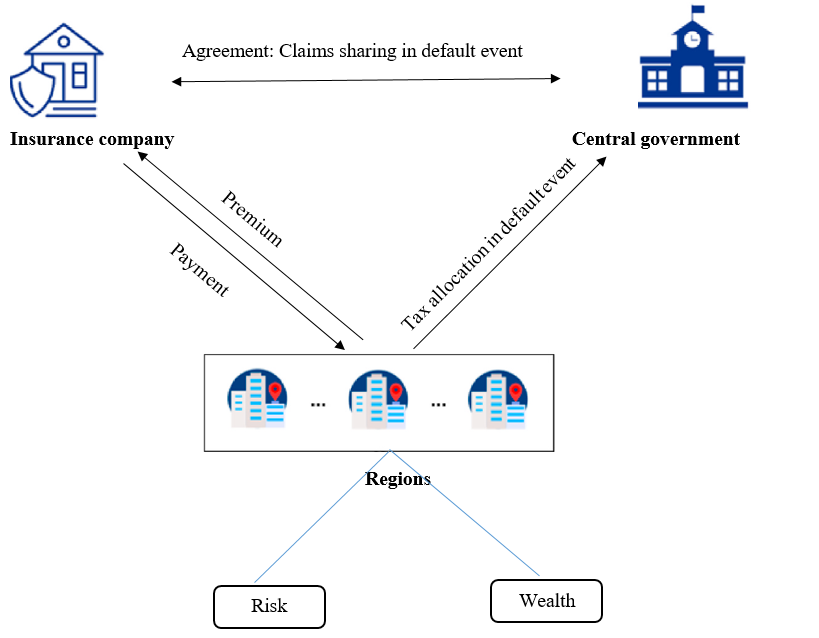}
        \caption{Public–private partnerships for multi-region catastrophe insurance (adapted from \cite{wang2023risk})}
        \label{fig:setup2_illustration}
    \end{center}
\end{figure}

\begin{figure}[h!]
    \centering
    \begin{subfigure}[h]{\textwidth}
        \centering
        \begin{tikzpicture}[scale=0.80]
\fill[BrickRed!20!white] (0,2) rectangle (0+2.5,3);
\fill[ForestGreen!20!white] (2.5,2) rectangle (2.5+2.5,3);
\fill[NavyBlue!20!white] (5,2) rectangle (5+2.5,3);
\fill[Goldenrod!20!white] (7.5,2) rectangle (7.5+2.5,3);
\fill[Plum!20!white] (10,2) rectangle (10+2.5,3);
\fill[Aquamarine!20!white] (12.5,2) rectangle (12.5+2.5,3);

\fill[BrickRed!50!white] (0,2) rectangle (0+2.5*.5,3);
\fill[ForestGreen!50!white] (2.5,2) rectangle (2.5+2.5*.6,3);
\fill[NavyBlue!50!white] (5,2) rectangle (5+2.5*.1,3);
\fill[Goldenrod!50!white] (7.5,2) rectangle (7.5+2.5*.7,3);
\fill[Plum!50!white] (10,2) rectangle (10+2.5*.4,3);
\fill[Aquamarine!50!white] (12.5,2) rectangle (12.5+2.5*.3,3);

\draw[black,  thick] (0,2) rectangle (0+2.5,3);
\draw[black,  thick] (2.5,2) rectangle (2.5+2.5,3);
\draw[black,  thick] (5,2) rectangle (5+2.5,3);
\draw[black,  thick] (7.5,2) rectangle (7.5+2.5,3);
\draw[black,  thick] (10,2) rectangle (10+2.5,3);
\draw[black,  thick] (12.5,2) rectangle (12.5+2.5,3);

\draw[BrickRed, <->, thick] (0,1.8)   -- (2.5*.5,1.8);
\node (A) at (2.5*.5/2,1.5) {\textcolor{BrickRed}{$X_1=w_1\tau_1$}};
\draw[ForestGreen, <->, thick] (2.5,1.8)   -- (2.5+2.5*.6,1.8);
\node (A) at (2.5+2.5*.6/2,1.5) {\textcolor{ForestGreen}{$X_2=w_2\tau_2$}};
\draw[NavyBlue, <->, thick] (5,1.8)   -- (5+2.5*.1,1.8);
\node (A) at (5.7+2.5*.1/2,1.5) {\textcolor{NavyBlue}{$X_3=w_3\tau_3$}};
\draw[Goldenrod, <->, thick] (7.5,1.8)   -- (7.5+2.5*.7,1.8);
\node (A) at (7.5+2.5*.7/2,1.5) {\textcolor{Goldenrod}{$X_4=w_4\tau_4$}};
\draw[Plum, <->, thick] (10,1.8)   -- (10+2.5*.4,1.8);
\node (A) at (10+2.5*.4/2,1.5) {\textcolor{Plum}{$X_5=w_5\tau_5$}};
\draw[Aquamarine, <->, thick] (12.5,1.8)   -- (12.5+2.5*.3,1.8);
\node (A) at (12.5+2.5*.3/2,1.5) {\textcolor{Aquamarine}{$X_6=w_6\tau_6$}};

\fill[BrickRed!50!white] (0,0) rectangle (0+2.5*.5,1);
\fill[ForestGreen!50!white] (2.5*.5,0) rectangle (2.5*.5+2.5*.6,1);
\fill[NavyBlue!50!white] (2.5*.5+2.5*.6,0) rectangle (2.5*.5+2.5*.6+2.5*.1,1);
\fill[Goldenrod!50!white] (2.5*.5+2.5*.6+2.5*.1,0) rectangle (2.5*.5+2.5*.6+2.5*.1+2.5*.7,1);
\fill[Plum!50!white] (2.5*.5+2.5*.6+2.5*.1+2.5*.7,0) rectangle (2.5*.5+2.5*.6+2.5*.1+2.5*.7+2.5*.4,1);
\fill[Aquamarine!50!white] (2.5*.5+2.5*.6+2.5*.1+2.5*.7+2.5*.4,0) rectangle (2.5*.5+2.5*.6+2.5*.1+2.5*.7+2.5*.4+2.5*.3,1);

\draw[black,  thick] (0,0) rectangle (15,1);
\draw[black,  thick] (0,0) rectangle (2.5*.5+2.5*.6+2.5*.1+2.5*.7+2.5*.4+2.5*.3,1);

\draw[pattern=north west lines] (5,0) rectangle (2.5*.5+2.5*.6+2.5*.1+2.5*.7+2.5*.4+2.5*.3,1);

\draw[black, <->, thick] (0,-.5)   -- (2.5*.5+2.5*.6+2.5*.1+2.5*.7+2.5*.4+2.5*.3,-.5);
\node (A) at (3.25,-.8) {\textcolor{black}{$S=X_1+\cdots+X_6$}};
\draw[black, - , thick] (5,0) -- (5,1);

\draw[black, - , thick] (4.2,0) -- (4.2,1);
\node (A) at (4.2,-.3) {\textcolor{black}{$k$}};


\draw[black, - , thick] (5,0) -- (5,1);
\node (A) at (5,-.3) {\textcolor{black}{$K$}};
\end{tikzpicture}
        \caption{Default scenarios ($\Omega_-$)}
        \label{fig:setup_illustration2}
    \end{subfigure}
    
    
    \begin{subfigure}[h]{\textwidth}
        \centering
         \begin{tikzpicture}[scale=0.80]
\fill[BrickRed!20!white] (0,2) rectangle (0+2.5,3);
\fill[ForestGreen!20!white] (2.5,2) rectangle (2.5+2.5,3);
\fill[NavyBlue!20!white] (5,2) rectangle (5+2.5,3);
\fill[Goldenrod!20!white] (7.5,2) rectangle (7.5+2.5,3);
\fill[Plum!20!white] (10,2) rectangle (10+2.5,3);
\fill[Aquamarine!20!white] (12.5,2) rectangle (12.5+2.5,3);

\fill[BrickRed!50!white] (0,2) rectangle (0+2.5*.5-.29,3);
\fill[ForestGreen!50!white] (2.5,2) rectangle (2.5+2.5*.6-.35,3);
\fill[NavyBlue!50!white] (5,2) rectangle (5+2.5*.1-.06,3);
\fill[Goldenrod!50!white] (7.5,2) rectangle (7.5+2.5*.7-.40,3);
\fill[Plum!50!white] (10,2) rectangle (10+2.5*.4-.23,3);
\fill[Aquamarine!50!white] (12.5,2) rectangle (12.5+2.5*.3-.17,3);

\fill[BrickRed!80!white] (0+2.5*.5,2) rectangle (0+2.5*.5-.29,3);
\fill[ForestGreen!80!white] (2.5+2.5*.6,2) rectangle (2.5+2.5*.6-.35,3);
\fill[NavyBlue!80!white] (5+2.5*.1,2) rectangle (5+2.5*.1-.06,3);
\fill[Goldenrod!80!white] (7.5+2.5*.7,2) rectangle (7.5+2.5*.7-.40,3);
\fill[Plum!80!white] (10+2.5*.4,2) rectangle (10+2.5*.4-.23,3);
\fill[Aquamarine!80!white] (12.5+2.5*.3,2) rectangle (12.5+2.5*.3-.17,3);

\draw[black,  thick] (0,2) rectangle (0+2.5,3);
\draw[black,  thick] (2.5,2) rectangle (2.5+2.5,3);
\draw[black,  thick] (5,2) rectangle (5+2.5,3);
\draw[black,  thick] (7.5,2) rectangle (7.5+2.5,3);
\draw[black,  thick] (10,2) rectangle (10+2.5,3);
\draw[black,  thick] (12.5,2) rectangle (12.5+2.5,3);

\draw[BrickRed, <->, thick] (0,1.8)   -- (2.5*.5,1.8);
\node (A) at (2.5*.5/2,1.5) {\textcolor{BrickRed}{$X_1=Y_1 +$}\textcolor{BrickRed!80!white}{$\varepsilon_1$}};
\draw[ForestGreen, <->, thick] (2.5,1.8)   -- (2.5+2.5*.6,1.8);
\node (A) at (2.5+2.5*.4,1.5) {\textcolor{ForestGreen}{$X_2=Y_2+ \varepsilon_2$}};

\fill[BrickRed!50!white] (0,0) rectangle (0+2.5*.5-.29,1);
\fill[ForestGreen!50!white] (2.5*.5-.29,0) rectangle (2.5*.5+2.5*.6-.64,1);
\fill[NavyBlue!50!white] (2.5*.5+2.5*.6-.64,0) rectangle (2.5*.5+2.5*.6+2.5*.1-.70,1);
\fill[Goldenrod!50!white] (2.5*.5+2.5*.6+2.5*.1-.70,0) rectangle (2.5*.5+2.5*.6+2.5*.1+2.5*.7-1.1,1);
\fill[Plum!50!white] (2.5*.5+2.5*.6+2.5*.1+2.5*.7-1.1,0) rectangle (2.5*.5+2.5*.6+2.5*.1+2.5*.7+2.5*.4-1.33,1);
\fill[Aquamarine!50!white] (2.5*.5+2.5*.6+2.5*.1+2.5*.7+2.5*.4-1.33,0) rectangle (2.5*.5+2.5*.6+2.5*.1+2.5*.7+2.5*.4+2.5*.3-1.5,1);

\fill[BrickRed!80!white] (5,0) rectangle (5+.29,1);
\fill[ForestGreen!80!white] (5+.29,0) rectangle (5+.64,1);
\fill[NavyBlue!80!white] (5+.64,0) rectangle (5+.7,1);
\fill[Goldenrod!80!white] (5+.7,0) rectangle (5+1.1,1);
\fill[Plum!80!white] (5+1.1,0) rectangle (5+1.33,1);
\fill[Aquamarine!80!white] (5+1.33,0) rectangle (5+1.5,1);

\draw[black,  thick] (0,0) rectangle (15,1);
\draw[black,  thick] (0,0) rectangle (2.5*.5+2.5*.6+2.5*.1+2.5*.7+2.5*.4+2.5*.3,1);

\draw[pattern=north west lines] (5,0) rectangle (2.5*.5+2.5*.6+2.5*.1+2.5*.7+2.5*.4+2.5*.3,1);

\draw[black, <->, thick] (0,-.2)   -- (5,-.2);
\node (A) at (2.5,-.5) {\textcolor{black}{$K=Y_1 + \dots Y_6$}};
\draw[black, <->, thick] (5.1,-.2)   -- (6.5,-.2);
\node (A) at (5.8,-.5) {\textcolor{black}{Taxation}};
\draw[black, - , thick] (5,0) -- (5,1);
\end{tikzpicture}
        \caption{Public-private risk-sharing}
        \label{fig:setup_illustration3}
    \end{subfigure}
    
    \caption{Illustration of default scenarios and public-private risk sharing}
    \label{fig:combined_scenario}
\end{figure}
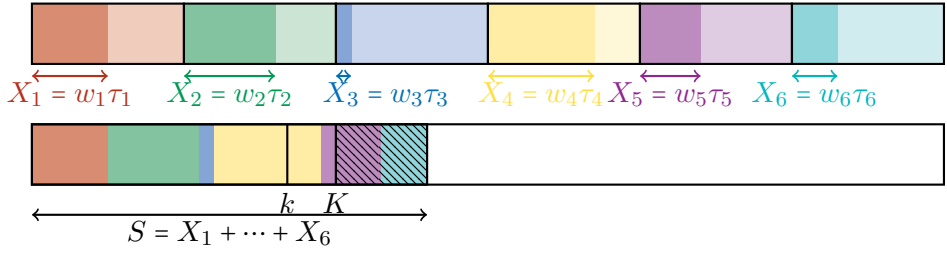
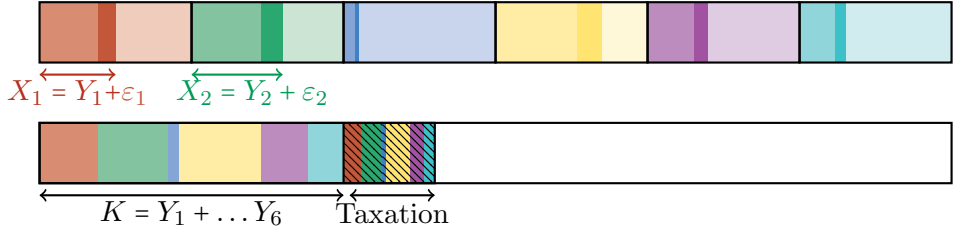

Figure \ref{fig:setup2_illustration} depicts public-private partnerships for multi-region catastrophe insurance. The government enters into a cooperation agreement with an insurance company to protect the regions against the risk of the insurer's default. This is achieved by distributing the residual claims among the regions through taxes. Figures \ref{fig:setup_illustration} to \ref{fig:combined_scenario} illustrate the intervention in the three types of scenarios. 
Figure \ref{fig:setup_illustration} depicts a favorable scenario ($\omega \in \Omega_+$) in which the total loss of wealth $S$ in the entire economy is lower than the total premium $k$ of the insurer. In this case, no government intervention is required, and the insurer reimburses all regional losses plus benefit sharing. An intermediate scenario ($\omega \in \Omega_=$) is represented in Figure \ref{fig:setup_illustration1bis} in which the total loss of wealth $S$ is greater than the total premium $k$ but the insurer remains solvent. Indeed, the residual part $(S-k)$, represented by the dotted part in Figure \ref{fig:setup_illustration1bis}, is then taken from its initial capital $k_0$ to fully reimburse the regions. Figure \ref{fig:combined_scenario} represents a default scenario ($\omega \in \Omega_-$)  where the total wealth loss is greater than the total capital of the insurer ($S > K$). The exceeding part $(S-K)$, denoted by the hatched area in Figure \ref{fig:combined_scenario}, corresponds to the residual claims.

\begin{figure}[h!]
    \centering
    \begin{tikzpicture}[scale=0.80]
\fill[BrickRed!20!white] (0,2) rectangle (0+2.5,3);
\fill[ForestGreen!20!white] (2.5,2) rectangle (2.5+2.5,3);
\fill[NavyBlue!20!white] (5,2) rectangle (5+2.5,3);
\fill[Goldenrod!20!white] (7.5,2) rectangle (7.5+2.5,3);
\fill[Plum!20!white] (10,2) rectangle (10+2.5,3);
\fill[Aquamarine!20!white] (12.5,2) rectangle (12.5+2.5,3);

\fill[BrickRed!50!white] (0,2) rectangle (0+2.5*.3,3);
\fill[ForestGreen!50!white] (2.5,2) rectangle (2.5+2.5*.1,3);
\fill[NavyBlue!50!white] (5,2) rectangle (5+2.5*.1,3);
\fill[Goldenrod!50!white] (7.5,2) rectangle (7.5+2.5*.2,3);
\fill[Plum!50!white] (10,2) rectangle (10+2.5*.1,3);
\fill[Aquamarine!50!white] (12.5,2) rectangle (12.5+2.5*.3,3);

\draw[black,  thick] (0,2) rectangle (0+2.5,3);
\draw[black,  thick] (2.5,2) rectangle (2.5+2.5,3);
\draw[black,  thick] (5,2) rectangle (5+2.5,3);
\draw[black,  thick] (7.5,2) rectangle (7.5+2.5,3);
\draw[black,  thick] (10,2) rectangle (10+2.5,3);
\draw[black,  thick] (12.5,2) rectangle (12.5+2.5,3);

\draw[BrickRed, <->, thick] (0,1.8)   -- (2.5*.3,1.8);
\node (A) at (2.5*.3/2,1.5) {\textcolor{BrickRed}{$X_1=w_1\tau_1$}};
\draw[ForestGreen, <->, thick] (2.5,1.8)   -- (2.5+2.5*.1,1.8);
\node (A) at (2.5+2.5*.1/2,1.5) {\textcolor{ForestGreen}{$X_2=w_2\tau_2$}};
\draw[NavyBlue, <->, thick] (5,1.8)   -- (5+2.5*.1,1.8);
\node (A) at (5+2.5*.1/2,1.5) {\textcolor{NavyBlue}{$X_3=w_3\tau_3$}};
\draw[Goldenrod, <->, thick] (7.5,1.8)   -- (7.5+2.5*.2,1.8);
\node (A) at (7.5+2.5*.2/2,1.5) {\textcolor{Goldenrod}{$X_4=w_4\tau_4$}};
\draw[Plum, <->, thick] (10,1.8)   -- (10+2.5*.1,1.8);
\node (A) at (10+2.5*.1/2,1.5) {\textcolor{Plum}{$X_5=w_5\tau_5$}};
\draw[Aquamarine, <->, thick] (12.5,1.8)   -- (12.5+2.5*.3,1.8);
\node (A) at (12.5+2.5*.3/2,1.5) {\textcolor{Aquamarine}{$X_6=w_6\tau_6$}};

\fill[BrickRed!50!white] (0,0) rectangle (0+2.5*.3,1);
\fill[ForestGreen!50!white] (2.5*.3,0) rectangle (2.5*.3+2.5*.1,1);
\fill[NavyBlue!50!white] (2.5*.3+2.5*.1,0) rectangle (2.5*.3+2.5*.1+2.5*.1,1);
\fill[Goldenrod!50!white] (2.5*.3+2.5*.1+2.5*.1,0) rectangle (2.5*.3+2.5*.1+2.5*.1+2.5*.2,1);
\fill[Plum!50!white] (2.5*.3+2.5*.1+2.5*.1+2.5*.2,0) rectangle (2.5*.3+2.5*.1+2.5*.1+2.5*.2+2.5*.1,1);
\fill[Aquamarine!50!white] (2.5*.3+2.5*.1+2.5*.1+2.5*.2+2.5*.1,0) rectangle (2.5*.3+2.5*.1+2.5*.1+2.5*.2+2.5*.1+2.5*.3,1);

\draw[black,  thick] (0,0) rectangle (15,1);
\draw[black,  thick] (0,0) rectangle (2.5*.3+2.5*.1+2.5*.1+2.5*.2+2.5*.1+2.5*.3,1);

\draw[black, <->, thick] (0,-.2)   -- (2.5*.3+2.5*.1+2.5*.1+2.5*.2+2.5*.1+2.5*.3,-.2);
\node (A) at (1.375,-.5) {\textcolor{black}{$S=X_1+\cdots+X_6$}};

\draw[black, - , thick] (4.2,0) -- (4.2,1);
\node (A) at (4.2,-.3) {\textcolor{black}{$k$}};

\draw[black, <->, thick] (2.5*.3+2.5*.1+2.5*.1+2.5*.2+2.5*.1+2.5*.3+0.1,-.2)   -- (4.1,-.2);
\node (A) at (3.5,-.95) {\textcolor{black}{Surplus}};

\draw[black, - , thick] (5,0) -- (5,1);
\node (A) at (5,-.3) {\textcolor{black}{$K$}};

\end{tikzpicture}
    \caption{Favorable scenarios ($\Omega_+$)}
    \label{fig:setup_illustration}
\end{figure}
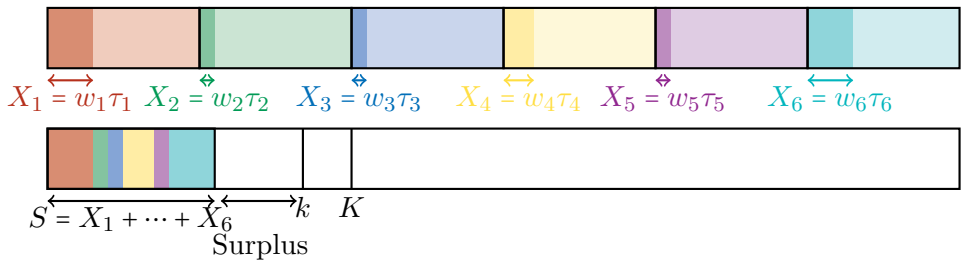

\begin{figure}[h!]
    \centering
    \begin{tikzpicture}[scale=0.80]
\fill[BrickRed!20!white] (0,2) rectangle (0+2.5,3);
\fill[ForestGreen!20!white] (2.5,2) rectangle (2.5+2.5,3);
\fill[NavyBlue!20!white] (5,2) rectangle (5+2.5,3);
\fill[Goldenrod!20!white] (7.5,2) rectangle (7.5+2.5,3);
\fill[Plum!20!white] (10,2) rectangle (10+2.5,3);
\fill[Aquamarine!20!white] (12.5,2) rectangle (12.5+2.5,3);

\fill[BrickRed!50!white] (0,2) rectangle (0+2.5*.5,3);
\fill[ForestGreen!50!white] (2.5,2) rectangle (2.5+2.5*.3,3);
\fill[NavyBlue!50!white] (5,2) rectangle (5+2.5*.1,3);
\fill[Goldenrod!50!white] (7.5,2) rectangle (7.5+2.5*.4,3);
\fill[Plum!50!white] (10,2) rectangle (10+2.5*.5,3);
\fill[Aquamarine!50!white] (12.5,2) rectangle (12.5+2.5*.1,3);


\draw[black,  thick] (0,2) rectangle (0+2.5,3);
\draw[black,  thick] (2.5,2) rectangle (2.5+2.5,3);
\draw[black,  thick] (5,2) rectangle (5+2.5,3);
\draw[black,  thick] (7.5,2) rectangle (7.5+2.5,3);
\draw[black,  thick] (10,2) rectangle (10+2.5,3);
\draw[black,  thick] (12.5,2) rectangle (12.5+2.5,3);

\draw[BrickRed, <->, thick] (0,1.8)   -- (2.5*.5,1.8);
\node (A) at (2.5*.3/2,1.5) {\textcolor{BrickRed}{$X_1=w_1\tau_1$}};
\draw[ForestGreen, <->, thick] (2.5,1.8)   -- (2.5+2.5*.3,1.8);
\node (A) at (2.5+2.5*.1/2,1.5) {\textcolor{ForestGreen}{$X_2=w_2\tau_2$}};
\draw[NavyBlue, <->, thick] (5,1.8)   -- (5+2.5*.1,1.8);
\node (A) at (5+2.5*.1/2,1.5) {\textcolor{NavyBlue}{$X_3=w_3\tau_3$}};
\draw[Goldenrod, <->, thick] (7.5,1.8)   -- (7.5+2.5*.4,1.8);
\node (A) at (7.5+2.5*.2/2,1.5) {\textcolor{Goldenrod}{$X_4=w_4\tau_4$}};
\draw[Plum, <->, thick] (10,1.8)   -- (10+2.5*.5,1.8);
\node (A) at (10+2.5*.1/2,1.5) {\textcolor{Plum}{$X_5=w_5\tau_5$}};
\draw[Aquamarine, <->, thick] (12.5,1.8)   -- (12.5+2.5*.1,1.8);
\node (A) at (12.5+2.5*.3/2,1.5) {\textcolor{Aquamarine}{$X_6=w_6\tau_6$}};

\fill[BrickRed!50!white] (0,0) rectangle (0+2.5*.5,1);
\fill[ForestGreen!50!white] (2.5*.5,0) rectangle (2.5*.5+2.5*.3,1);
\fill[NavyBlue!50!white] (2.5*.5+2.5*.3,0) rectangle (2.5*.5+2.5*.3+2.5*.1,1);
\fill[Goldenrod!50!white] (2.5*.5+2.5*.3+2.5*.1,0) rectangle (2.5*.5+2.5*.3+2.5*.1+2.5*.4,1);
\fill[Plum!50!white] (2.5*.5+2.5*.3+2.5*.1+2.5*.4,0) rectangle (2.5*.5+2.5*.3+2.5*.1+2.5*.4+2.5*.5,1);
\fill[Aquamarine!50!white] (2.5*.5+2.5*.3+2.5*.1+2.5*.4+2.5*.5,0) rectangle (2.5*.5+2.5*.3+2.5*.1+2.5*.4+2.5*.5+2.5*.1,1);

\draw[black,  thick] (0,0) rectangle (15,1);
\draw[black,  thick] (0,0) rectangle (2.5*.5+2.5*.3+2.5*.1+2.5*.4+2.5*.5+2.5*.1,1);

\draw[pattern=crosshatch dots] (2.5*.5+2.5*.3+2.5*.1+2.5*.4+2.5*.5+2.5*.1,0) rectangle (4.2,1);

\draw[black, <->, thick] (0,-.5)   -- (2.5*.5+2.5*.3+2.5*.1+2.5*.4+2.5*.5+2.5*.1,-.5);
\node (A) at (2.0,-.8) {\textcolor{black}{$S=X_1+\cdots+X_6$}};

\draw[black, - , thick] (4.2,0) -- (4.2,1);
\node (A) at (4.2,-.3) {\textcolor{black}{$k$}};


\draw[black, - , thick] (5,0) -- (5,1);
\node (A) at (5,-.3) {\textcolor{black}{$K$}};

\end{tikzpicture}
    \caption{Intermediate scenarios ($\Omega_=$)}
    \label{fig:setup_illustration1bis}
\end{figure}
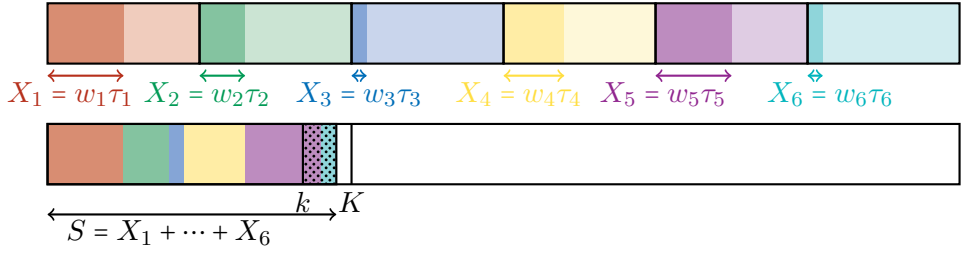

Such a residual risk-sharing mechanism is particularly pertinent given rising disaster losses and the financial stress on public insurers, as highlighted in the U.S. flood insurance context by \cite{michel2011redesigning}. Their recommendation for a hybrid approach, pairing private sector risk pricing with public guarantees and post-event financing, motivates our hierarchical structure involving both insurers and tax-funded government backstops.
Figure \ref{fig:setup_illustration3} illustrates the public-private risk-sharing system in this case of default. The residual claims of the region $i$ are the difference between the loss amount $X_i$ and the compensation $Y_i=\frac{X_i}{S} K$ provided by the insurance, i.e. 
\begin{equation}
    \varepsilon_{i}= (X_i - \frac{X_i}{S} K)^+ \quad \text{on} \quad \Omega_-.
    \label{res}
\end{equation}
The central government sets up a pool for the regions to share the residual claims, which are given by
\begin{equation}
   \boldsymbol{\varepsilon} = (\varepsilon_1, \varepsilon_2, \dots,\varepsilon_n).
\end{equation}
The total residual claims is denoted by
\begin{equation}
    S_{\varepsilon}= \sum^{n}_{i=1} \varepsilon_i= S - K.
\end{equation}
The government then asks all regions to pay an additional tax to cover the residual claims $\boldsymbol{\varepsilon}$.
Hence, combining the ex-ante risk transfer and the ex-post  risk-sharing, the final payoff of region $i$ is given by:
\begin{align}\label{eq:final_formula}
     [X_i + c_i(k - S)]\mathbf{1}_{\Omega_+} + X_i\mathbf{1}_{\Omega_=} + (X_i - T_i(\boldsymbol{\varepsilon})) \mathbf{1}_{\Omega_-} \nonumber\\
    = X_i + c_i(k - S) \mathbf{1}_{\Omega_+} - T_i(\boldsymbol{\varepsilon}) \mathbf{1}_{\Omega_-} 
\end{align}
where $T_i(\boldsymbol{\varepsilon})$ is the tax contribution of region $i$ for the residual claims. The main contribution of the paper is to determine optimally (in a sense that will be precise hereafter) the taxation level  $T_i(\boldsymbol{\varepsilon})$ of each region.  In our framework, the government tax is risk-based, meaning that the tax contribution of each region $i$ depends on its disaster risk $\tau_i$ and its wealth level $w_i$. 
       
\section{Optimal taxation rule \label{sect3}}
In a disaster, the government provides an unlimited guarantee to cover each region's losses, financed through taxation. 
The problem of allocating disaster losses across regions through taxation resembles the optimal transfer problem in fiscal unions studied by \cite{farhi2017fiscal}. In their model, interregional transfers serve to smooth idiosyncratic shocks and maximize aggregate utility, subject to incentive and budget constraints. We adopt a similar welfare-maximization approach to the context of disaster financing, where taxation is used ex-post to share residual losses across regions with heterogeneous risk aversions.
The total loss of region $i$ is the sum of the premium  $\pi_i$ paid at the time 0, and its tax contribution to the residual claims $\boldsymbol{\varepsilon}$, denoted by $T_i(\boldsymbol{\varepsilon})$, taking values in $[0,w_i - \pi_i]$, and paid at the end of the period in case of a default scenario.
Therefore the final wealth of region $i$ is given by $$w_i-\pi_i - T_i(\boldsymbol{\varepsilon}) \mathbf{1}_{\Omega_-} + c_i (S-k)^+\mathbf{1}_{\Omega_+}.$$ 
Each region's risk aversion is modeled by a utility function $u_i$, which ranks its risky outcome. 
The expected utility of region $i$ is the expected value of $u_i$ evaluated at $w_i-\pi_i - T_i(\boldsymbol{\varepsilon}) \mathbf{1}_{\Omega_-} + c_i (S-k)^+ \mathbf{1}_{\Omega_+}$. 
This paper aims to determine the level of taxation $T_i(\boldsymbol{\varepsilon})$ for each region $i$, that is Pareto optimal and satisfying a fairness constraint, as specified in the forthcoming Definitions \ref{deffairness} and \ref{def23}. To focus on the taxation variable $T_i(\boldsymbol{\varepsilon})$ that has to be minimized to achieve Pareto optimality,  the optimization problem for region $i$ is formulated using the strictly increasing and convex random disutility functions $v_i$ that have to be minimized:
 \begin{equation}\label{valuefunctioni}
 v_i(y,\omega):=-u_i\left(w_i-\pi_i + c_i (S-k)^+ \mathbf{1}_{\omega \in \Omega_+} - y \mathbf{1}_{\omega \in \Omega_-}\right),
 \end{equation}
 or equivalently
$$
 v_i(y,\omega)=-u_i\left(w_i-\pi_i\right)\mathbf{1}_{\omega \in \Omega_=}-u_i\left(w_i-\pi_i + c_i (S-k)^+\right)\mathbf{1}_{\omega \in \Omega_+}-u_i\left(w_i-\pi_i - y \right)\mathbf{1}_{\omega \in \Omega_-}.
$$
 Hence, the disutility of region $i$ with taxation rule $T_i$ is given by $v_i(T_i(\boldsymbol{\varepsilon}), \omega)$. On $\Omega_-$, which are the scenarios of interest in which taxation occurs, the disutility function $v_i(y, \omega)=-u_i\left(w_i-\pi_i - y \right)$ is defined on $[0, w_i-\pi_i] \mapsto(-\infty,+\infty)$. To shorten the notation, we will now omit the dependency in $\omega$. We will assume here that there is no participation cost. If participation in the taxation-based backstop involves administrative or political costs, the individual rationality constraint must be tightened. Regions will only join if their expected utility net of entry costs exceeds autarky, possibly leading to incomplete coalition formation and lower overall risk sharing.\\
Furthermore,  we will assume the  following  hypothesis on the utility $u_i$:
\begin{assumption}
For each $i \in \{1, \dots, n\}$, the function $u_i$ is strictly increasing, strictly concave,  continuously differentiable, and satisfies the  following condition
\begin{equation}
 u^\prime_i(w_i-\pi_i) > 0. 
\end{equation}
\label{H1}
\end{assumption}
\noindent The condition $u^\prime_i(w_i-\pi_i) > 0$ is equivalent to $v^\prime_i(0) > 0$.
In the literature, it is common practice to assume Inada conditions to ensure that the marginal utility function has a well-defined reciprocal bijection over the entire domain. However, when introducing a shift such as $(w_i - \pi_i)$, applying an Inada condition to this shift becomes restrictive. If Inada conditions are imposed, the methods presented in \cite{niakh2024fixed} can be applied. Here we aim to adopt a more general approach. The condition in Assumption \ref{H1} allows for a broad class of utility functions, including the commonly used exponential and power functions, while avoiding the need to impose a specific condition on $w_i - \pi_i$. The condition in Assumption \ref{H1} is needed to ensure the existence of the taxation rule studied in this paper. The inverse marginal disutility of region $i$ is denoted by $I_i(.):=(v_i^{\prime})^{-1}(.)$ defined on $ [v^\prime_i(0), v^\prime_i(w_i-\pi_i)]$ and with values in $ [0, w_i-\pi_i]$.\\

It is important to note that the maximum total loss is bounded above by the sum of all regions' wealth $\sum_{i=1}^{n}w_i$, while $K=k+k_0$ is the total capital of the insurance company. Therefore, the support of the total residual claims $S_\varepsilon$ is given by the interval
\begin{equation}
     A:=\left[0, \sum_{i=1}^{n}w_i- K \right].
    \label{formule24}
\end{equation}
The government aims to maximize social welfare, which is the sum of the expected utilities of all regions, by imposing catastrophe taxes on all regions exposed to catastrophe risk. This objective is consistent with the literature on optimal fiscal transfers, such as \cite{boadway1996efficiency}, which shows that efficiency gains can be achieved when redistribution is aligned with risk-sharing needs.\\

\begin{definition}[Risk-sharing rule]
A risk-sharing rule $\mathbf{h}$ is a mapping  $$\mathbf{h}: \left(L^{\infty}_{+}(\Omega, \mathcal{F}, \mathbb{P})\right)^n \to \left(L^{\infty}_{+}(\Omega, \mathcal{F}, \mathbb{P})\right)^n$$ which transforms any pool $\boldsymbol{\varepsilon} = (\varepsilon_1, \varepsilon_2, \dots,\varepsilon_n)$ into another random vector $\mathbf{h}(\boldsymbol{\varepsilon})=\left(h_1(\boldsymbol{\varepsilon}), h_2(\boldsymbol{\varepsilon}), \dots, h_n(\boldsymbol{\varepsilon})\right)$ where the deterministic functions $h_i$ are such that
\begin{equation}
    \sum_{i=1}^{n} h_{i}(\boldsymbol{\varepsilon})=S_\varepsilon,~\text{a.s.}.
    \label{def1}
\end{equation}
\end{definition}
\noindent An important subclass of risk-sharing rules consists of \emph{aggregate risk-sharing rule} as called by \cite{denuit_risk-sharing_2022} or  \emph{`Non-olet' risk-sharing rule} as called by \cite{feng_unified_2022}, namely rules that are functions only  of the total residual claims $S_\varepsilon$: 
$$\left(h_1(\boldsymbol{\varepsilon}), h_2(\boldsymbol{\varepsilon}), \dots, h_n(\boldsymbol{\varepsilon})\right)=\left(g_1(S_\varepsilon), g_2(S_\varepsilon), \dots, g_n(S_\varepsilon)\right),~\text{a.s.},$$
for some functions $g_1, \dots, g_n:$ $L^{\infty}_{+}(\Omega, \mathcal{F}, \mathbb{P}) \to L^{\infty}_{+}(\Omega, \mathcal{F}, \mathbb{P})$.  In the sequel, we will focus exclusively on \emph{aggregate} or \emph{'non-olet'} risk-sharing rules. This is motivated by the fact   that under mild assumptions, participants' optimal risk-sharing depends only on  the residual aggregate loss $S_\varepsilon$, as   established  by \cite{borch_equilibrium_1962}.\bigskip

\begin{definition}[Taxation rule]
A taxation rule $\mathbf{T}$ is an aggregate risk-sharing rule for which we impose the following constraints :
\begin{itemize}
    \item (Principle of Indemnity) No one should profit from others’ losses, i.e. $T_i(S_\varepsilon) \geq 0$,~\text{a.s.},  for all $i=1, \dots n.$
    \item (Limited Liability) Each region is held accountable up to some tax-bearing capacity, i.e., $T_i(S_\varepsilon) \leq w_i-\pi_i$,~\text{a.s.}, for all $i=1, \dots n.$
\end{itemize}
\end{definition}

\begin{definition}[Actuarially fair taxation rules]\label{deffairness}
    A taxation rule $\mathbf{T}$ for the total residual claims $S_\varepsilon$ is \textbf{actuarially fair} if participants do neither gain nor lose from collective risk-sharing through taxation, in the sense that  their expected tax is equal to their expected loss when the insurer's default, i.e.,
            \begin{equation}
                \mathbb{E}\left[T_{i}(S_\varepsilon)\right]=\mathbb{E}[\varepsilon_i] \quad\forall \ i \ \in \{1, \dots, n \}.
                \label{formule32}
            \end{equation}
\end{definition}

\begin{definition}[Pareto Optimal taxation rules]
        A taxation rule $\left(T_{1}, \ldots, T_{n}\right)$ is\textbf{ Pareto optimal} for $S_\varepsilon$ if there does not exist a taxation rule $\left(\tilde{T}_{1}, \ldots, \tilde{T}_{n}\right)$ such that 
        \begin{align*}
        & 
        \mathbb{E}\left[v_i(\tilde{T}_i(S_\varepsilon))\right] \leq \mathbb{E}\left[v_i(T_i(S_\varepsilon))\right]
        \ \forall \ i =1,\cdots, n  \\
        & \text{and} \ \exists \ j \quad  \text{s.t. } \quad
        \mathbb{E}\left[v_j(\tilde{T}_j(S_\varepsilon))\right] < \mathbb{E}\left[v_j(T_j(S_\varepsilon))\right].
        \end{align*} 
\label{def23}
\end{definition}

\begin{definition}[Actuarially Fair Pareto Optimal taxation rules]
    An actuarially fair Pareto optimal (AFPO) taxation rule is a Pareto optimal taxation rule that, in addition, satisfies the actuarially fairness condition \eqref{formule32}.
\end{definition}
\medskip
This approach resonates with the cooperative framework introduced by \cite{buhlmann1980economic}, in which insurance contracts and risk-sharing rules are determined by maximizing joint utility under individual rationality constraints. In our context, this manifests through a taxation rule that is both Pareto optimal and actuarially fair, extending Bühlmann’s principles to multi-agent public-private disaster financing. Such AFPO taxation rule implies internal stability (in the sense that no participant has an incentive to unilaterally leave the risk-sharing agreement, because they would do worse on their own) of the coalition asunder reasonable assumptions (individual rationality and risk-averse agents with quasi-concave utilities), as in \cite{buhlmann1980economic}.
Define by $\Delta^n_+:=\{\boldsymbol{a} \in \mathbb{R}^n_{+,*}: \sum^{n}_{i=1}a_i=1\}$ the open unit simplex. Pareto optimal taxation rules can be characterized as follows, using the Kuhn-Tucker Theorem (see also \cite{Buhlmann1978,buhlmann1980economic}). For the sake of completeness, the proof is given in Appendix \ref{secA1}.\\

\begin{theorem}
    A taxation rule $\mathbf{T}=\left(T_1, \cdots, T_n\right)$ is Pareto optimal for any given total residual risk $S_\varepsilon$ taking values in the domain $A$ if and only if there exists $(\alpha_{1}, \ldots, \alpha_{n}) \in \Delta^n_+$  and a function $\Lambda: A \rightarrow \mathbb{R}_{+}$ such that
        $$
        \begin{array}{ll}
        \alpha_i v_i^{\prime}\left(T_i(s_\varepsilon)\right)=\Lambda(s_\varepsilon) & \text { if } 0<T_i(s_\varepsilon)<w_i-\pi_i \\
        \alpha_i v_i^{\prime}\left(T_i(s_\varepsilon)\right) \geq \Lambda(s_\varepsilon) & \text { if } T_i(s_\varepsilon)=0 \\
        \alpha_i v_i^{\prime}\left(T_i(s_\varepsilon)\right) \leq \Lambda(s_\varepsilon) & \text { if } T_i(s_\varepsilon)=w_i-\pi_i.
        \end{array}
        $$
        \label{thm2}
\end{theorem}
\noindent According to \cite{gerber1978pareto} for a given vector of weights $(\alpha_{1}, \ldots, \alpha_{n}) \in \Delta^n_+$, there is exactly one taxation rule that is Pareto optimal.  $(\alpha_{1}, \ldots, \alpha_{n})$ indicates the relative weights of each region in the collective risk-sharing taxation: the larger the weight $\alpha_i$, the smaller the contribution $T_i$ (all other things being equal). Note that $\Lambda(s_\varepsilon)$ is an intermediate function pertaining to taxes $T_i(s_\varepsilon)$. One shall first set a fixed value $\Lambda(s_\varepsilon)=\lambda$, which works as a parameter that determines the Pareto optimal taxation rule. For any $(\alpha_{1}, \ldots, \alpha_{n})$, Theorem \ref{thm2} allows us to easily visualize and compute the shape of the Pareto optimal taxation rules $T_i(s_\varepsilon)$. 

\begin{figure}
\centering
\begin{subfigure}{.32\textwidth}
  \centering
  \includegraphics[width=\linewidth]{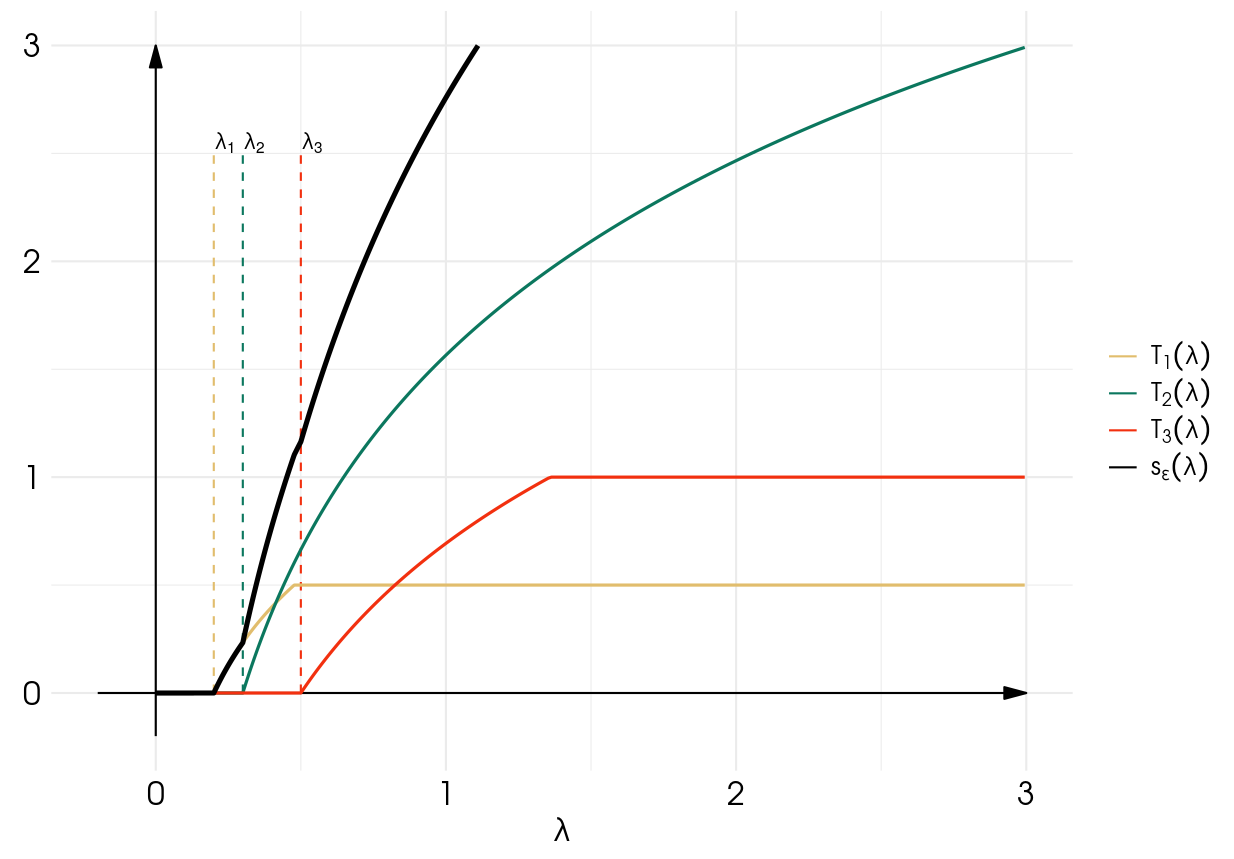}
  \caption{$T_i(\lambda)$ and $s_{\varepsilon}(\lambda)$}
  \label{fig:sub1}
\end{subfigure}%
\begin{subfigure}{.32\textwidth}
  \centering
  \includegraphics[width=\linewidth]{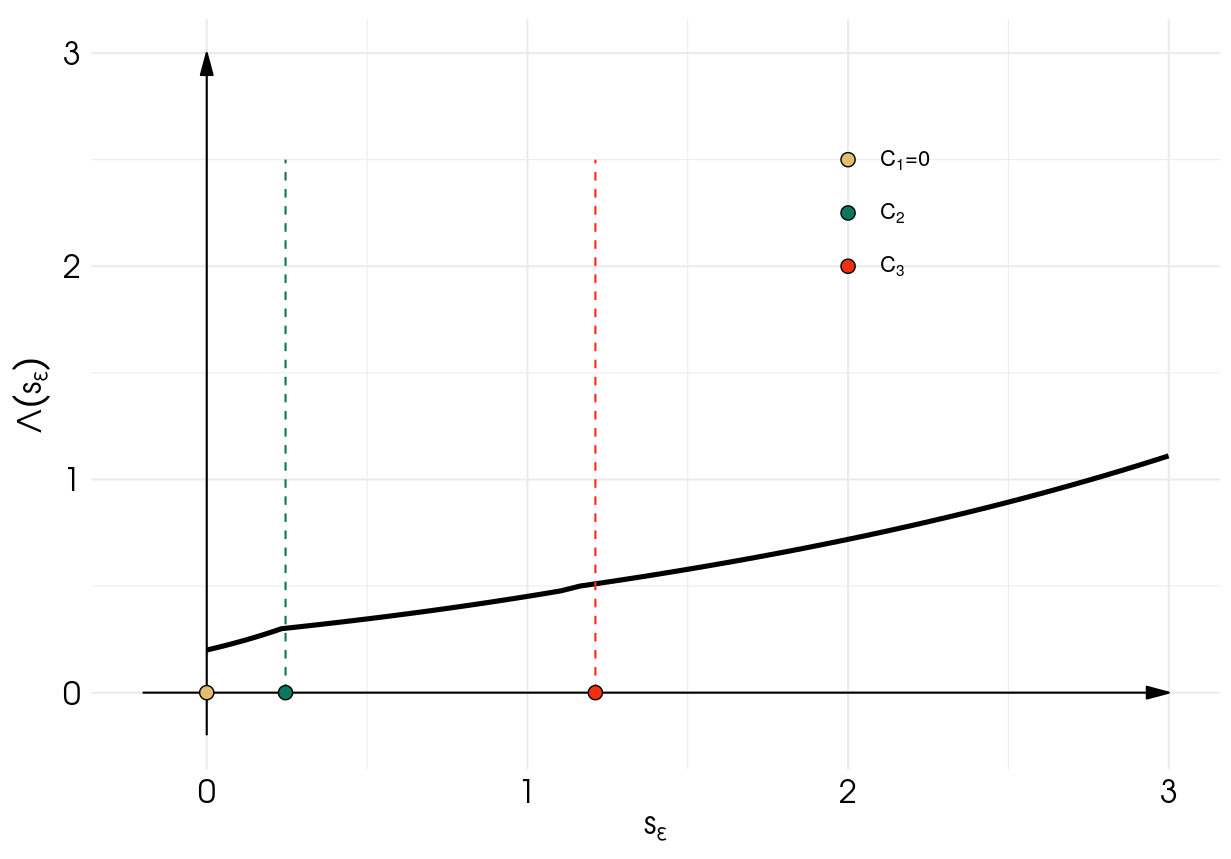}
  \caption{$\Lambda(s_{\varepsilon})$}
  \label{fig:sub2}
\end{subfigure}%
\begin{subfigure}{.32\textwidth}
  \centering
  \includegraphics[width=\linewidth]{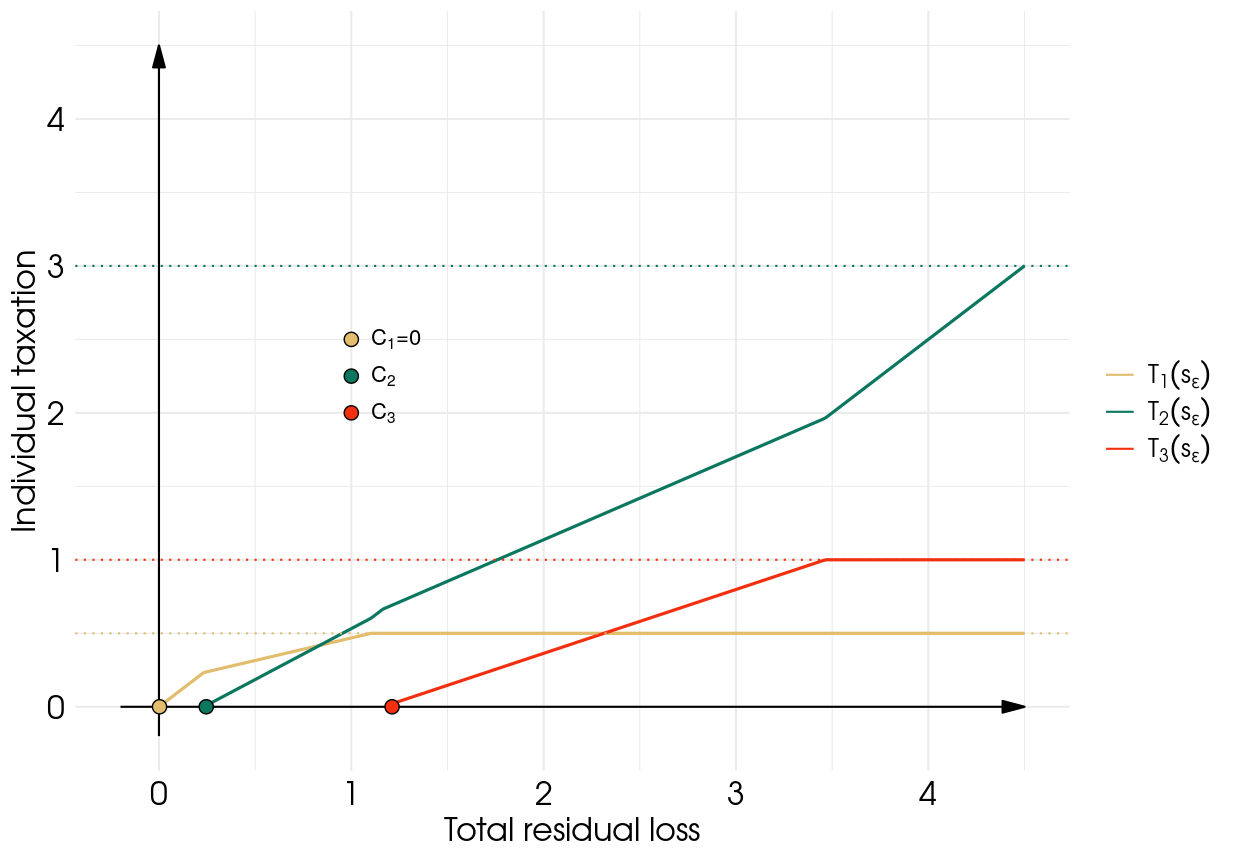}
  \caption{Individual contributions}
  \label{fig:sub3}
\end{subfigure}
\caption{Visualization of the sharing mechanism with three participants with different levels of wealth, risk, and disutility functions. A region can not share more than the total value of its assets, which will result in a ``{\em flatlining}'' of the sharing capabilities once the point is reached. The left pane illustrates the taxation $T_i(\lambda)$ and the total residual losses $s_{\varepsilon}(\lambda)$ as a function of the parameter $\lambda$. Here, the red participant is joining the sharing mechanism later (and therefore only for larger losses). Center pane, invert the curve of $\lambda \rightarrow s_\varepsilon(\lambda)$ to have $s_\varepsilon  \rightarrow \Lambda(s_\varepsilon)$. Right pane, taxation for individual regions, based on the same set of parameters as in the left pane, with maximum wealth indicated on $y$-axis. Note that the levels $C_1,C_2,C_3$ on the right pane correspond to the intersections of $\lambda_1, \lambda_2, \lambda_3$ with $\Lambda(s_\varepsilon)$ projected on the $x$-axis in the left pane}
\label{fig:test}
\end{figure}

To analyze how these taxes are distributed, let us first focus on each region $i$,  characterized by the parameters $\lambda_i=\alpha_iv_i^{\prime}\left(0\right)$ and $\lambda^{\max}_i=\alpha_iv_i^{\prime}\left(w_i-\pi_i\right)$. The position of $\lambda=\Lambda(s_\varepsilon)$ with respect to $\lambda_i$ and $\lambda^{\max}_i$ determines the contribution $T_i = \mathcal{T}_i(\lambda)$ of  region $i$:  $\lambda \leq \lambda_i$ implies $T_i=0$, $\lambda_i < \lambda < \lambda^{\max}_i$ implies $\alpha_i v_i^{\prime}\left(T_i \right) = \lambda$ (and thus $0 < T_i < w_i-\pi_i$), finally  $\lambda \geq \lambda^{\max}_i$ implies $T_i= w_i-\pi_i$. Without loss of generality, the regions are numbered in ascending order of their parameters $\lambda_i$, that is, $\lambda_1 \leq \dots \leq \lambda_n$. This ordering determines the sequence in which regions start to contribute to the residual claims as $\Lambda(s_\varepsilon) = \lambda$ increases, while 
 the position of  $\lambda^{\max}_i$ determines when the region $i$ reaches its maximum contribution. This is illustrated in Figure \ref{fig:sub1}. Then, by parametrically increasing $\lambda$, we pass through different layers of taxation in which regions successively participate, while some of them progressively reach their maximum contribution.   
Furthermore, since at each level $\lambda$ the values of the $T_i=\mathcal{T}_i(\lambda)$  are given explicitly,  the total residual claims can be computed as $s_\varepsilon=\sum_{i=1}^{n}\mathcal{T}_i(\lambda)$ by ``horizontal addition". The heavy black line in Figure \ref{fig:sub1} represents the resulting parametric curve of $\lambda \rightarrow s_\varepsilon(\lambda)$,  inverting it provides then $s_\varepsilon  \rightarrow \Lambda(s_\varepsilon)$ in Figure \ref{fig:sub2}.  Finally, plugging this value $\Lambda(s_\varepsilon)$ in  $\mathcal{T}_i(\lambda)$ gives the optimal taxation $T_i(s_\varepsilon)$ for each region, as shown in  Figure \ref{fig:sub3}.
The layering constants $C_i:=s_\varepsilon(\lambda_i)$ determine at which residual claims value region $i$ starts to participate.

Adding the fairness constraint \eqref{eq:fairness} on the Pareto optimal rules leads to the following characterization of AFPO taxation rules.\\

\begin{corollary}[Characterization of AFPO taxation rules]

A taxation rule $\mathbf{T}=\left(T_1, \cdots, T_n\right)$ is AFPO if and only if there exists $(\alpha_{1}, \ldots, \alpha_{n}) \ \in \Delta^n_+$ and a function $\Lambda: A \rightarrow \mathbb{R}_{+}$ such that 
        $$
        \begin{array}{ll}
        \alpha_i v_i^{\prime}\left(T_i(s_\varepsilon)\right)=\Lambda(s_\varepsilon) & \text { if } 0<T_i(s_\varepsilon)<w_i-\pi_i \\
        \alpha_i v_i^{\prime}\left(T_i(s_\varepsilon)\right) \geq \Lambda(s_\varepsilon) & \text { if } T_i(s_\varepsilon)=0 \\
        \alpha_i v_i^{\prime}\left(T_i(s_\varepsilon)\right) \leq \Lambda(s_\varepsilon) & \text { if } T_i(s_\varepsilon)=w_i-\pi_i\\
        \text{and} \ \mathbb{E}\left[T_{i}(S_\varepsilon)\right]=\mathbb{E}[\varepsilon_i] & \text{ for all} \ i=1, \dots, n .
        \end{array}
        $$
    \label{prob2}
\end{corollary}
\noindent The existence and unicity of AFPO taxation rules are given in \cite{Buhlmann1978,buhlmann_optimal_1979}. We refer to Theorem 2 in \cite{Buhlmann1978} for the proof.\\
\begin{theorem}
Under Assumption \ref{H1}, the AFPO taxation rule characterized in Corollary \ref{prob2} exists and is unique. 
\end{theorem}

\section{Analytical study for the case of  two regions \label{sect4}}
This section is dedicated to a complete analytic study of two regions ($n=2$), each with a constant absolute risk aversion. The procedure for determining the AFPO taxation rule is detailed, which provides an explicit solution for this particular case.

\subsection{Analytic solution under CARA disutilities}\label{sec:subsec:analytical}
This case study is based on \cite{feng_decentralized_2023} (Chapter 6, Example 6.7). The economy under consideration consists of two regions, with their initial wealth of $w_1$ and $w_2$. For simplicity, let us assume that $\pi_1=\pi_2=0$, which means no insurer exists, and the regions directly share their losses. After a catastrophe, the loss occurring to a region is given by  $X_i=\tau_iw_i$, where $\tau_i$ is the random proportion of wealth that region $i$ loses due to catastrophes. The aggregate wealth loss to be shared between the two regions is  $S_\varepsilon = S = X_1+X_2$. The disutility functions of the two regions are assumed to be with constant absolute risk aversion (CARA), that is   $v_1(x)=\gamma_1 e^{\frac{x}{\gamma_1}}$ and $v_2(x)=\gamma_2 e^{\frac{x}{\gamma_2}}$, where $\gamma_i$ is the risk tolerance parameter of region $i$. The larger the value of $\gamma_i$, the more risk-tolerant is region $i$. Without loss of generality, we order the region by increasing value of $\frac{w_i}{\gamma_i}$.\\

\begin{assumption}\label{hyp:ex}
The setting of the  2-regions example is the following 
\begin{itemize}
\item $v_1(x)=\gamma_1 e^{\frac{x}{\gamma_1}}$ and $v_2(x)=\gamma_2 e^{\frac{x}{\gamma_2}}$
\item $\frac{w_1}{\gamma_1} \leq \frac{w_2}{\gamma_2}$
\item $S_\varepsilon=X_1+X_2$ follows a mixed distribution: with probability $p_0$ it is exactly $0$ (a Dirac mass at $0$), and with probability $1-p_0$ it is uniformly distributed on the interval $[0, w_1+w_2]$ with mean $\mu_1+\mu_2$ where $\mu_i:= \mathbb E [X_i]$.
\end{itemize}
\end{assumption}

\noindent Proposition \ref{prop1} computes explicitly the optimal taxation rule in this setting.\\

\begin{proposition}  Under Assumption~\ref{hyp:ex}, we define the following quantities
\[
\gamma = \frac{\gamma_2}{\gamma_1}, \quad
\mu = \frac{\mu_2}{\mu_1}, \quad
 \zeta = \ln\Bigl(\frac{\alpha_2}{\alpha_1}\Bigr),\quad
c = (1+\gamma)w_1 - \gamma_2\,\zeta .
\]
We introduce the sets of the parameters $(\mu_i,\gamma_i,w_i)_{i=1,2}$, characterized through the quantity
$$M = \left(\frac{2}{1-p_0}\right)^2\left(\mu+1\right)\left(\frac{\mu_1}{\gamma_1}\right)^2 : $$
\begin{eqnarray*}
A_1&:=&\Bigl\{\,(\mu_i,\gamma_i,w_i)_{i=1,2} \; / \; M \in \left[\frac{\gamma^2}{\mu}\left(\frac{w_2}{\gamma_2}\right)^2, \frac{\gamma}{\mu}\left(\frac{w_1}{\gamma_1}\right)^2+\frac{\gamma^2}{\mu}\left(\frac{w_2}{\gamma_2}\right)^2 \right] \Bigr\};  \\
A_2&:=&\Bigl\{\,(\mu_i,\gamma_i,w_i)_{i=1,2} \; / \; M \in \left[\left(\gamma + 1\right)  \left(\frac{w_1}{\gamma_1}\right)^2, \left(1-\gamma \right)  \left(\frac{w_1}{\gamma_1}\right)^2 + 2\gamma\frac{w_1}{\gamma_1}\frac{w_2}{\gamma_2} \right] \Bigr\};  \\
A_3&:=&\Bigl\{\,(\mu_i,\gamma_i,w_i)_{i=1,2} \; / \; M \in \left[ \left(\frac{w_1}{\gamma_1}\right)^2  , \left(\gamma + 1\right)  \left(\frac{w_1}{\gamma_1}\right)^2\right] \Bigr\}.
\end{eqnarray*}
The optimal weight  \(\zeta\),  the AFPO taxation rule \(T_1(s_\varepsilon)\) and  $T_2(s_\varepsilon)=s_\varepsilon-T_1(s_\varepsilon)$ are determined on each set as follows:\\
{\bf Case 1}: The set $A_1$ corresponds to the weight $\zeta  \in [0, \frac{w_1}{\gamma_1} ]$. Then 
\[
\zeta  = \frac{w_1}{\gamma_1}
  - \sqrt{ \frac{M\mu}{\gamma} - \gamma\bigl(\frac{w_2}{\gamma_2}\bigr)^2},
\quad
T_1(s_\varepsilon)=
\begin{cases}
s_\varepsilon, & 0\leq s_\varepsilon \leq \gamma_1\,\zeta ,\\
\frac{\gamma_1}{\gamma_1+\gamma_2}\,s_\varepsilon
 +\frac{\gamma_1\gamma_2}{\gamma_1+\gamma_2}\,\zeta ,
 & \gamma_1\zeta  \leq s_\varepsilon \leq c,\\
w_1, & c \leq s_\varepsilon \leq w_1+w_2.
\end{cases}
\]
{\bf Case 2}: The set $A_2$ corresponds to the weight $\zeta  \in [\frac{w_1}{\gamma_1}-\frac{w_2}{\gamma_2}, 0 ]$. Then
\[
\zeta  = \frac{1}{\gamma_2}\Bigl[
  \frac{(1+\gamma)w_1}{2}
  +\frac{2(\mu_1+\mu_2)}{1-p_0}\Bigl(\frac{\mu_1}{(1-p_0)w_1}-1\Bigr)
\Bigr],
\quad
T_1(s_\varepsilon)=
\begin{cases}
0, & 0\leq s_\varepsilon \leq -\gamma_2\zeta ,\\
\frac{\gamma_1}{\gamma_1+\gamma_2}\,s_\varepsilon
 +\frac{\gamma_1\gamma_2}{\gamma_1+\gamma_2}\,\zeta ,
 & -\gamma_2\zeta  \leq s_\varepsilon \leq c,\\
w_1, & c\leq s_\varepsilon \leq w_1+w_2.
\end{cases}
\]
{\bf Case 3}: The set $A_3$ corresponds to the weight $\zeta  \in [-\frac{w_2}{\gamma_2}, \frac{w_1}{\gamma_1}-\frac{w_2}{\gamma_2}]$. Then
\[
\zeta  = -\frac{w_2}{\gamma_2}
  + \sqrt{\frac{M}{\gamma}- \frac{1}{\gamma}\bigl(\frac{w_1}{\gamma_1}\bigr)^2},
\quad
T_1(s_\varepsilon)=
\begin{cases}
0, & 0 \leq s_\varepsilon\leq-\gamma_2\zeta ,\\
\frac{\gamma_1}{\gamma_1+\gamma_2}\,s_\varepsilon
 -\frac{\gamma_1\gamma_2}{\gamma_1+\gamma_2}\,\zeta ,
 & -\gamma_2\zeta  \leq s_\varepsilon\leq\frac{(\gamma+1)(w_1+w_2)-c}{\gamma},\\
s_\varepsilon-w_2, & \frac{(\gamma+1)(w_1+w_2)-c}{\gamma} \leq s_\varepsilon\leq w_1+w_2.
\end{cases}
\]

\label{prop1}
\end{proposition}
\noindent The proof of Proposition \ref{prop1} is postponed in Appendix \ref{proof}, and we comment below on this analytical result and in particular the impact of the parameter $\zeta = \ln (\frac{\alpha_2}{\alpha_1})$ on the risk sharing mechanism. Note that, with all else the same, an increase in $\alpha_i$ raises the threshold $\lambda_i$ at which region $i$ enters in the redistribution of loss, thereby reducing its burden in the taxation, whereas a decrease in $\alpha_i$ lowers this threshold, causing region $i$ to assume a larger share of losses.
 \\
For a CARA‐disutility case with 2 regions, each layer of loss is shared in \emph{quota‐share} fashion between the two regions.  In the middle quota‐share layer, both taxes are an increasing affine function of 
\(s_{\varepsilon}\), with slope respectively
\[
T_1'(s_{\varepsilon})
= \frac{\gamma_1}{\gamma_1 + \gamma_2},
\qquad
T_2'(s_{\varepsilon})
= \frac{\gamma_2}{\gamma_1 + \gamma_2}.
\]
Note that $\zeta$ determines the threshold at which region 1 enters and exits the quota-share layer; any change in $\zeta$ that shifts region 1’s thresholds also produces the corresponding shifts in region 2’s thresholds.

\begin{itemize}
  \item \textbf{Case 1 (\(\zeta \in [0,\frac{w_{1}}{\gamma_{1}}]\)):}  
    As \(\zeta\) increases, region 1’s deductible threshold \(\gamma_{1}\zeta\) becomes larger (i.e., sharing begins only after a higher loss) and its cap \(c=(1+\gamma)w_{1}-\gamma_{2}\zeta\) becomes smaller (i.e., the flat‐tax region starts earlier in terms of loss). Consequently, the shared interval \([\gamma_{1}\zeta,\,c]\) contracts. From region 2’s perspective, this means region 2 also shares over a narrower interval: it starts sharing only after a larger loss retained by region 1, and reaches its own cap sooner. Conversely, reducing \(\zeta\) enlarges the shared layer for both regions.
  
  \item \textbf{Case 2 (\(\zeta \in [\frac{w_{1}}{\gamma_{1}} - \frac{w_{2}}{\gamma_{2}},\,0]\)):}  
    Both breakpoints, region 2’s deductible (\(-\gamma_{2}\zeta\)) and region 1’s cap \(c=(1+\gamma)w_{1}-\gamma_{2}\zeta\), depend affinely on \(\zeta\) in the same way, so their difference remains constant at \((1+\gamma)w_{1}\). Hence, the width of the quota‐share layer does not change; rather, changing \(\zeta\) shifts the entire interval of sharing to occur at different loss levels without altering its size. From region 2’s viewpoint, increasing \(\zeta\) causes both its deductible and cap to correspond to smaller loss values (so sharing happens earlier), while decreasing \(\zeta\) corresponds to larger loss values (sharing happens later), but the span of losses over which sharing occurs remains fixed for both regions.
  
  \item \textbf{Case 3 (\(\zeta \in [-\frac{w_{1}}{\gamma_{1}},\frac{w_{1}}{\gamma_{1}}-\frac{w_{2}}{\gamma_{2}}]\)):}  
    Decreasing \(\zeta\) causes region 2’s deductible (\(-\gamma_{2}\zeta\)) to correspond to a higher loss level (so region 2 begins sharing only after a larger loss) and simultaneously causes region 1’s cap \(c=(1+\gamma)w_{1}-\gamma_{2}\zeta\) to correspond to a higher loss level (so region 1’s flat‐tax region begins later). Thus, the shared interval \([-\gamma_{2}\zeta,\,c]\) expands in length, meaning both regions share over a wider range of losses. Increasing \(\zeta\) within this case compresses that interval, so sharing occurs over a narrower range for both regions.
\end{itemize}

One can note that \(\zeta\) not only selects the applicable set \(A_i\) and controls the width of the quota‐share layer, but also induces additive shifts in the intercepts of the affine taxation rules: the slopes \(\tfrac{\gamma_1}{\gamma_1+\gamma_2}\) and \(\tfrac{\gamma_2}{\gamma_1+\gamma_2}\) remain unchanged inside that layer, while the deductible and cap thresholds move according to \(\zeta\).

\subsection{Sensitivity analysis}\label{sec:subsec:sensitivity}
\begin{figure}
\centering
\begin{subfigure}{.48\textwidth}
  \centering
  \includegraphics[width=\linewidth]{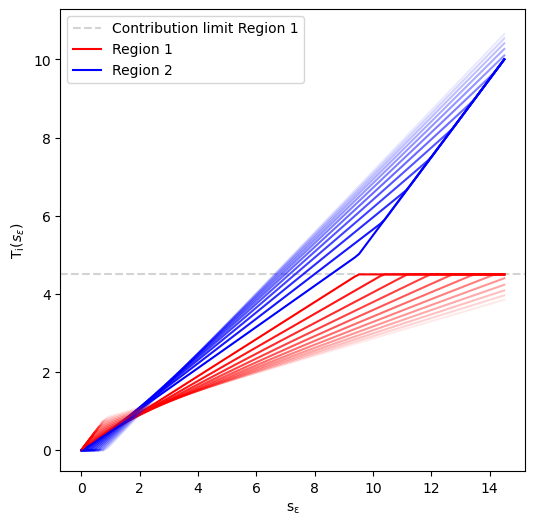}
  \caption{Case 1}
  \label{fig:sub_gammachange_alphaA3}
\end{subfigure}%
\begin{subfigure}{.48\textwidth}
  \centering
  \includegraphics[width=\linewidth]{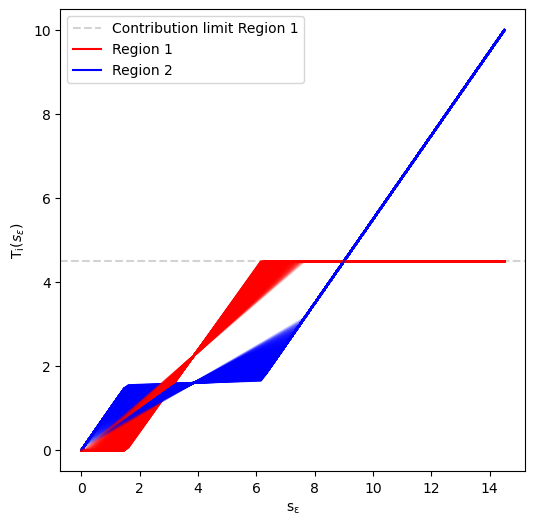}
  \caption{Case 2}
  \label{fig:sub_gammachange_taxA3}
\end{subfigure}
\caption{Sensitivity analysis of $T_i$ with respect to the ratio $\frac{\gamma_2}{\gamma_1}$ based on the analytical solution for two regions with $w_1 = 4.5$, $w_2 = 10$, $\mu_2=4, \mu_1=(1-p_0)\frac{w_1+w_2}{2}-\mu_2$, $\gamma_1 = 1$, and $p_0 = 0.1$.}
\label{fig:sensitivity_analysisfin}
\end{figure}

To assess the robustness of the AFPO taxation rule under CARA disutility, we conduct two one‐dimensional sensitivity experiments around the baseline parameterization 
\[
w_{1}=4.5,\; w_{2}=10,\; p_{0}=0.2,\; \gamma_{1}=1,\; \gamma_{2}=2,\; \mu_{2}=4,\; \mu_{1}=(1-p_{0})\tfrac{w_{1}+w_{2}}{2}-\mu_{2}.
\]
First, for each value of the relative expected‐loss ratio \(\mu=\frac{\mu_{2}}{\mu_{1}}\) in \([0.1,10]\), we compute the three candidate expressions for the weight \(\alpha_{1}\) arising from Cases~1--3, then determine which case’s bound condition is met and retain the unique solution \(\alpha_{1}\) satisfying that case. Figure~\ref{fig:sensitivity_analysis} displays the resulting curve \(\alpha_{1}\) versus \(\mu\) (solid) and \(\alpha_{2}=1-\alpha_{1}\) (dashed), with segment colors (\textcolor{red}{red}/\textcolor{blue}{blue}/\textcolor{ForestGreen}{green}) indicating the active case; a secondary axis shows the log‐odds \(\zeta=\ln\ \bigl(\frac{(1-\alpha_{1})}{\alpha_{1}}\bigr)\), which uniquely determines both the deductible threshold \(\gamma_{1}\,\zeta\) and the cap shift \(c=(1+\gamma)\,w_{1}-\gamma_{2}\,\zeta\). Second, we vary the relative risk‐tolerance ratio \(\gamma=\frac{\gamma_{2}}{\gamma_{1}}\) over \([0.1,10]\), enforcing the feasibility \(\tfrac{w_{1}}{\gamma_{1}}\le \tfrac{w_{2}}{\gamma_{2}}\), and similarly identify the applicable case at each \(\gamma\) to plot \(\alpha_{1}\) and \(\alpha_{2}\) with respect to \(\gamma\) in Figure~\ref{fig:sensitivity_analysisfin}. In this second experiment, only two of the three cases arise within the chosen parameter range; for completeness, we include further sensitivity plot in Appendix \ref{addsen}. Overall, one observes that a larger \(\mu\) shifts the overall risk burden toward Region~2, while a smaller \(\gamma\) leads Region~2 to assume a greater share of small, frequent losses and a smaller share of large losses. Moreover, the log‐odds \(\zeta\) faithfully tracks how the pure‐retention, quota‐share and pure‐limit layers widen, narrow or translate as parameters vary, even though the within‐layer sharing rates \(\frac{\gamma_{1}}{(\gamma_{1}+\gamma_{2})}\) and \(\frac{\gamma_{2}}{(\gamma_{1}+\gamma_{2})}\) remain constant.

\begin{figure}
    \centering
    \includegraphics[width=.8\linewidth]{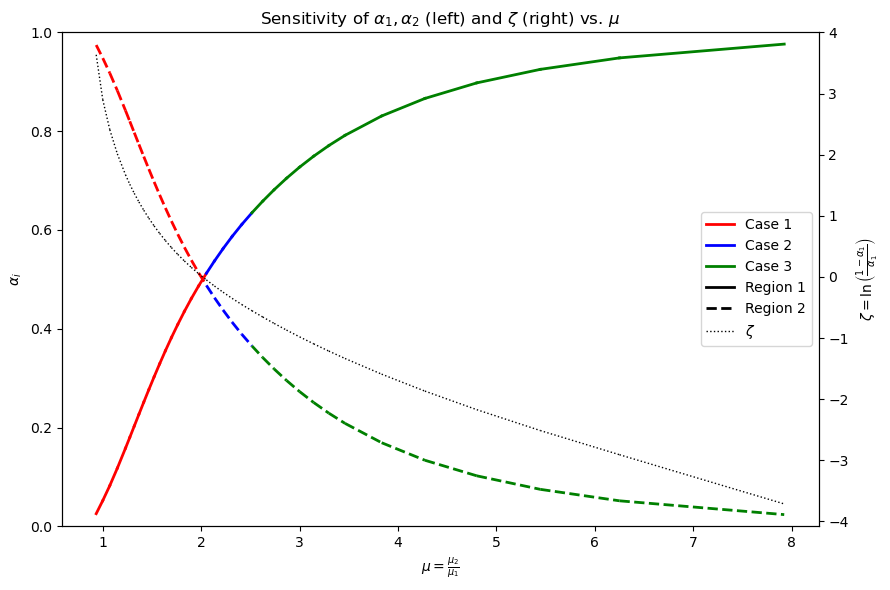}
    \caption{Sensitivity of $\zeta  = \ln \left(\frac{1-\alpha_1}{\alpha_1}\right)$ (dotted) and the optimal weights \(\alpha_1\) (solid) and \(\alpha_2 = 1 - \alpha_1\) (dashed), with respect to the relative riskiness ratio $\mu=\frac{\mu_2}{\mu_1}$.  
    Curve segments are colored by the active case: \textcolor{red}{Case 1}, \textcolor{blue}{Case 2}, and \textcolor{ForestGreen}{Case 3}, determined by the parametric regions.  
    Parameters are fixed to: \(w_1 = 4.5\), \(w_2 = 10\), \(\gamma_1 = 1\), \(\gamma_2 = 2\), and \(p_0 = 0.2\).}
    \label{fig:sensitivity_analysis}
\end{figure}

\begin{figure}
\centering
\begin{subfigure}{.32\textwidth}
  \centering
  \includegraphics[width=\linewidth]{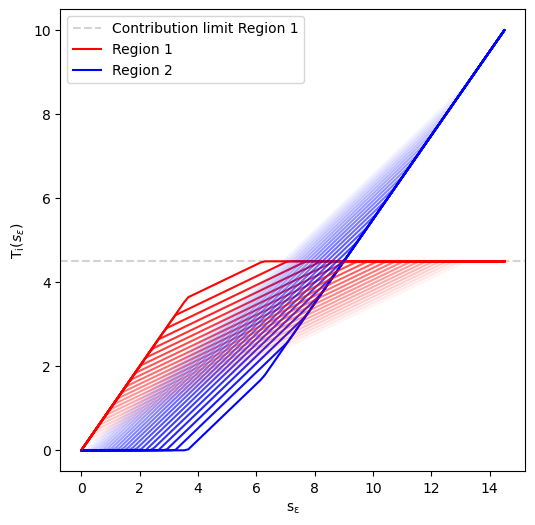}
  \caption{Case 1}
  \label{fig:sub_muchange_taxA1}
\end{subfigure}%
\begin{subfigure}{.32\textwidth}
  \centering
  \includegraphics[width=\linewidth]{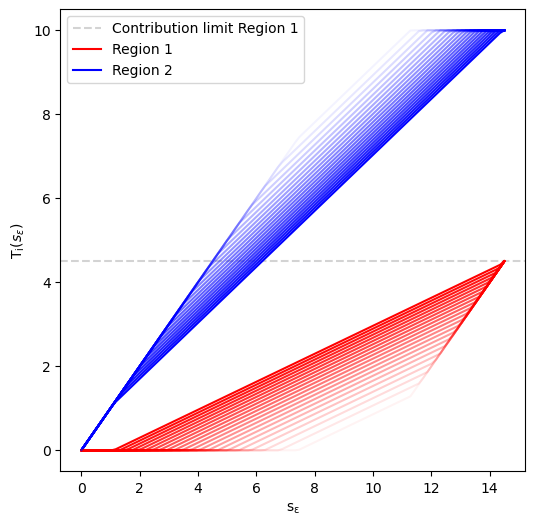}
  \caption{Case 2}
  \label{fig:sub_muchange_taxA2}
\end{subfigure}%
\begin{subfigure}{.32\textwidth}
  \centering
  \includegraphics[width=\linewidth]{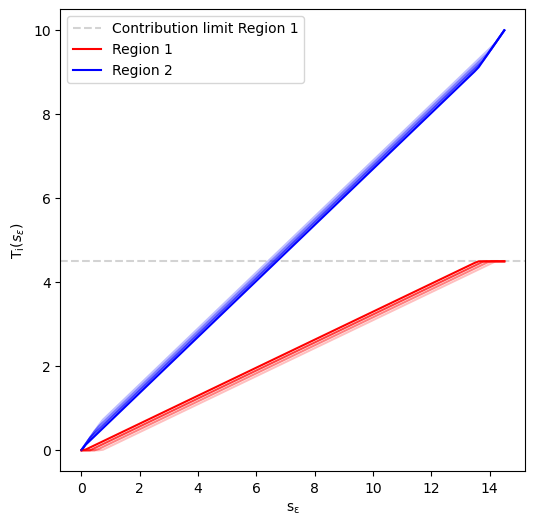}
  \caption{Case 3}
  \label{fig:sub_muchange_taxA3}
\end{subfigure}
\caption{Sensitivity analysis of $T_i$ with respect to the ratio $\frac{\mu_2}{\mu_1}$ based on the analytical solution for two regions with $w_1 = 4.5, w_2=10$, $p_0=0.2$, $\gamma_1=1, \gamma_2=2$.}
\label{fig:sensitivity_analysistax}
\end{figure}

\begin{figure}
    \centering
    \includegraphics[width=.8\linewidth]{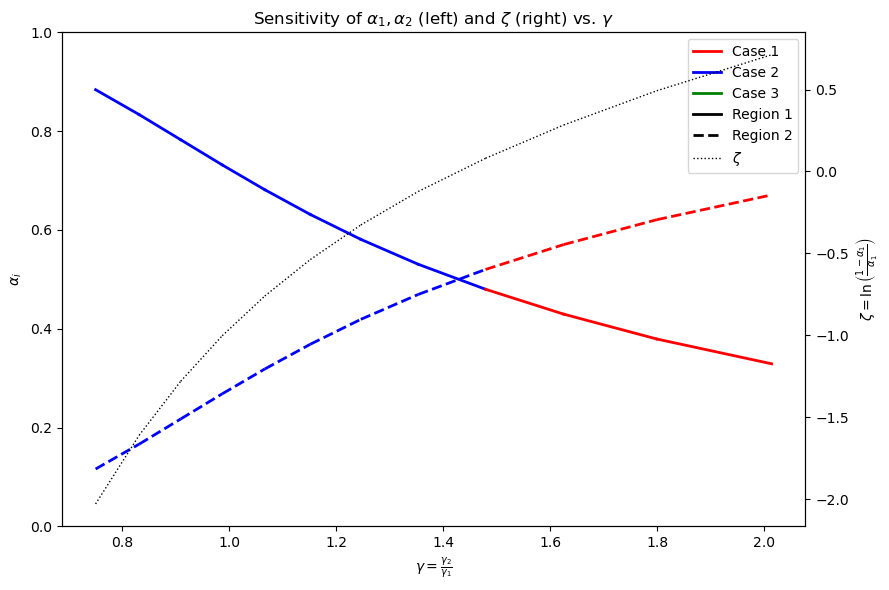}
    \caption{Sensitivity of $\zeta  = \ln \left(\frac{1-\alpha_1}{\alpha_1}\right)$ (dotted) and the optimal weights \(\alpha_1\) (solid) and \(\alpha_2 = 1 - \alpha_1\) (dashed), with respect to the relative risk tolerance ratio $\gamma=\frac{\gamma_2}{\gamma_1}$.  
    Curve segments are colored by the active case: \textcolor{red}{Case 1} and \textcolor{blue}{Case 2}, determined by the parametric regions.  
    Parameters are fixed to: $w_1 = 4.5$, $w_2 = 10$, $\mu_2=4, \mu_1=(1-p_0)\frac{w_1+w_2}{2}-\mu_2$, $\gamma_1 = 1$, and $p_0 = 0.1$.}
    \label{fig:sensitivity_analysis}
\end{figure}

\section{Numerical study for the general case of $n$ regions}\label{sec:numerical_algo}
Although the analytical approach and Corollary \ref{prob2} shows how a solution can be found, there might not always exists an explicit expression thereof. For example, as the analytical solution relies on the characterization of the loss distribution, simulation-based approaches of risks would not be compatible with our mechanism. Further, as the number of participants in the system increases, an explicit solution becomes harder to find. An alternative is to use a numerical algorithm to find the AFPO taxation rule. 
The purpose of this section is to illustrate how the actuarially fair Pareto optimal (AFPO) taxation rule performs in a realistic multi-region environment. We aim to investigate (i) how heterogeneity in risk aversion and exposure shapes tax contributions, (ii) how ex-post losses are distributed under different default scenarios, and (iii) how the AFPO rule compares to alternative burden-sharing rules. Thus,
we present an approach to solve the general $n$-region problem, which includes the two-region problem as a special case. With a numerical approach, the advantage is that it applies to any situation where it is possible to find a taxation rule, that is, where the participants are risk-averse. Further, the numerical estimation method also naturally allows for imposing constraints, such as the condition that no region may be held liable for more than its total wealth, easing the handling of certain hypotheses.

\subsection{General approach}
To find a solution, we can make use of the fact that the sharing mechanism can be understood as a root-finding process, where we look for a vector $\boldsymbol{\alpha}^*=(\alpha^*_1,\dots,\alpha^*_n) \in \Delta^+_n$ such that
\begin{align}\label{eq:fairness}
    \mathbb{E}[T_i(S_{\varepsilon}, \boldsymbol{\alpha}^*)] = \mathbb{E}[\varepsilon_i]\enspace.
\end{align}
Relation \eqref{eq:fairness} simply enforces the actuarial fairness condition, that is, the expected taxation given the joint loss is equal to the expected individual loss. Given that the vector $(\alpha_1,\dots,\alpha_n) $ regulates the relative participation of each region, and that any sharing rule covers the entire joint loss, deviations from the fairness condition in Equation \eqref{eq:fairness} are always offset across participants. As an illustrative example, consider the case where $n=2$, then the search for the optimal parameter is reduced to finding a scalar in $]0,1[$. If we choose an initial scalar $\alpha_1^0 \ne \alpha_1^*$, then $\mathbb{E}\left[T_i(S_\varepsilon ; \alpha_1^0, 1-\alpha_1^0)\right]$ for $i=1, 2$ will be too high for one participant and too low for the other. This opens up a straightforward approach to solve the problem numerically with two steps:
\begin{itemize}
    \item Initialize a random $\alpha_1^0 \in ]0,1[$ and calculate $\mathbb{E}\left[T_i(S_\varepsilon ; \alpha_1^0,  1-\alpha_1^0)\right]$. If necessary, renumber the region with $\mathbb{E}\left[T_i(S_\varepsilon ; \alpha_1^0, 1-\alpha_1^0)\right] > \mathbb{E}\left[\varepsilon_i\right]$ as region 2, and the other as region 1.  
    \item Increase $\alpha_1^0$ until $\mathbb{E}\left[T_2(S_\varepsilon ; \alpha_1^0, 1-\alpha_1^0)\right]$ decreases to $\mathbb{E}\left[\varepsilon_2\right]$ (and $\mathbb{E}\left[T_1(S_\varepsilon ; \alpha_1^0, 1-\alpha_1^0)\right]$ increases to $\mathbb{E}\left[\varepsilon_1\right]$).
\end{itemize}
The procedure then only requires an efficient method to calculate the expected taxation given a set of taxation weights $\boldsymbol{\alpha}$ (or for the special case where $n=2$, a scalar $\alpha_1$). Below, we outline an algorithm that implements these steps in an iterative procedure. 

\subsection{Numerical algorithm}
As the numerical procedure relies in essence on two parts, namely a tax function given parameters $\boldsymbol{\alpha}$ and the estimation of the expected taxation under a given loss model, we outline both parts separately. First, the expected taxation can be estimated via a standard histogram-based approach as outlined in Algorithm \ref{algo:rel_participation}. The advantage here is that numerical samples can be obtained from both parametric distributions and empirical data. Given an arbitrary sharing rule $\boldsymbol{T}$, parametrized by $\boldsymbol{\alpha}$, its corresponding $\lambda$ and samples of residual losses (that is, losses not covered by an insurance contract) $S_\varepsilon$, the goal is to estimate the participation of each region in this sample. To find these values, we either use empirical data or obtain samples from a loss model, and the values are then converted into a standard histogram. For numerical stability, we will often use the \emph{relative} participation of each region, referred to as $t^{(m)}_i$, which allows us to calculate a relative participation within a small bin of a histogram. Multiplying the relative participation value by the probability of a loss arising in a given bin and the average loss within said bin, then leads to the absolute (or euro value) required from each participant. Note that $n[j]$ in Algorithm \ref{algo:rel_participation} refers to the number of observations found in the given histogram bin. 

\begin{algorithm}
\caption{Expected Participation}\label{algo:rel_participation}
\begin{algorithmic}[1]
\Require Sample of $S_\varepsilon$, current sharing rule $\boldsymbol{T}(S_\varepsilon;\boldsymbol{\alpha}^{(m)})$ and $\lambda$
\Ensure $\mathbb{E}[T_i(S_\varepsilon,\boldsymbol{\alpha}^{(m)}]$ for all $i=1,\dots,n$

\State hist, edges $\gets$ histogram$(S_\varepsilon)$ \Comment{Estimate histogram}

\For{$i$ in Regions} 
        \State $t_i^{(m)}(\lambda) \mathrel{/}=  s_\varepsilon(\lambda)$ \Comment{Relative participation per region}

        $\mu_i \gets 0$
        
        $\overline{s} \gets 0$
        
    \For{$j$ in range(edges)} 
            \State bin\_middle $\gets \frac{1}{2}(\text{edges}_j + \text{edges}_{j+1})$
            \State $\mu_i \mathrel{+}=  t_i^{(m)}(\text{bin\_middle})*\text{bin\_middle}*n[j]$ \Comment{Average over histogram} 
            \State $\overline{s} \mathrel{+}= n[j]$
    \EndFor
        
    $\mu_i \mathrel{/}= \overline{s}$
\EndFor
\end{algorithmic}
\end{algorithm}

To obtain the taxation rule, we follow the same approach as laid out in \cite{feng_decentralized_2023}, Example 6.4. However, as in our general model, the loss distribution is generally unknown, we need to rely on an iterative procedure to find the optimal AFPO taxation rule. To find a solution to the taxation problem, we proceed as follows: we start with an initial guess of $\boldsymbol{\alpha}$, denoted as $\boldsymbol{\alpha}^0$, and then nudge the resulting taxation function closer to the actuarially fair solution across a series of iterations by following the logic laid out in \cite{feng_decentralized_2023} but with the actuarial fairness condition as an optimization goal. The procedure is outlaid in detail in Algorithm \ref{algo:general_algo}, where in step 7, we include Algorithm \ref{algo:rel_participation} to estimate the expected value under the current taxation scheme. Once convergence is achieved on the parameters, we obtain the AFPO taxation rule.

\begin{algorithm}
\caption{Optimal AFPO taxation}\label{algo:general_algo}
\begin{algorithmic}[1]
\Require Samples of $(\varepsilon_1, \dots, \varepsilon_n)$, initial wealth sets $(w_1,\dots,w_n)$, set of premiums $(\pi_1,\dots,\pi_n)$, set of disutility functions $\mathcal{V}=(v_1,\dots,v_n)$, ${\delta}$-tolerance, stepsize $\gamma$
\Ensure Optimal taxation rule $\boldsymbol{T}(s_\varepsilon,\boldsymbol{\alpha})=(T_1(s_\varepsilon),\dots,T_n(s_\varepsilon))$ for any residual claim

\Initialize{$m \gets 0$ \\ \strut$\alpha_i^{(m)} \gets$ \text{uniform sampling on the unit simplex}}

\Repeat
    \State $\boldsymbol{\alpha}^{(m)} \gets \boldsymbol{\alpha}^{(m)}$/ $|\boldsymbol{\alpha}^{(m)}|$ \Comment{Set $\boldsymbol{\alpha}^{(m)}$ in the unit simplex: $\alpha_i^{(m)}=\frac{\alpha_i^{(m)}}{\sum^{n}_{j=1} \alpha_j^{(m)}}.$}
    \State Re-index $\boldsymbol{\alpha}^{(m)}, \mathcal{V}$ according to $\lambda_1 \leq \lambda_2 \leq \cdots \leq  \lambda_n \leq \lambda_{n+1}$ \enspace ,
    
    where $\lambda_i = \alpha_i v_i^\prime(0)$ for $i=1,\dots,n$ and ${ \lambda_{n+1}=\infty}$

    \State For $\lambda_j^{(m)}<\lambda \leq \lambda_{j+1}^{(m)}$, $j=1,\dots,n$, set
    $$
    T_i^{(m)}(\lambda)= \begin{cases}I_i\left(\frac{\lambda}{\alpha_i^{(m)}}\right), & i=1,2, \cdots, j \\ 0, & i=j+1, \cdots, n .
    \end{cases}
    $$

    \State Calculate:
    $$
    s_\varepsilon(\lambda) = \sum_{i=1}^n T_i^{(m)}(\lambda), \quad \Lambda(s_\varepsilon)=s_\varepsilon^{-1}(\lambda)
    $$
    \State With $\lambda^{\text{max}}=\Lambda\left(\sum_{i=1}^n (w_i-\pi_i)\right)$, set:
    
    $t_i^{(m)}(\lambda) \gets t_i^{(m)}(\lambda) / {s_\varepsilon(\lambda^{\text{max}}})$ 

    \State Verify:
    $$
   w_i \geq t_i^{m}(\lambda) ~\forall ~\lambda
    $$

    \State
    
    $\eta_i = \mathbb{E}[\varepsilon_i] - \mathbb{E}[T_i(S_\varepsilon;\boldsymbol{\alpha}^{(m)})]$
    
    \State Recover $\alpha_l, \alpha_h$ such that $l=\arg\min_{i}\eta_i, h=\arg\max_{i}\eta_i$

    Set $\alpha_l^{(m+1)} = \alpha_l^{(m)}+\gamma \min \lbrace |\eta_l|, |\eta_h| \rbrace, \quad \alpha_h^{(m+1)} = \alpha_h^{(m)}-\gamma\min \lbrace |\eta_l|, |\eta_h| \rbrace$

    $\alpha_j^{(m+1)}=\alpha_j^{(m)}, j\neq l,h$

    $m\mathrel{+}=1$

\Until{$||\boldsymbol{\alpha}^{(m+1)}- \boldsymbol{\alpha}^{(m)}||_2 < \delta$}

\end{algorithmic}
\end{algorithm}
 
As the sharing mechanism applies to residual losses of \emph{any} insurance scheme, including the no-insurance case or the case outlined above in Section \ref{sec:gov_setup},  Algorithm \ref{algo:general_algo} is also capable of finding the pure-risk sharing scenario. A difficulty arises in the general problem as the errors in the fairness condition are not necessarily distributed in a 1:1 ratio across participants. That is, if a single participant in the scheme is overcharged, the corresponding undercharges may be distributed across multiple different participants who are \emph{undercharged}. If only the most extreme observations are chosen in Step 10, the algorithm might get stuck in an endless loop. A simple fix for this problem is to consider not only the two observations with the largest discrepancies but the $M$ observations with the largest discrepancies that is, the observations corresponding to the largest deviations from $\mathbb{E}[\varepsilon_i] - \mathbb{E}[T_i(S_\varepsilon, \boldsymbol{\alpha}^{(m)})]$ and then nudge the corresponding $\boldsymbol{\alpha}$-values simultaneously.

\section{Application to simulated storm disasters}\label{sec:application}
With the general algorithm developed in the previous section, we now study how the proposed AFPO taxation mechanism can be applied in a stylized real-world setting. This application is designed to demonstrate how the model handles large-scale regional heterogeneity and spatially correlated catastrophic losses—key features that motivated the theoretical framework introduced in Sections 2 through 5. While we do not aim to provide the most realistic empirical model, we use a simplified storm damage scenario in Western Europe to highlight the scalability and practical behavior of the mechanism. The toy example is grounded in actual economic data (regional GDPs) and serves purely as an illustrative case to show how our framework can operate under realistic spatial and economic heterogeneity. To the best of our knowledge, implementing such a fine-grained, utility-based allocation across numerous participants would not be feasible with existing closed-form approaches. The simulation framework allows us to incorporate analytically intractable loss dependencies while maintaining a strong link to the economic principles of our taxation rule.

\subsection{Data}

\begin{figure}
    \centering
    \includegraphics[width=0.8\textwidth]{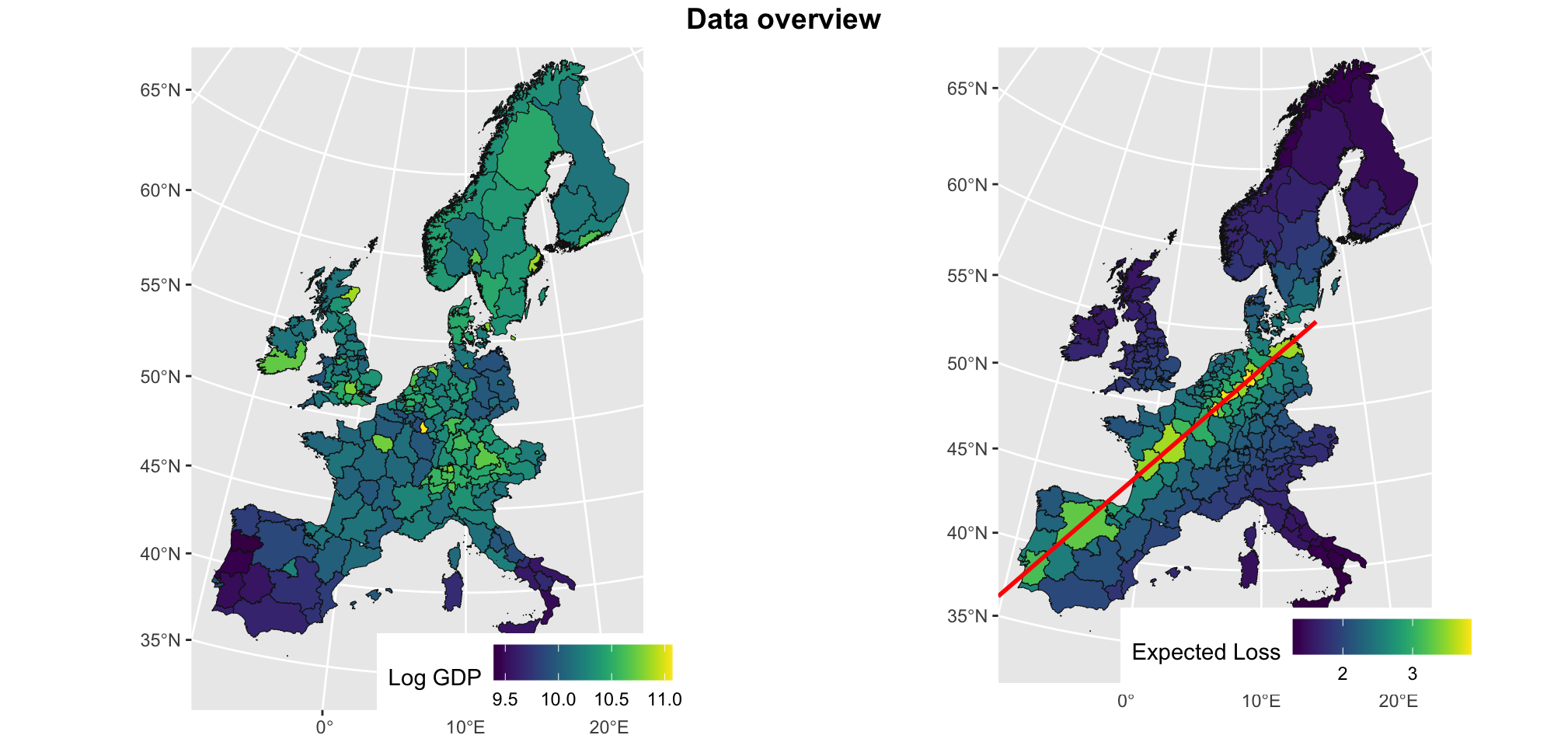}
    \caption{Left pane, log GDP per Nuts II region. Right pane: a simulation of a storm affecting different regions to differing degrees of destruction, the coloration corresponds to the average loss across the 100,000 simulations.}
    \label{fig:full_maps}
\end{figure}

We consider a dataset consisting of 212 geographic regions, based on the European Union's  (nomenclature of territorial units for statistics\footnote{See the official documentation for more information on \href{https://ec.europa.eu/eurostat/web/nuts}{NUTS II}}) regions, which represent large parts of Western and Northern Europe and the United Kingdom. The individual wealth, in our notation $w_i$ for region $i$, is represented by the log-GDP in millions of euros of each region, depicted in the left pane of Figure \ref{fig:full_maps}. To simulate storm damage, we abstract from a realistic model and instead follow a simplified scenario, which allows us to focus on the main aspects of our mechanism. Note, however, that using more advanced models of storm damages, such as those developed by, for example, \cite{koks2020high}, could easily be integrated into the mechanism as it simply affects the simulated losses. \\
In this simplified reality, storms only take place along a single line that traverses the region. Although the path of the storms is fixed, we allow them to vary in intensity across simulations and even spatially within a single simulation. For that, we define the expected intensity, that is, the expected destruction rate $\tau_i$ for region $i$, as a function of the minimum distance $d_i$ of the region's centroid to the storm path. In particular we focus on destruction rates $\tau_i$ sampled from a beta distribution according to $\beta(\frac{0.1}{d_i + 0.3}, 0.5)$. Losses are then spatially correlated through the use of a Gaussian Copula, where each entry in the variance-covariance matrix of the multivariate normal distribution corresponds to 1 minus the normalized distance between the centroids of each region (which results in a variance of $1$ for the normal distribution). We then sample the destruction rates from this model and calculate the per-region loss as $X_i=\tau_i w_i$. Averaging over many simulations, we then receive the expected losses for each region, depicted in the right pane of Figure \ref{fig:full_maps}, which also includes the storm path.

\begin{figure}
    \centering
    \includegraphics[width=0.75\linewidth]{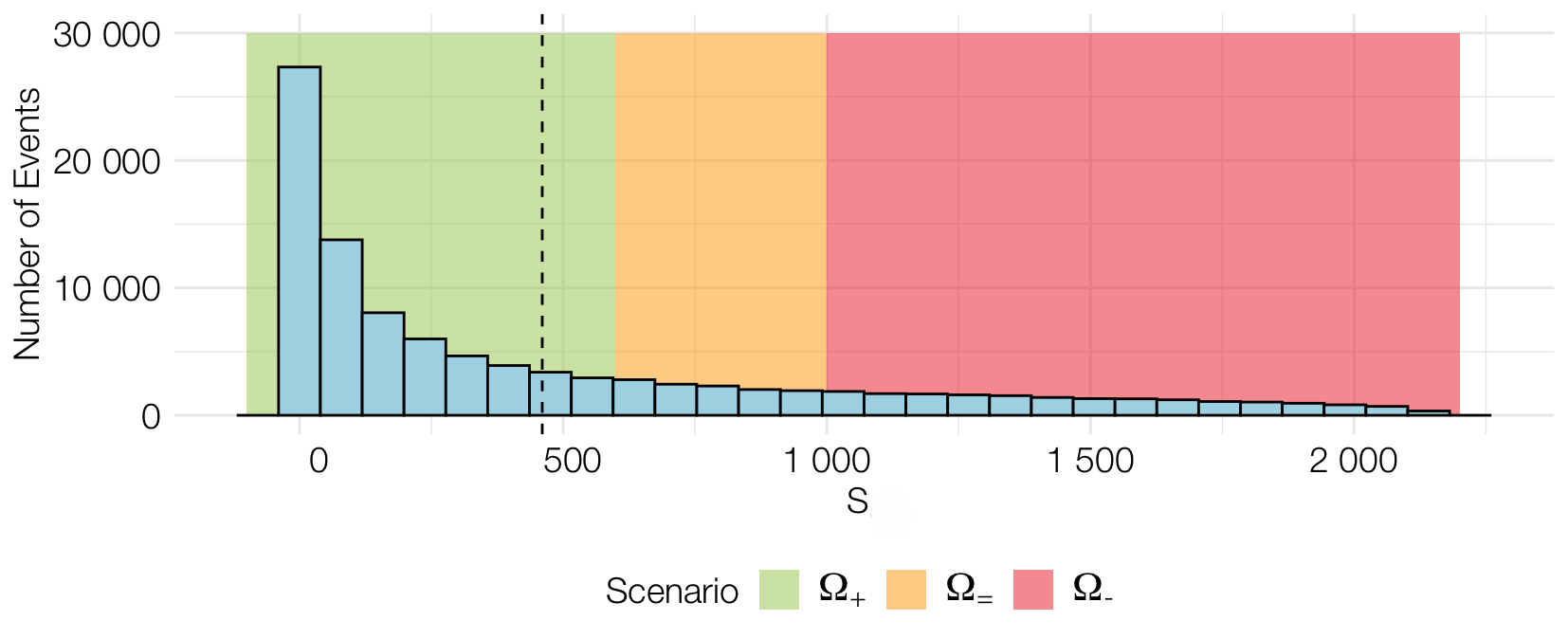}
    \caption{Histogram of total losses ($S$) in million euros across the 100,000 scenarios, the colors indicate where scenario losses are situated. The dashed black line represents the expected value of the aggregated losses.}
    \label{fig:total_losses}
\end{figure}

To evaluate our proposed scheme and compare it to other risk-sharing mechanisms, we simulate 100,000 disasters from the damage model and compare the resulting (dis)-utility when employing one of the four scenarios:
\begin{enumerate}
    \item The baseline case, no insurance or taxation scheme is involved, and the disutility of each region stems from the incurred damages.
    \item The traditional insurance case. Here, regions cover their losses through insurance for a premium, but the payments are subject to default. Hence, disutility is incurred through the premium which needs to be paid, but also any costs which are uncovered in the case of the insurer's default. 
    \item The pure risk-sharing scheme, where no traditional insurance is used. Here, the disutility of each region stems from payments that need to be made under the taxation regime. 
    \item The hybrid scheme, where traditional insurance is used for most claims, but the redistributive taxation scheme is used to cover the remaining claims in case of an insurer's default. Here, disutility is incurred through the payments of the insurance premium and the possible taxation transfers.     
\end{enumerate}
As each mechanism can excel in different situations, for example, a pure risk sharing mechanism would perform well when the losses are below their expected values, as insurance premiums are based on them, and would impose higher than materialized costs. To analyze how each mechanism performs across the simulations, we categorize our 100,000 simulations into the three scenarios introduced earlier, namely the \emph{favorable scenario} ($\Omega_+$), the \emph{intermediate scenario} ($\Omega_=$) and the \emph{default scenario} ($\Omega_-$). Roughly speaking, these scenarios correspond to situations where the damages incurred have been mild, significant, and catastrophic. Figure \ref{fig:total_losses} visualizes how the simulated events are classified, where most aggregated losses incurred are rather small, but the distribution has a thick right tail. We then compare the mechanisms between each other and across the three scenarios. 

\subsection{Expected disutility incurred}\label{sec:app:utilities}

\begin{figure}[htbp]    
\begin{subfigure}{\textwidth}
    \centering
    \includegraphics[width=0.7\textwidth]{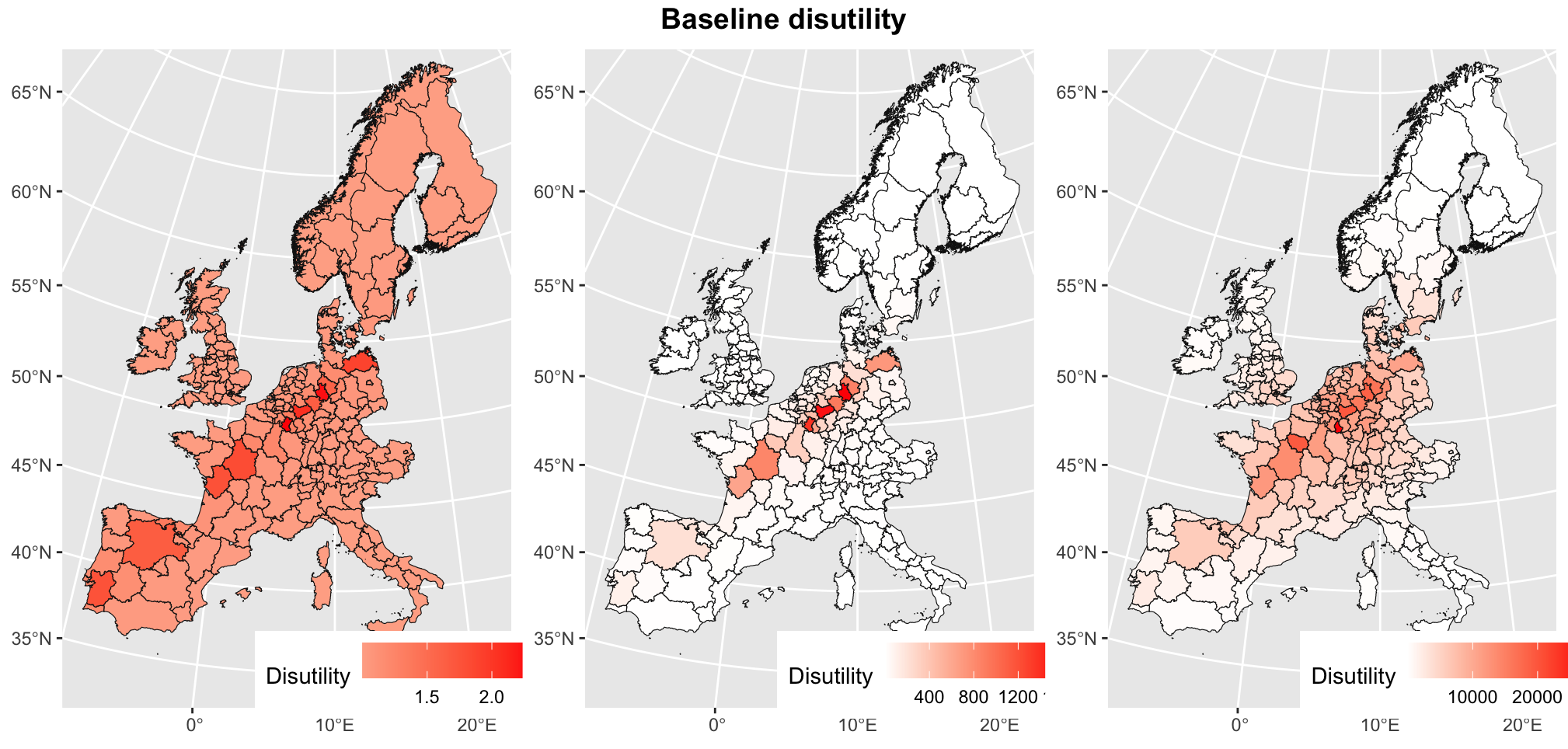}
    \label{fig:baseline}
\end{subfigure}
\begin{subfigure}{\textwidth}
    \centering
    \includegraphics[width=0.7\textwidth]{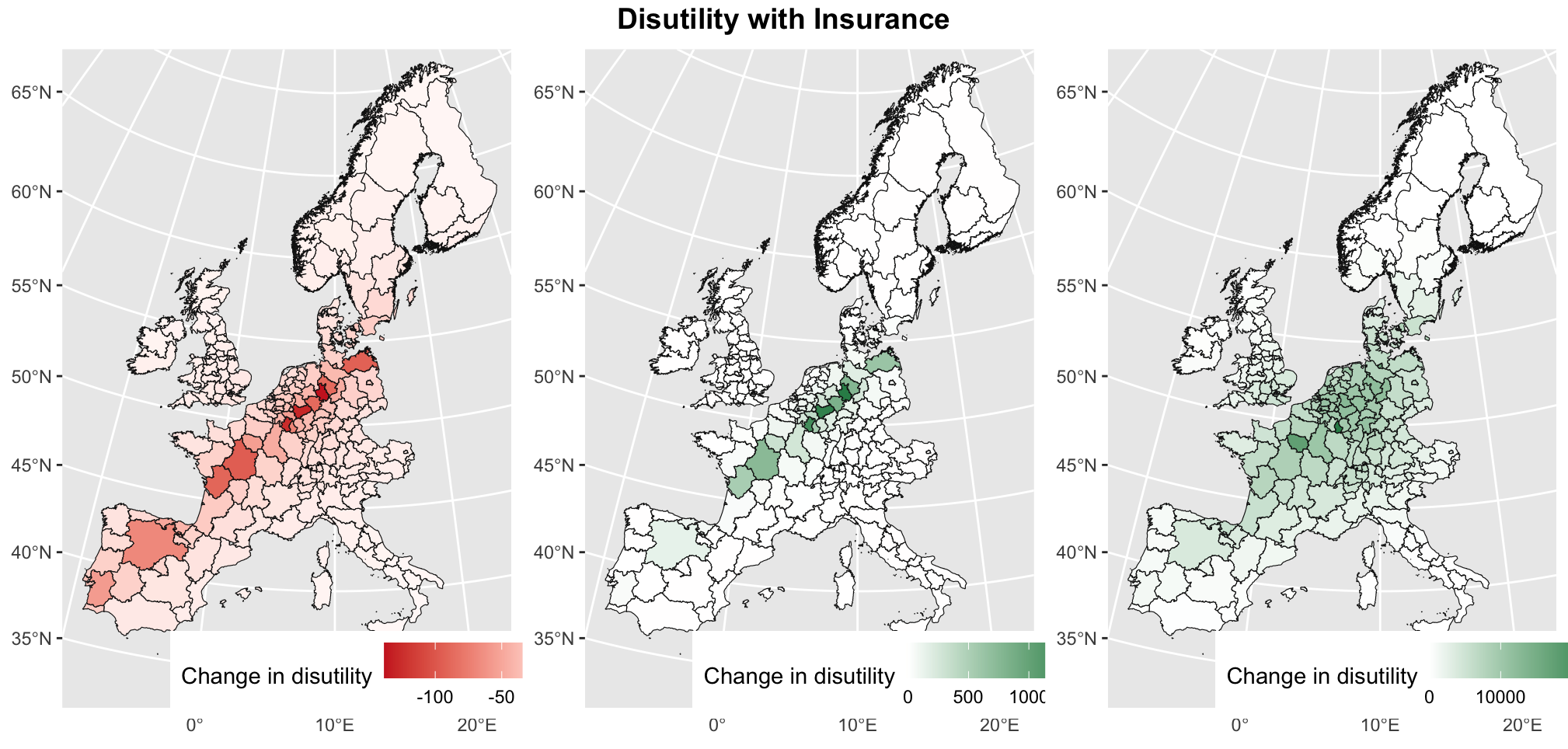}
    \caption{Difference baseline disutility against insurance disutility}
    \label{fig:insurance}
\end{subfigure}
\begin{subfigure}{\textwidth}
    \centering
    \includegraphics[width=0.7\textwidth]{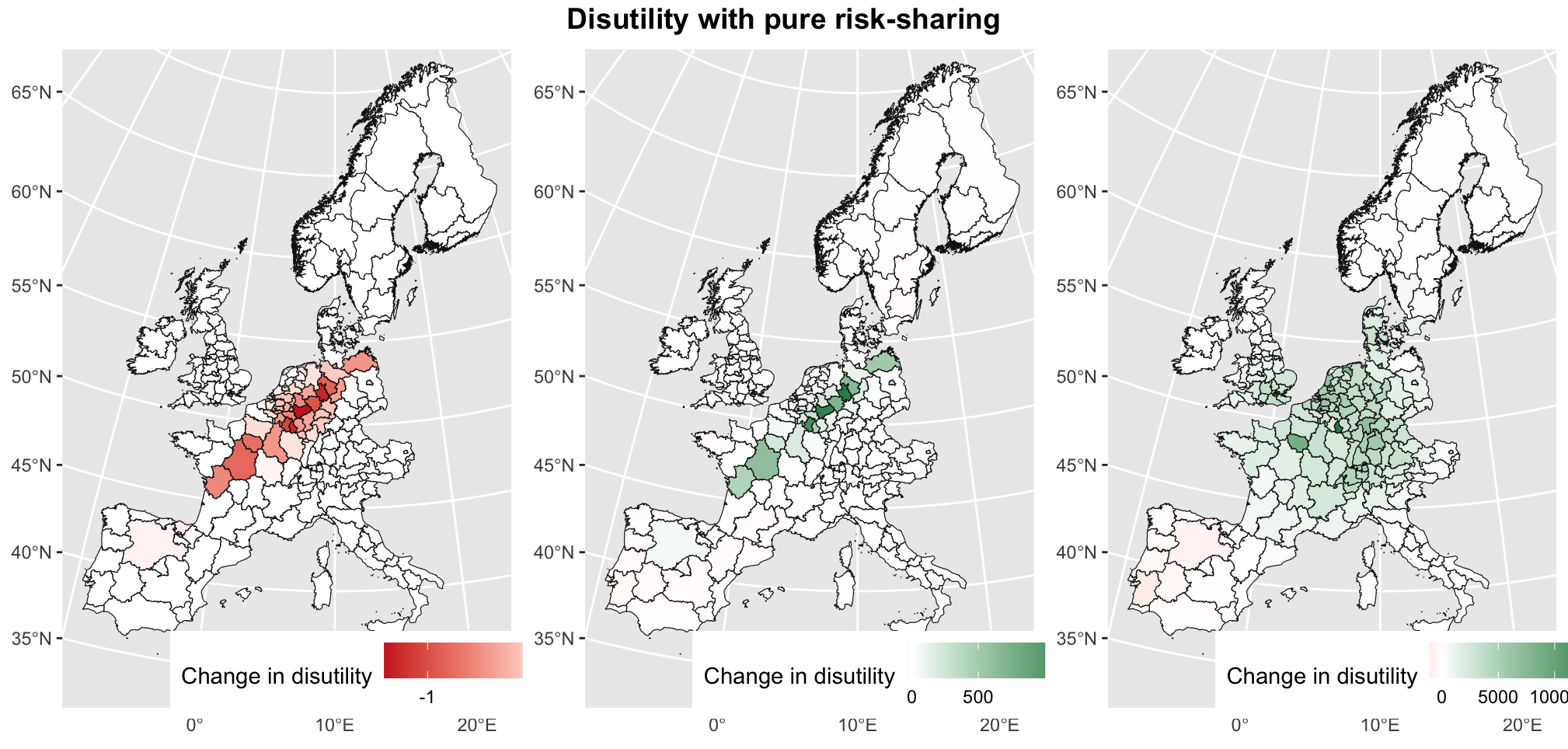}
    \caption{Difference baseline disutility against pure risk-sharing disutility}
    \label{fig:pure}
\end{subfigure}
\begin{subfigure}{\textwidth}
    \centering
    \includegraphics[width=0.7\textwidth]{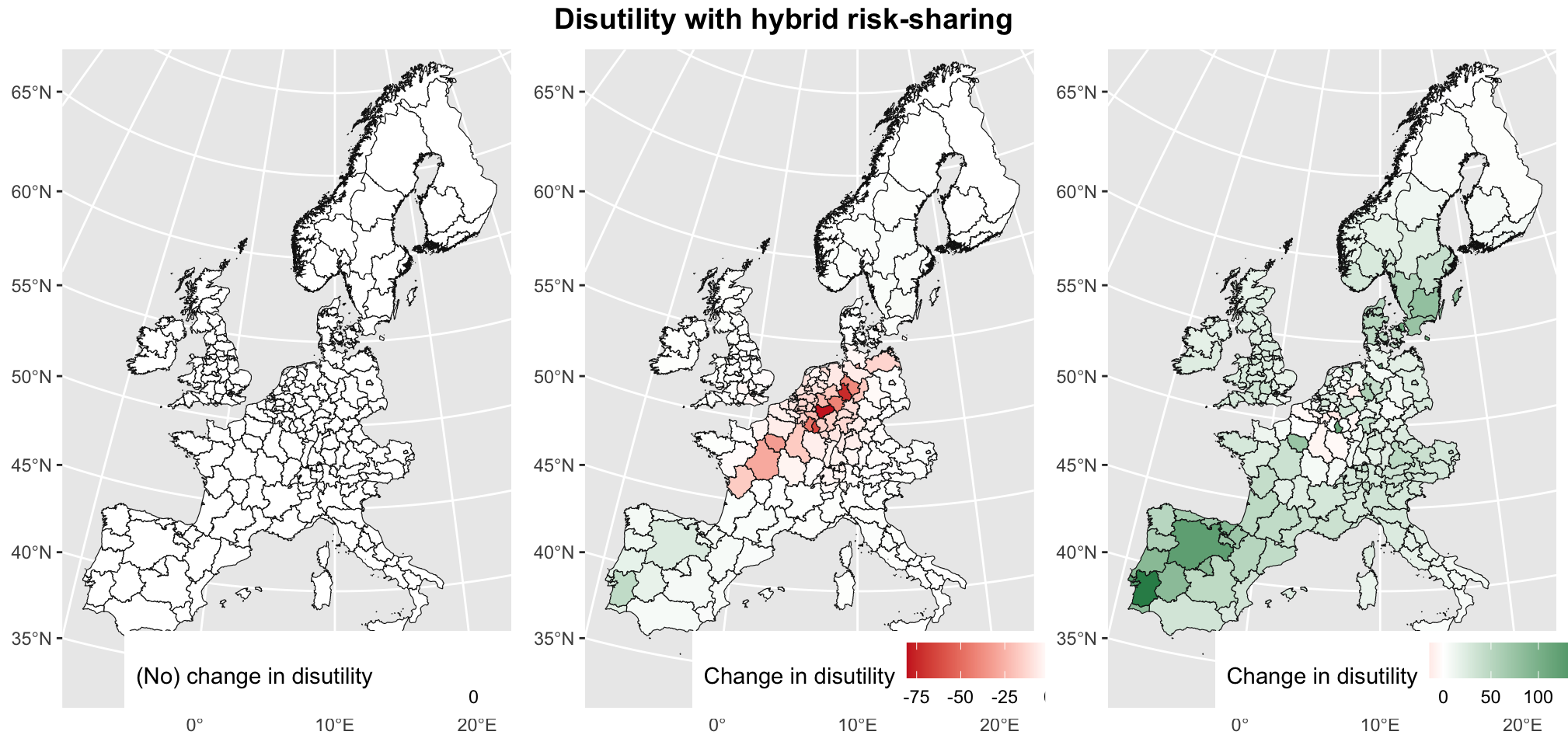}
    \caption{Difference insurance disutility against hybrid-mechanism disutility}
    \label{fig:hyb}
\end{subfigure}
\caption{Evaluation of disutility across the different mechanisms.}
\label{fig:results_visual}
\end{figure}

The numerical illustrations of our mechanism are split into two separate parts, although the underlying geographical and economic data remain constant across both. In a first part in Subsection~\ref{sec:app:utilities}, we largely fix considerations about an insurance premiums and a surplus sharing mechanism to investigate utilities under the baseline case, that is, no risk sharing and no traditional insurance against both pure risk-sharing through taxation and traditional insurance, as well as the mechanism which combines both elements as outlined in Equation \eqref{eq:final_formula}. We then focus our attention on a more individual transfer level in Subsection~\ref{sec:app:transfers}, where we analyze the insurance premiums $\pi_i$, initial capital allocations, and individual transfers under different scenarios.

To calculate the expected disutility across the mechanisms and scenarios, we fix our parameters as follows. The insurer's setup from Equation \eqref{eq:insurancepremium} is fixed at $\theta=0.3, \eta=0$. We assume that the insurer can accurately estimate the expected damage incurred by each region, which results in the collected premium $k$ amounting to around 465 million euros; the corresponding default value $K$ is set up at $K=1000$ million euros. In this setup, the insurer should be able to cover around 81\% of the simulated scenarios. In the remaining scenarios, $K$ is split among the regions according to the proportional rule (that is, $\frac{X_i}{S}$). We assume that all regions have the same exponential disutility function $v(l)=e^{l}$, which is calculated on the payments made, either through a premium, risk-sharing, self-covered losses, or a combination thereof. The expected disutilities are then calculated for each scenario, that is $\Omega_{+}, \Omega_{=}$ and $\Omega_{-}$ separately, to achieve an understanding of the impact of the mechanism in each separate scenario. This allows us to consider scenarios that correspond to situations where the region as a whole faces catastrophic losses (that is, a high $S$) or faces a situation where globally not many losses occur (that is, a low $S$), which can incur either through many small losses or catastrophic losses in a subpart of the whole region. Results are summarized visually in Figure \ref{fig:results_visual}. Note here that we compare \emph{differences} in disutilities between mechanisms. This allows a more straightforward comparison between two mechanisms. We compare each mechanism against the baseline (no-insurance, no taxation) case, with the exception of the hybrid mechanism, which is compared to the traditional insurance case.

The results help us to summarize the underlying advantages of each insurance scheme. First, as the traditional insurance is based on expected values, scenarios with lower losses (that is, $\Omega_{+}$ or the first column in Figure \ref{fig:results_visual}) result in a higher disutility for the regions, as opposed to the non-insurance case. This situation inverts as the losses become larger. Second, as the pure risk-sharing scheme is also based on expected values, the effects are more concentrated but overall less negative than traditional insurance, as riskier regions would need to provide higher transfers in years where the whole geographical area is less affected than its average. However, as the risk-sharing scheme only applies to actual incurred losses and not to the expected values, the incurred disutilities should be smaller overall. Finally, as the pure risk sharing or the hybrid scheme can absorb larger losses, they are beneficial for situations where a traditional insurance provider would default on the payments (right column in Figure \ref{fig:results_visual}). Unsurprisingly, the differences between the hybrid mechanism and the traditional insurance setup are equal to zero in the case where the insurer is not at default. Similar to the cases above, this changes as the losses get large and the insurer is in default, as the sharing mechanism can then substitute for the reallocation. Note here that the baseline for the comparison has changed according to the image descriptions. 

\subsection{Individual transfers}\label{sec:app:transfers}

Whereas the last subsection investigated the overall disutilities across many simulations, we now focus on disutilities incurred after a single event. To visualize the mechanism in more detail, suppose that there is no insurer present, that is, $S=S_\varepsilon$. This example follows the same setup as in the previous section, with expected losses simulated and the corresponding insurance risk sharing system implemented. 

\begin{figure}[htbp]
\centering
\begin{subfigure}{.32\textwidth}
  \centering
  \includegraphics[width=\linewidth]{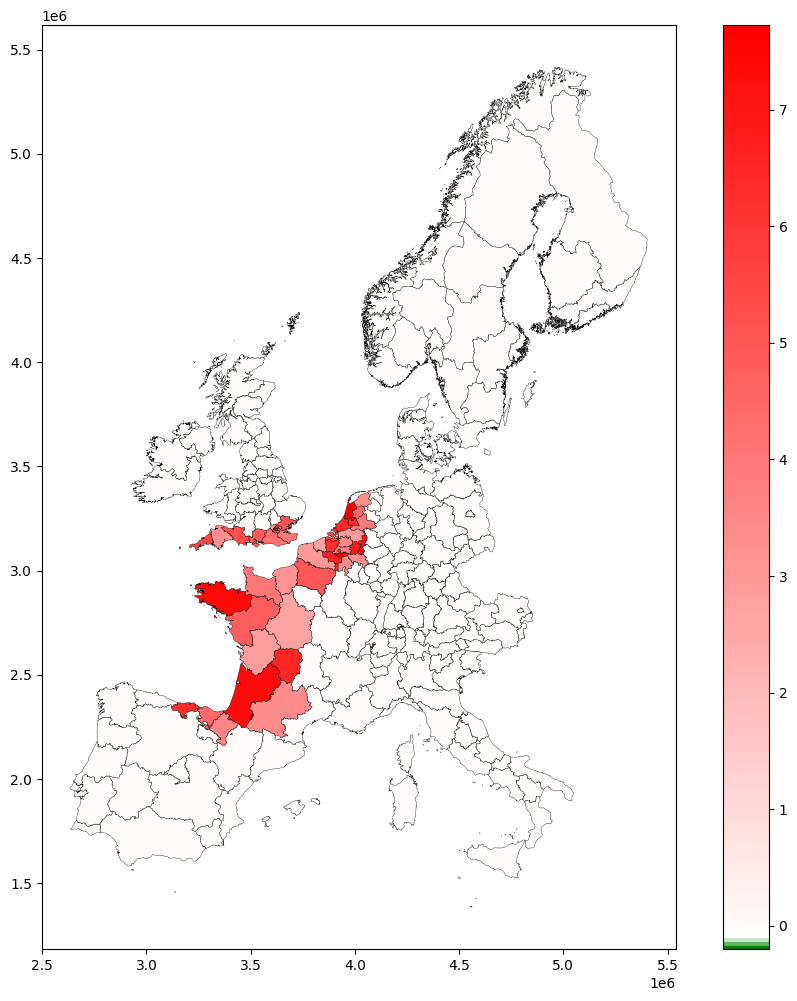}
  \caption{Incurred Losses after a single disaster}
  \label{fig:losses_coast}
\end{subfigure}%
\begin{subfigure}{.32\textwidth}
  \centering
  \includegraphics[width=\linewidth]{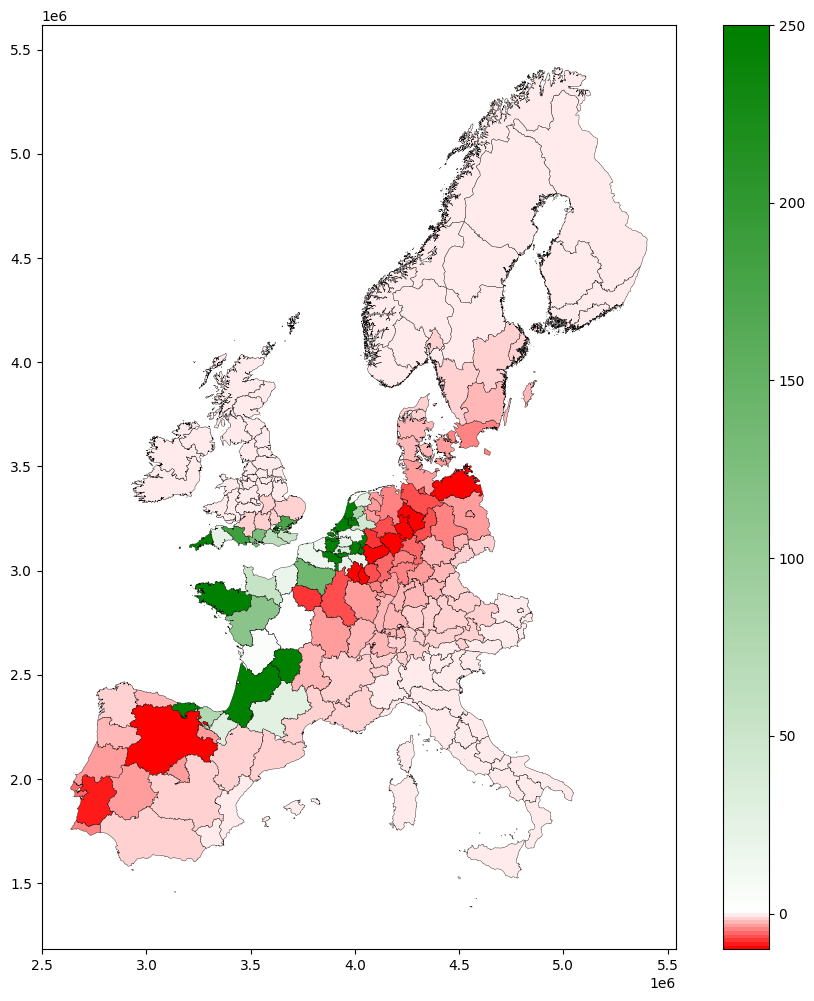}
  \caption{Differences in disutility post-sharing}
  \label{fig:disut_change_coast}
\end{subfigure}%
\begin{subfigure}{.32\textwidth}
  \centering
  \includegraphics[width=\linewidth]{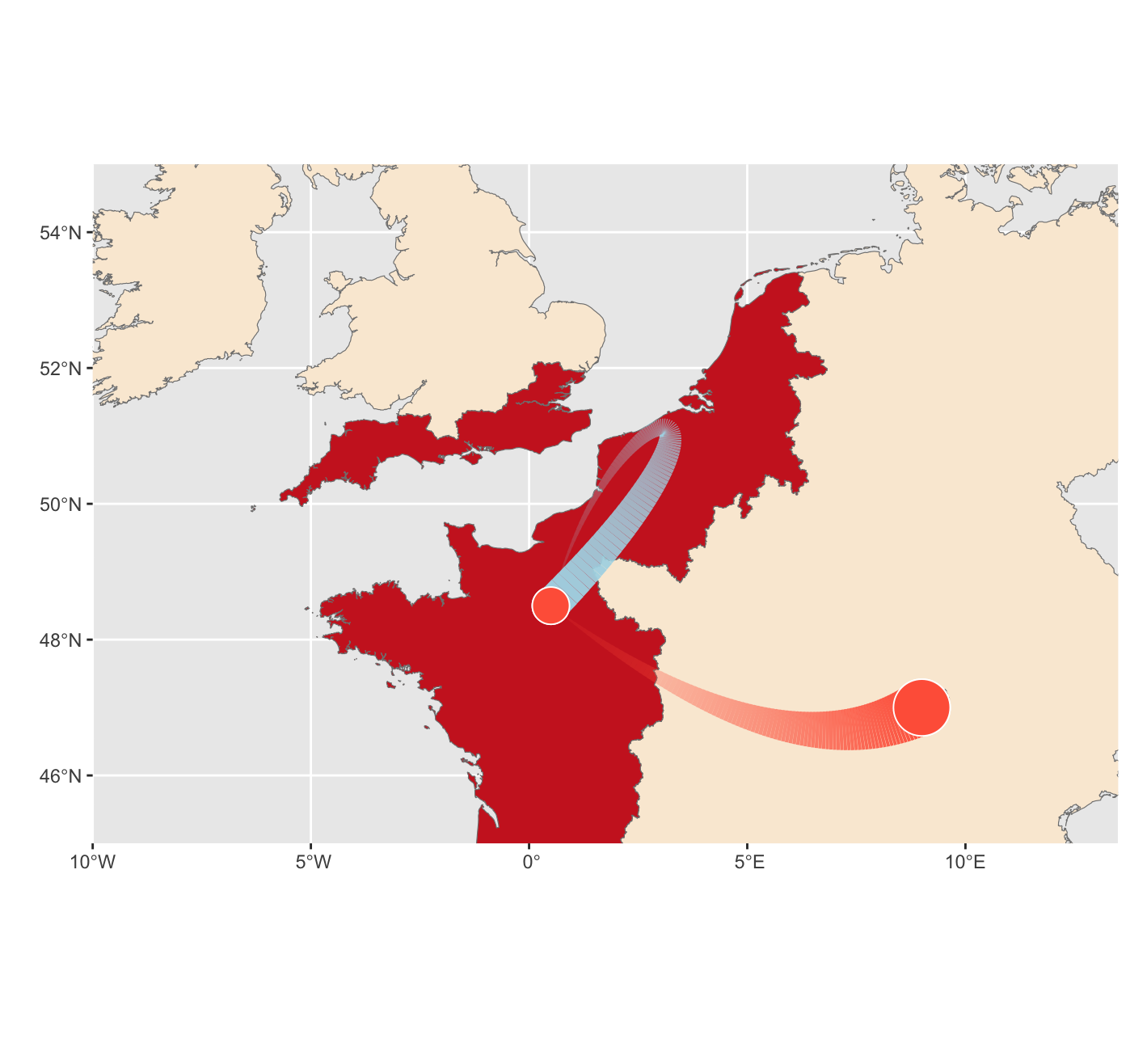}
  \caption{Example transfer of incurred losses}
  \label{fig:sharing_coast}
\end{subfigure}%
\caption{Regional transfers between coastal regions of Europe and participants in the risk sharing scheme. }
\label{fig:individual_coast}
\end{figure}

As an example, consider a storm which batters the Atlantic coastline of large parts of France as well as the area around the English Channel and parts of the North Sea, incurred losses are illustrated in the left panel of Figure \ref{fig:individual_coast}. 
The risk-sharing mechanism can then spring into action. According to what was seen, the regions around the storm's path are most at risk, although, in this specific instance, many of these regions did not report any losses. Nevertheless, as the risk-sharing mechanism works on the \emph{expected} loss and not on an individual event, these regions will still be expected to participate in such an event. This is visible in the middle panel of Figure \ref{fig:individual_coast}, where the disutilities differ.

As transfers occur ex-post of the event, but are based on the expected losses across many events, regions might experience different outcomes after individual events, that is, after a particular storm. For that, we also analyze the overall transfers between regions after this special case. This is depicted in the right panel of Figure \ref{fig:individual_coast}. Overall, the redistribution mechanism primarily helps to cover the cost of the affected areas, but we can also observe that this leads the affected region to assume some of the incurred losses itself (depicted as the blue transfer). Overall, if losses are high enough and not isolated to extremely small geographical sub-regions, the mechanism results in each region taking on a part of the incurred losses, which is of particular importance for catastrophic losses. 

\section{Conclusion}\label{sec:conc}

We introduced a risk-sharing framework that incorporates both actuarial fairness and individual rationality. Given the changing environment in catastrophic risk management, we studied the situation where the traditional insurance market can absorb most small losses, but where substantial losses are no longer covered by it. Instead, we adapted a risk-sharing mechanism based on government taxation, which could, in theory, provide almost unlimited coverage. We presented an algorithm that extends the simple two-player scenarios commonly studied to the general case where the scheme allows for $n$ participants. Through a sensitivity analysis, we were able to study the effects that changes in either the riskiness or utility function of regions have on the final result. We then applied the mechanism in a simulation, inspired by real-world wealth, where we studied how strong geospatial correlations with differing expected losses play together. We found that the mechanism behaves largely as expected, with regions more affected by catastrophic losses being more inclined to adopt a risk-sharing scheme, as traditional insurance costs them dearly in good years. Although our results are largely in line with previous research on risk sharing, they provide the unique opportunity to actually be implementable with modern loss simulation systems. Further, by having an adaptive sharing mechanism, it can easily be integrated into existing schemes that provide some kind of regional diversification. With events related to climate change proving difficult to handle by traditional schemes, it provides a first step into an alternative form of protection.

\pagebreak

\section*{Declaration of interests}

All authors declare that they have no conflict of interest.

\section*{Declaration of generative AI in scientific writing}

During the writing process of this paper, we used ChatGPT to improve the readability and language of the manuscript. We carefully reviewed the subsequent content and take full responsibility and accountability for the content of the publication.

\bibliographystyle{elsarticle-harv} 
\bibliography{P2P_risk_renewable}

\begin{thebibliography}{38}
\expandafter\ifx\csname natexlab\endcsname\relax\def\natexlab#1{#1}\fi
\providecommand{\url}[1]{\texttt{#1}}
\providecommand{\href}[2]{#2}
\providecommand{\path}[1]{#1}
\providecommand{\DOIprefix}{doi:}
\providecommand{\ArXivprefix}{arXiv:}
\providecommand{\URLprefix}{URL: }
\providecommand{\Pubmedprefix}{pmid:}
\providecommand{\doi}[1]{\href{http://dx.doi.org/#1}{\path{#1}}}
\providecommand{\Pubmed}[1]{\href{pmid:#1}{\path{#1}}}
\providecommand{\bibinfo}[2]{#2}
\ifx\xfnm\relax \def\xfnm[#1]{\unskip,\space#1}\fi
\bibitem[{{Associated Press}(2024)}]{CBS_calif}
\bibinfo{author}{{Associated Press}}, \bibinfo{year}{2024}.
\newblock \bibinfo{title}{State {Farm} won’t renew homeowners coverage for 72,000 {California} homes and apartments}.
\newblock \bibinfo{journal}{CBS News} \URLprefix \url{https://www.cbsnews.com/news/state-farm-california-homeowners-coverage-wont-renew-72000-homes-apartments/}.
\bibitem[{Awondo(2019)}]{awondo2019efficiency}
\bibinfo{author}{Awondo, S.N.}, \bibinfo{year}{2019}.
\newblock \bibinfo{title}{Efficiency of region-wide catastrophic weather risk pools: Implications for {African} {Risk} {Capacity} {Insurance} {Program}}.
\newblock \bibinfo{journal}{Journal of Development Economics} \bibinfo{volume}{136}, \bibinfo{pages}{111--118}.
\bibitem[{Boadway and Keen(1996)}]{boadway1996efficiency}
\bibinfo{author}{Boadway, R.}, \bibinfo{author}{Keen, M.}, \bibinfo{year}{1996}.
\newblock \bibinfo{title}{Efficiency and the optimal direction of federal-state transfers}.
\newblock \bibinfo{journal}{International Tax and Public Finance} \bibinfo{volume}{3}, \bibinfo{pages}{137--155}.
\bibitem[{Borch(1962)}]{borch_equilibrium_1962}
\bibinfo{author}{Borch, K.}, \bibinfo{year}{1962}.
\newblock \bibinfo{title}{Equilibrium in a {reinsurance} {market}}.
\newblock \bibinfo{journal}{Econometrica} , \bibinfo{pages}{424--444}.
\bibitem[{B{\"u}hlmann(1980)}]{buhlmann1980economic}
\bibinfo{author}{B{\"u}hlmann, H.}, \bibinfo{year}{1980}.
\newblock \bibinfo{title}{An economic premium principle}.
\newblock \bibinfo{journal}{ASTIN Bulletin: The Journal of the IAA} \bibinfo{volume}{11}, \bibinfo{pages}{52--60}.
\bibitem[{Bühlmann and Jewell(1978)}]{Buhlmann1978}
\bibinfo{author}{Bühlmann, H.}, \bibinfo{author}{Jewell, W.S.}, \bibinfo{year}{1978}.
\newblock \bibinfo{title}{Unicity of fair {Pareto} optimal risk exchanges}.
\newblock \bibinfo{journal}{University of California, Berkeley Operations Research Center} .
\bibitem[{Bühlmann and Jewell(1979)}]{buhlmann_optimal_1979}
\bibinfo{author}{Bühlmann, H.}, \bibinfo{author}{Jewell, W.S.}, \bibinfo{year}{1979}.
\newblock \bibinfo{title}{Optimal risk exchanges}.
\newblock \bibinfo{journal}{{ASTIN} Bulletin} \bibinfo{volume}{10}, \bibinfo{pages}{243--262}.
\bibitem[{Cai and Tan(2007)}]{cai2007optimal}
\bibinfo{author}{Cai, J.}, \bibinfo{author}{Tan, K.S.}, \bibinfo{year}{2007}.
\newblock \bibinfo{title}{Optimal retention for a stop-loss reinsurance under the {VaR} and {CTE} risk measures}.
\newblock \bibinfo{journal}{ASTIN Bulletin: The Journal of the IAA} \bibinfo{volume}{37}, \bibinfo{pages}{93--112}.
\bibitem[{{Caisse Centrale de Réassurance}(2024)}]{ccr2024}
\bibinfo{author}{{Caisse Centrale de Réassurance}}, \bibinfo{year}{2024}.
\newblock \bibinfo{title}{Catastrophes naturelles : Lancement de l’observatoire de l’assurabilité}.
\newblock \URLprefix \url{https://www.ccr.fr/-/lancement-observatoire-assurabilite}.
\bibitem[{Charpentier and Le~Maux(2014)}]{charpentier_natural_2014}
\bibinfo{author}{Charpentier, A.}, \bibinfo{author}{Le~Maux, B.}, \bibinfo{year}{2014}.
\newblock \bibinfo{title}{Natural catastrophe insurance: How should the government intervene?}
\newblock \bibinfo{journal}{Journal of Public Economics} \bibinfo{volume}{115}, \bibinfo{pages}{1--17}.
\bibitem[{Coculescu and Delbaen(2022)}]{coculescu_fairness_2022}
\bibinfo{author}{Coculescu, D.}, \bibinfo{author}{Delbaen, F.}, \bibinfo{year}{2022}.
\newblock \bibinfo{title}{Fairness principles for insurance contracts in the presence of default risk}.
\newblock \bibinfo{journal}{Mathematical Finance} \bibinfo{volume}{32}, \bibinfo{pages}{595--626}.
\bibitem[{Courbage and Oros(2024)}]{courbage2024effects}
\bibinfo{author}{Courbage, C.}, \bibinfo{author}{Oros, C.}, \bibinfo{year}{2024}.
\newblock \bibinfo{title}{On the effects of public subsidies for severe and mild dependency on long-term care insurance}.
\newblock \bibinfo{journal}{Insurance: Mathematics and Economics} \bibinfo{volume}{119}, \bibinfo{pages}{106--118}.
\bibitem[{DeLisa(2023)}]{farmers_florid}
\bibinfo{author}{DeLisa, C.}, \bibinfo{year}{2023}.
\newblock \bibinfo{title}{Farmers {Insurance}, {AIG} scale back insurance offerings in {Florida}}.
\newblock \bibinfo{journal}{The Capitolist} \URLprefix \url{https://thecapitolist.com/farmers-insurance-aig-scale-back-insurance-offerings-in-florida/}.
\bibitem[{Denuit et~al.(2022)Denuit, Dhaene and Robert}]{denuit_risk-sharing_2022}
\bibinfo{author}{Denuit, M.}, \bibinfo{author}{Dhaene, J.}, \bibinfo{author}{Robert, C.Y.}, \bibinfo{year}{2022}.
\newblock \bibinfo{title}{Risk-sharing rules and their properties, with applications to peer-to-peer insurance}.
\newblock \bibinfo{journal}{Journal of Risk and Insurance} \bibinfo{volume}{89}, \bibinfo{pages}{615--667}.
\bibitem[{Dinan(2017)}]{dinan2017projected}
\bibinfo{author}{Dinan, T.}, \bibinfo{year}{2017}.
\newblock \bibinfo{title}{Projected increases in hurricane damage in the {United} {States}: The role of climate change and coastal development}.
\newblock \bibinfo{journal}{Ecological Economics} \bibinfo{volume}{138}, \bibinfo{pages}{186--198}.
\bibitem[{Farhi and Werning(2017)}]{farhi2017fiscal}
\bibinfo{author}{Farhi, E.}, \bibinfo{author}{Werning, I.}, \bibinfo{year}{2017}.
\newblock \bibinfo{title}{Fiscal unions}.
\newblock \bibinfo{journal}{American Economic Review} \bibinfo{volume}{107}, \bibinfo{pages}{3788--3834}.
\bibitem[{Feng(2023)}]{feng_decentralized_2023}
\bibinfo{author}{Feng, R.}, \bibinfo{year}{2023}.
\newblock \bibinfo{title}{Decentralized Insurance: Technical Foundation of Business Models}.
\newblock Springer Actuarial, \bibinfo{publisher}{Springer International Publishing}, \bibinfo{address}{Berlin}.
\bibitem[{Feng et~al.(2024)Feng, Liu and Zhang}]{feng_unified_2022}
\bibinfo{author}{Feng, R.}, \bibinfo{author}{Liu, M.}, \bibinfo{author}{Zhang, N.}, \bibinfo{year}{2024}.
\newblock \bibinfo{title}{A unified theory of decentralized insurance}.
\newblock \bibinfo{journal}{Insurance: Mathematics and Economics} \bibinfo{volume}{119}, \bibinfo{pages}{157--178}.
\bibitem[{{Florida Office of Insurance Regulation }(2011)}]{florida_risk_2011}
\bibinfo{author}{{Florida Office of Insurance Regulation }}, \bibinfo{year}{2011}.
\newblock \bibinfo{title}{Risk analysis report}.
\bibitem[{Gao and Shi(2022)}]{gao2022leveraging}
\bibinfo{author}{Gao, L.}, \bibinfo{author}{Shi, P.}, \bibinfo{year}{2022}.
\newblock \bibinfo{title}{Leveraging high-resolution weather information to predict hail damage claims: A spatial point process for replicated point patterns}.
\newblock \bibinfo{journal}{Insurance: Mathematics and Economics} \bibinfo{volume}{107}, \bibinfo{pages}{161--179}.
\bibitem[{Gerber(1978)}]{gerber1978pareto}
\bibinfo{author}{Gerber, H.U.}, \bibinfo{year}{1978}.
\newblock \bibinfo{title}{Pareto-optimal risk exchanges and related decision problems}.
\newblock \bibinfo{journal}{ASTIN Bulletin: The Journal of the IAA} \bibinfo{volume}{10}, \bibinfo{pages}{25--33}.
\bibitem[{Gurenko(2004)}]{gurenko2004catastrophe}
\bibinfo{author}{Gurenko, E.N.}, \bibinfo{year}{2004}.
\newblock \bibinfo{title}{Catastrophe {risk} and {reinsurance}: A {country} {risk} {management} {perspective}}.
\newblock \bibinfo{publisher}{World Bank Publications}, \bibinfo{address}{London}.
\bibitem[{Ishiwatari(2012)}]{ishiwatari2012chapter}
\bibinfo{author}{Ishiwatari, M.}, \bibinfo{year}{2012}.
\newblock \bibinfo{title}{{Chapter 2 Government Roles in Community-Based Disaster Risk Reduction}}, in: \bibinfo{booktitle}{Community-Based Disaster Risk Reduction}. \bibinfo{publisher}{Emerald Group Publishing Limited}, \bibinfo{address}{Leeds}, pp. \bibinfo{pages}{19--33}.
\bibitem[{Koks and Haer(2020)}]{koks2020high}
\bibinfo{author}{Koks, E.}, \bibinfo{author}{Haer, T.}, \bibinfo{year}{2020}.
\newblock \bibinfo{title}{A high-resolution wind damage model for {Europe}}.
\newblock \bibinfo{journal}{Scientific Reports} \bibinfo{volume}{10}, \bibinfo{pages}{6866}.
\bibitem[{Langreney et~al.(2024)Langreney, Cozannet and Merad}]{merad2024adapter}
\bibinfo{author}{Langreney, T.}, \bibinfo{author}{Cozannet, G.L.}, \bibinfo{author}{Merad, M.}, \bibinfo{year}{2024}.
\newblock \bibinfo{title}{{Adapting} {the} {French} {insurance} {system} {to} {changing} {climate} {risks}}.
\newblock \bibinfo{type}{Technical Report}. {CNRS ; Universit{\'e} Dauphine ; BRGM}.
\newblock \URLprefix \url{https://lesateliersdufutur.org/wp-content/uploads/2024/09/Eng-Rapport_final_22122023.pdf}.
\bibitem[{Matthews et~al.(1999)Matthews, Sheffield, Andre, Lafayette, Roethen and Dobkin}]{matthews1999insolvency}
\bibinfo{author}{Matthews, P.}, \bibinfo{author}{Sheffield, M.}, \bibinfo{author}{Andre, J.}, \bibinfo{author}{Lafayette, J.}, \bibinfo{author}{Roethen, J.}, \bibinfo{author}{Dobkin, E.}, \bibinfo{year}{1999}.
\newblock \bibinfo{title}{Insolvency: will historic trends return}.
\newblock \bibinfo{journal}{Best’s Review—Property/Casualty Edition} \bibinfo{volume}{1999--03}.
\bibitem[{Meyer and Kunreuther(2017)}]{meyer2017ostrich}
\bibinfo{author}{Meyer, R.}, \bibinfo{author}{Kunreuther, H.}, \bibinfo{year}{2017}.
\newblock \bibinfo{title}{The Ostrich paradox: Why we underprepare for disasters}.
\newblock \bibinfo{publisher}{University of Pennsylvania Press}, \bibinfo{address}{Philadelphia}.
\bibitem[{Michel-Kerjan and Kunreuther(2011)}]{michel2011redesigning}
\bibinfo{author}{Michel-Kerjan, E.}, \bibinfo{author}{Kunreuther, H.}, \bibinfo{year}{2011}.
\newblock \bibinfo{title}{Redesigning flood insurance}.
\newblock \bibinfo{journal}{Science} \bibinfo{volume}{333}, \bibinfo{pages}{408--409}.
\bibitem[{Niakh(2024)}]{niakh2024fixed}
\bibinfo{author}{Niakh, F.}, \bibinfo{year}{2024}.
\newblock \bibinfo{title}{A fixed point approach for computing actuarially fair pareto optimal risk-sharing rules}.
\newblock \bibinfo{journal}{European Actuarial Journal} , \bibinfo{pages}{1--38}.
\bibitem[{Qiu et~al.(2023)Qiu, Jin and Li}]{qiu2023optimal}
\bibinfo{author}{Qiu, M.}, \bibinfo{author}{Jin, Z.}, \bibinfo{author}{Li, S.}, \bibinfo{year}{2023}.
\newblock \bibinfo{title}{Optimal risk sharing and dividend strategies under default contagion: A semi-analytical approach}.
\newblock \bibinfo{journal}{Insurance: Mathematics and Economics} \bibinfo{volume}{113}, \bibinfo{pages}{1--23}.
\bibitem[{Roth~Jr(1998)}]{roth1998earthquake}
\bibinfo{author}{Roth~Jr, R.J.}, \bibinfo{year}{1998}.
\newblock \bibinfo{title}{Earthquake insurance protection in California}.
\newblock \bibinfo{publisher}{Joseph Henry Press}, \bibinfo{address}{Washington, DC:}.
\bibitem[{Rummukainen(2012)}]{rummukainen2012changes}
\bibinfo{author}{Rummukainen, M.}, \bibinfo{year}{2012}.
\newblock \bibinfo{title}{Changes in climate and weather extremes in the 21st century}.
\newblock \bibinfo{journal}{Wiley Interdisciplinary Reviews: Climate Change} \bibinfo{volume}{3}, \bibinfo{pages}{115--129}.
\bibitem[{Shi et~al.(2024)Shi, Zhang and Shi}]{shi2024leveraging}
\bibinfo{author}{Shi, P.}, \bibinfo{author}{Zhang, W.}, \bibinfo{author}{Shi, K.}, \bibinfo{year}{2024}.
\newblock \bibinfo{title}{Leveraging weather dynamics in insurance claims triage using deep learning}.
\newblock \bibinfo{journal}{Journal of the American Statistical Association} \bibinfo{volume}{119}, \bibinfo{pages}{825--838}.
\bibitem[{Tadayon and Ghanbarzadeh(2024)}]{tadayon2024spatial}
\bibinfo{author}{Tadayon, V.}, \bibinfo{author}{Ghanbarzadeh, M.}, \bibinfo{year}{2024}.
\newblock \bibinfo{title}{Spatial copula-based modeling of claim frequency and claim size in third-party car insurance: A {Poisson}-mixed approach for predictive analysis}.
\newblock \bibinfo{journal}{Insurance: Mathematics and Economics} \bibinfo{volume}{119}, \bibinfo{pages}{119--129}.
\bibitem[{Thirawat et~al.(2017)Thirawat, Udompol and Ponjan}]{thirawat2017disaster}
\bibinfo{author}{Thirawat, N.}, \bibinfo{author}{Udompol, S.}, \bibinfo{author}{Ponjan, P.}, \bibinfo{year}{2017}.
\newblock \bibinfo{title}{Disaster risk reduction and international catastrophe risk insurance facility}.
\newblock \bibinfo{journal}{Mitigation and Adaptation Strategies for Global Change} \bibinfo{volume}{22}, \bibinfo{pages}{1021--1039}.
\bibitem[{Wang et~al.(2023)Wang, Zhou and Shao}]{wang2023risk}
\bibinfo{author}{Wang, X.}, \bibinfo{author}{Zhou, M.}, \bibinfo{author}{Shao, J.}, \bibinfo{year}{2023}.
\newblock \bibinfo{title}{A risk-sharing mechanism for multi-region catastrophe insurance with government subsidies}.
\newblock \bibinfo{journal}{International Journal of Disaster Risk Reduction} \bibinfo{volume}{86}, \bibinfo{pages}{103558}.
\bibitem[{Wu(2015)}]{wu2015reexamining}
\bibinfo{author}{Wu, Y.C.}, \bibinfo{year}{2015}.
\newblock \bibinfo{title}{Reexamining the feasibility of diversification and transfer instruments on smoothing catastrophe risk}.
\newblock \bibinfo{journal}{Insurance: Mathematics and Economics} \bibinfo{volume}{64}, \bibinfo{pages}{54--66}.
\bibitem[{Wu(2020)}]{wu2020equilibrium}
\bibinfo{author}{Wu, Y.C.}, \bibinfo{year}{2020}.
\newblock \bibinfo{title}{Equilibrium in natural catastrophe insurance market under disaster-resistant technologies, financial innovations and government interventions}.
\newblock \bibinfo{journal}{Insurance: Mathematics and Economics} \bibinfo{volume}{95}, \bibinfo{pages}{116--128}.

\end{thebibliography}

\pagebreak
\begin{appendices}

\section{Pareto‐optimal solutions via weighted minimization \label{secA1}}

\noindent
We recall that a Pareto‐optimal allocation among \(n\) regions can be obtained by solving, for any weight vector \(\boldsymbol\alpha=(\alpha_{1},\dots,\alpha_{n})\) in the open unit simplex \(\Delta^n_{+}=\{\,\alpha_i>0,\ \sum_i\alpha_i=1\}\), the following weighted minimization problem over taxation rules \((T_{1}(\cdot),\dots,T_{n}(\cdot))\):
\[
\min_{(T_{1},\dots,T_{n})}\;\; 
\sum_{i=1}^n \alpha_i \,\mathbb{E}\bigl[v_i\bigl(T_i(S_\varepsilon)\bigr)\bigr]
\]
subject to, for each realized total loss \(s_{\varepsilon}\in\Omega_{-}\),
\[
\left\{
\begin{aligned}
  &\sum_{i=1}^n T_i(s_{\varepsilon}) \;=\; s_{\varepsilon}, 
    \quad\text{(full allocation condition)}\\
  &0 \;\le\; T_i(s_{\varepsilon}) \;\le\; w_i - \pi_i, 
    \quad i=1,\dots,n.
\end{aligned}
\right.
\]
Because the objective is additive over states \(s_{\varepsilon}\), it suffices to solve, for each fixed \(s_{\varepsilon}\), the finite‐dimensional problem
\[
\min_{(T_{1},\dots,T_{n}) \in \mathbb{R}^n}\;\; 
\sum_{i=1}^n \alpha_i \,v_i\bigl(T_i(s_{\varepsilon})\bigr)
\quad\text{subject to}\quad
\sum_{i=1}^n T_i(s_{\varepsilon}) = s_{\varepsilon}, 
\ \ 0 \le T_i(s_{\varepsilon}) \le w_i - \pi_i.
\]
We now derive the necessary optimality conditions via the Karush–Kuhn–Tucker (KKT) framework.

\medskip

\noindent\textbf{Lagrangian and first‐order conditions.}
Introduce multipliers 
\(\Lambda(s_{\varepsilon})\in\mathbb{R}\) for the equality \(\sum_i T_i = s_{\varepsilon}\), 
and for each \(i\) multipliers \(\eta_i(s_{\varepsilon})\ge0\) for the upper bound \(T_i \le w_i - \pi_i\), and \(\mu_i(s_{\varepsilon})\ge0\) for the lower bound \(T_i \ge 0\).  The Lagrangian for the pointwise problem at \(s_{\varepsilon}\) is
\[
\begin{aligned}
\mathcal{L}\bigl((T_i),\,\Lambda,\;(\eta_i),\,(\mu_i)\bigr)
&=
\sum_{i=1}^n \alpha_i\,v_i\bigl(T_i(s_{\varepsilon})\bigr)
\;-\;\Lambda(s_{\varepsilon})\Bigl(\sum_{i=1}^n T_i(s_{\varepsilon}) - s_{\varepsilon}\Bigr)\\
&\quad
+\sum_{i=1}^n \eta_i(s_{\varepsilon})\,\bigl(T_i(s_{\varepsilon}) - (w_i - \pi_i)\bigr)
\;-\;\sum_{i=1}^n \mu_i(s_{\varepsilon})\,T_i(s_{\varepsilon}).
\end{aligned}
\]
The first‐order conditions with respect to \(T_i(s_{\varepsilon})\) read:
\[
\frac{\partial \mathcal{L}}{\partial T_i(s_{\varepsilon})}
\;=\;
\alpha_i \,v_i'\bigl(T_i(s_{\varepsilon})\bigr)
\;-\;\Lambda(s_{\varepsilon})
\;+\;\eta_i(s_{\varepsilon})
\;-\;\mu_i(s_{\varepsilon})
\;=\;0,
\quad i=1,\dots,n,
\]
or equivalently
\[
\alpha_i \,v_i'\bigl(T_i(s_{\varepsilon})\bigr)
\;=\;\Lambda(s_{\varepsilon})
\;-\;\eta_i(s_{\varepsilon})
\;+\;\mu_i(s_{\varepsilon}).
\]
Complementary slackness and feasibility for each \(i\) impose:
\[
\left\{
\begin{aligned}
  &\eta_i(s_{\varepsilon}) \;\ge\; 0, \quad 
   T_i(s_{\varepsilon}) \;\le\; w_i - \pi_i, \quad
   \eta_i(s_{\varepsilon})\,\bigl(T_i(s_{\varepsilon}) - (w_i - \pi_i)\bigr)=0,\\
  &\mu_i(s_{\varepsilon}) \;\ge\; 0, \quad
   T_i(s_{\varepsilon}) \;\ge\; 0, \quad
   \mu_i(s_{\varepsilon})\,T_i(s_{\varepsilon}) = 0.
\end{aligned}
\right.
\]

\vspace{1ex}
\noindent\textbf{Case analysis.}
Depending on whether \(T_i(s_{\varepsilon})\) is at the lower bound (\(T_i = 0\)), at the upper bound (\(T_i = w_i - \pi_i\)), or strictly in the interior (\(0 < T_i < w_i - \pi_i\)), we obtain:

\begin{itemize}
    \item \emph{If \(T_i(s_{\varepsilon})=0\).} Then \(\mu_i>0\), \(\eta_i=0\).  By the first order conditions,
  \[
  \alpha_i\,v_i'(0) \;=\; \Lambda - 0 + \mu_i \;\ge\; \Lambda, \ \text{i.e.} \ \alpha_i\,v_i'(0)\ge \Lambda(s_{\varepsilon}).
  \]
  
  \item \emph{If \(T_i(s_{\varepsilon})=w_i-\pi_i\).} Then \(\eta_i>0\), \(\mu_i=0\).  The first order conditions give
  \[
  \alpha_i\,v_i'(w_i-\pi_i) \;=\; \Lambda \;-\;\eta_i \;\le\;\Lambda, \ \text{i.e.} \ \alpha_i\,v_i'(w_i-\pi_i)\le \Lambda(s_{\varepsilon}).
  \]

  \item \emph{If \(0< T_i(s_{\varepsilon})<w_i-\pi_i\).} Then \(\eta_i=\mu_i=0\).  The first order conditions yield
  \[
  \alpha_i\,v_i'\bigl(T_i(s_{\varepsilon})\bigr) \;=\; \Lambda(s_{\varepsilon}).
  \]
Thus in any interior region the weighted marginal disutilities must be equalized across all participating \(i\).
\end{itemize}
Finally, the multiplier \(\Lambda(s_{\varepsilon})\) is determined by the aggregate constraint \(\sum_{i}T_i(s_{\varepsilon})=s_{\varepsilon}\). 

\medskip

\noindent\textbf{Conclusion.}
The KKT conditions show that, at each loss level \(s_{\varepsilon}\), regions fall into three categories: those that pay zero tax (pure retention), those whose tax lies strictly between 0 and the cap and for which 
\(\alpha_i\,v_i'(T_i)=\Lambda(s_{\varepsilon})\), and those that reach their cap \(T_i = w_i - \pi_i\) (pure limit). The weight vector \(\boldsymbol\alpha\) governs the common multiplier \(\Lambda\) and hence the boundaries between these layers; varying \(\boldsymbol\alpha\) traces out the Pareto frontier.  Therefore, solving the weighted minimization for each \(\boldsymbol\alpha\in\Delta^n_{+}\) indeed yields all Pareto‐optimal taxation rules.

\section{Proof of the analytical solution for the case of two regions in Section \ref{sec:subsec:analytical} \label{proof}}
We consider the case of two regions with CARA disutility functions satisfying Assumption~\ref{hyp:ex}.  Under this setup, write
\[
\gamma = \frac{\gamma_2}{\gamma_1}, 
\quad
\mu = \frac{\mu_2}{\mu_1}, 
\quad
\zeta = \ln\ \!\Bigl(\frac{\alpha_2}{\alpha_1}\Bigr),
\quad
c = (1+\gamma)w_1 - \gamma_2\,\zeta,
\quad
M = \Bigl(\tfrac{2}{1-p_0}\Bigr)^2(\mu+1)\Bigl(\tfrac{\mu_1}{\gamma_1}\Bigr)^2.
\]
\begin{itemize}
    \item \textbf{Preliminary Step }: Set interval endpoints in the range of $\lambda$.\\
The marginal exponential utility of both regions is given by,  
$$ \alpha_1 v_1^{\prime}\left(x\right)=\alpha_1 e^{\frac{x}{\gamma_1}}, \quad \mbox{ and } \quad
 \alpha_2 v_2^{\prime}\left(x \right)=\alpha_2 e^{\frac{x}{\gamma_2}}
$$

with $\alpha_1 \in ]0,1[$ and $\alpha_2=1-\alpha_1$. We can distinguish three  cases according to the orders of magnitude of $\lambda_1=\alpha_1$ and $\lambda_2=1-\alpha_1$ on the one hand, and $\lambda_1^{\max}=\alpha_1 e^{\frac{w_1}{\gamma_1}}$ and $\lambda_2^{\max}= (1-\alpha_1) e^{\frac{w_2}{\gamma_2}}$ on the other hand. Namely, since the region are ordered such that $\frac{w_1}{\gamma_1} \leq \frac{w_2}{\gamma_2}$ (Assumption \ref{hyp:ex}), the following 3 cases are possible:
\begin{itemize}
    \item Case 1:
    $\lambda_1\leq\lambda_2\leq\lambda_1^{\max}\leq\lambda_2^{\max}$ which is equivalent to $0 \leq \zeta \leq \frac{w_1}{\gamma_1}$;
    \item Case 2:
    $\lambda_2\leq\lambda_1<\lambda_1^{\max}\leq\lambda_2^{\max}$ which is equivalent to $ \frac{w_1}{\gamma_1}- \frac{w_2}{\gamma_2}\leq\zeta \leq 0$;
   \item Case 3: $\lambda_2\leq\lambda_1\leq\lambda_2^{\max}\leq\lambda_1^{\max}$ which is equivalent to $-\frac{w_2}{\gamma_2} \leq \zeta \leq \frac{w_1}{\gamma_1}-\frac{w_2}{\gamma_2}$.
\end{itemize}
We investigate the optimal AFPO taxation rule in each case and characterize them in terms of the model parameters.

\end{itemize}

\paragraph{Case 1:  $0 \leq \zeta \leq \frac{w_1}{\gamma_1}$}\bigskip

\begin{itemize}
\item \textbf{Step 1}: Find $\mathcal{T}_i(\lambda)$ as piecewise-defined functions. \bigskip

By Theorem \ref{thm2}, the Pareto optimal taxation can be determined by first solving the equations $\alpha_i v_i^{\prime}\left(T_i(s_\varepsilon)\right)= \Lambda(s_\varepsilon)=\lambda$ for $i=1,2$. We shall represent $T_1(s_\varepsilon), T_2(s_\varepsilon)$ as functions of $\lambda$ using the notations {$\mathcal{T}_i(\lambda)$}. Observe that
$$
\begin{aligned}
& \mathcal{T}_1(\lambda)= \begin{cases}0 & 0\leq\lambda\leq\alpha_1 \\
\gamma_1 \ln \left(\frac{\lambda}{\alpha_1}\right) & \alpha_1\leq\lambda\leq\alpha_1 e^{\frac{w_1}{\gamma_1}} \\
w_1, & \alpha_1 e^{\frac{w_1}{\gamma_1}}\leq\lambda\end{cases} \\
& \mathcal{T}_2(\lambda)= \begin{cases}0 & 0\leq\lambda\leq 1-\alpha_1 \\
\gamma_2 \ln \left(\frac{\lambda}{1-\alpha_1}\right), & 1-\alpha_1\leq\lambda\leq(1-\alpha_1) e^{\frac{w_2}{\gamma_2}}\end{cases}
\end{aligned}
$$

\item \textbf{Step 2}: Find $s_{\varepsilon}(\lambda)$ and $\Lambda(s_\varepsilon)$ as piecewise-defined functions.\bigskip

It consists in inverting the function $s_\varepsilon(\lambda):=\mathcal{T}_1(\lambda)+\mathcal{T}_2(\lambda)$.
$$
s_{\varepsilon}(\lambda)= \begin{cases}\gamma_1 \ln \left(\frac{\lambda}{\alpha_1}\right), & \alpha_1\leq\lambda\leq 1-\alpha_1 \\ \ln \left(\frac{\lambda^{\gamma_1+\gamma_2}}{\alpha_1^{\gamma_1}\left(1-\alpha_1\right)^{\gamma_2}}\right), & 1-\alpha_1\leq\lambda\leq\alpha_1 e^{\frac{w_1}{\gamma_1}} \\ w_1+\gamma_2 \ln \left(\frac{\lambda}{1-\alpha_1}\right), & \alpha_1 e^{\frac{w_1}{\gamma_1}}\leq\lambda\leq(1-\alpha_1) e^{\frac{w_2}{\gamma_2}}\end{cases}
$$

Or one has $w_1>\zeta$ and $c>\zeta$ since by assumption $\frac{1}{\left(1+e^{\frac{w_1}{\gamma_1}}\right)}\leq\alpha_1$. Solving for the inverse function $\Lambda\left(s_{\varepsilon}\right)$ gives

$$
\Lambda\left(s_{\varepsilon}\right)= \begin{cases}\alpha_1 e^{\frac{s_{\varepsilon}}{\gamma_1}}, & 0\leq s_{\varepsilon}\leq \gamma_1 \zeta \\ \left(\alpha_1^{\gamma_1}\left(1-\alpha_1\right)^{\gamma_2} e^{s_{\varepsilon}}\right)^{\frac{1}{\left(\gamma_1+\gamma_2\right)}}, & \gamma_1 \zeta\leq s_{\varepsilon}\leq c \\ \left(1-\alpha_1\right) e^{\frac{\left(s_{\varepsilon}-w_1\right)}{\gamma_2}}, & c\leq s_{\varepsilon} \leq w_1+w_2\end{cases}
$$

\item \textbf{Step 3}: Find $T_i(s_\varepsilon)$ as a breakdown of $s_\varepsilon$.\bigskip

Inserting $\Lambda(s_\varepsilon)$ into $\mathcal{T}_i(\lambda)$ for $i=1,2$ yields the solution
$$
\begin{aligned}
& T_1\left(s_{\varepsilon}\right)=\mathcal{T}_1\left(\Lambda\left(s_{\varepsilon}\right)\right)= \begin{cases}s_{\varepsilon}, & 0\leq s_{\varepsilon}\leq\gamma_1 \zeta \\
\left(\frac{\gamma_1}{\left(\gamma_1+\gamma_2\right)}\right) s_{\varepsilon}+\left(\frac{\gamma_1 \gamma_2}{\left(\gamma_1+\gamma_2\right)}\right) \zeta, & \gamma_1 \zeta\leq s_{\varepsilon}\leq c \\
w_1, & c\leq s_{\varepsilon}\leq w_1+w_2\end{cases} \\
& T_2\left(s_{\varepsilon}\right)=\mathcal{T}_2\left(\Lambda\left(s_{\varepsilon}\right)\right)= \begin{cases}0, & 0\leq s_{\varepsilon}\leq \gamma_1 \zeta \\
\left(\frac{\gamma_2}{\left(\gamma_1+\gamma_2\right)}\right) s_{\varepsilon}-\left(\frac{\gamma_1 \gamma_2}{\left(\gamma_1+\gamma_2\right)}\right) \zeta, & \gamma_1 \zeta\leq s_{\varepsilon}\leq c \\
s_{\varepsilon}-w_1, & c\leq s_{\varepsilon}\leq w_1+w_2\end{cases}
\end{aligned}
$$

\item \textbf{Step 4}: Find the unique AFPO taxation rule by adding the fairness constraint.\bigskip
$$
\begin{aligned}
\mathbb{E}\left[T_1(S_\varepsilon ; \zeta)\right]= & (1-p_0) \left( \int_0^{\gamma_1 \zeta} \frac{s_\varepsilon}{w_1 + w_2}   d s_\varepsilon+\int_{\gamma_1 \zeta}^c\frac{1}{w_1 + w_2} \left(\left(\frac{\gamma_1}{\left(\gamma_1+\gamma_2\right)}\right) s_{\varepsilon}+ \left(\frac{\gamma_1 \gamma_2}{\left(\gamma_1+\gamma_2\right)} \right) \zeta\right) d s_\varepsilon \right)\\
& +(1-p_0) \left(\frac{1}{w_1+w_2} \int_c^{w_1+w_2} w_1  d s_\varepsilon \right)\\
= & (1-p_0) \left(\frac{w_1+w_2}{2}-\frac{1}{2(w_1+w_2)} \left(w_2^2 - (c-w_1)^2 + \frac{\gamma_2}{\gamma_1+\gamma_2} (c-\gamma_1 \zeta)^2\right) \right); \\
\mathbb{E}\left[T_2(S_\varepsilon ; \zeta)\right]= & (1-p_0) \left(\frac{1}{w_1+w_2} \int_{\gamma_1 \zeta}^c\left(\left(\frac{\gamma_2}{\left(\gamma_1+\gamma_2\right)}\right) s_{\varepsilon}-\left(\frac{\gamma_1 \gamma_2}{\left(\gamma_1+\gamma_2\right)}\right) \zeta\right)  d s_\varepsilon\right) \\
& +(1-p_0) \left(\frac{1}{w_1+w_2} \int_c^{w_1+w_2}\left(s_\varepsilon-w_1\right) d s_\varepsilon \right). \\
= & (1-p_0) \left(\frac{1}{2(w_1+w_2)} \left(w_2^2 - (c-w_1)^2 + \frac{\gamma_2}{\gamma_1+\gamma_2} (c-\gamma_1 \zeta)^2\right)\right),
\end{aligned}
$$
where they sum up to $(1-p_0)\frac{w_1+w_2}{2}$ as expected. By the actuarial fairness condition, $\zeta$ should be determined to satisfy
$$
\mathbb{E}\bigl[T_1(S_\varepsilon; \zeta)\bigr] = \mathbb{E}[X_1] = \mu_1,\quad
\mathbb{E}\bigl[T_2(S_\varepsilon; \zeta)\bigr] = \mathbb{E}[X_2] = \mu_2.
$$
It follows from the equation $\mathbb{E}\left[T_1(S_\varepsilon ; \zeta)\right]=\mu_1$ that

\begin{equation}
   \zeta  = \frac{w_1}{\gamma_1}
  - \sqrt{ \frac{M\mu}{\gamma} - \gamma\bigl(\frac{w_2}{\gamma_2}\bigr)^2}.
    \label{fixpoint}
\end{equation}
Equation \eqref{fixpoint} gives an explicit solution for the weight $\zeta$. One can see that $\zeta$ depends on the wealths $w_1, w_2$, the risk tolerance of regions modeled by $\gamma_1, \gamma_2$, and the distribution of losses of regions through $\mu_1, \mu_2$. The set of these parameters that leads to this solution $\zeta$ with $0 \leq \zeta\leq \frac{w_1}{\gamma_1}$ is given by: 
$$A_1:=\Bigl\{\,(\mu_i,\gamma_i,w_i)_{i=1,2} \; / \; M \in \left[\frac{\gamma^2}{\mu}\left(\frac{w_2}{\gamma_2}\right)^2, \frac{\gamma}{\mu}\left(\frac{w_1}{\gamma_1}\right)^2+\frac{\gamma^2}{\mu}\left(\frac{w_2}{\gamma_2}\right)^2 \right] \Bigr\}.$$

\end{itemize}

\paragraph{Case 2:  $ \frac{w_1}{\gamma_1}- \frac{w_2}{\gamma_2}\leq \zeta \leq 0$} \bigskip

\begin{itemize}
\item \textbf{Step 1}: Find $\mathcal{T}_i(\lambda)$ as piecewise-defined functions. 
$$
\begin{aligned}
& \mathcal{T}_1(\lambda)= \begin{cases}0 & 0\leq\lambda\leq\alpha_1 \\
\gamma_1 \ln \left(\lambda / \alpha_1\right) & \alpha_1\leq\lambda\leq\alpha_1 e^{w_1 / \gamma_1} \\
w_1, & \alpha_1 e^{w_1 / \gamma_1}\leq\lambda\end{cases} \\
& \mathcal{T}_2(\lambda)= \begin{cases}0 & 0\leq\lambda\leq1-\alpha_1 \\
\gamma_2 \ln \left(\lambda / 1-\alpha_1\right), & 1-\alpha_1\leq\lambda\end{cases}
\end{aligned}
$$

\item \textbf{Step 2}: Find $s_{\varepsilon}(\lambda)$ and $\Lambda(s_\varepsilon)$ as piecewise-defined functions.
$$
s_{\varepsilon}(\lambda)= \begin{cases}\gamma_2 \ln \left(\lambda / 1-\alpha_1\right), & 1-\alpha_1\leq\lambda\leq\alpha_1 \\ \ln \left(\lambda^{\gamma_1+\gamma_2} / \alpha_1^{\gamma_1}\left(1-\alpha_1\right)^{\gamma_2}\right), & \alpha_1\leq\lambda\leq\alpha_1 e^{w_1 / \gamma_1} \\ w_1+\gamma_2 \ln \left(\lambda / 1-\alpha_1\right), & \alpha_1 e^{w_1 / \gamma_1}\leq\lambda\end{cases}
$$
Solving for the inverse function $\Lambda\left(s_{\varepsilon}\right)$ gives

$$
\Lambda\left(s_{\varepsilon}\right)= \begin{cases}(1-\alpha_1) e^{s_{\varepsilon} / \gamma_2}, & 0\leq s_{\varepsilon}\leq-\gamma_2 \zeta \\ \left(\alpha_1^{\gamma_1}\left(1-\alpha_1\right)^{\gamma_2} e^{s_{\varepsilon}}\right)^{1 /\left(\gamma_1+\gamma_2\right)}, & -\gamma_2 \zeta\leq s_{\varepsilon}\leq c \\ \left(1-\alpha_1\right) e^{\left(s_{\varepsilon}-w_1\right) / \gamma_2}, & c\leq s_{\varepsilon}\end{cases}
$$

\item \textbf{Step 3}: Find $T_i(s_\varepsilon)$ as a breakdown of $s_\varepsilon$.
$$
\begin{aligned}
& T_1\left(s_{\varepsilon}\right)=\mathcal{T}_1\left(\Lambda\left(s_{\varepsilon}\right)\right)= \begin{cases}0, & 0\leq s_{\varepsilon}\leq-\gamma_2 \zeta \\
\left(\gamma_1 /\left(\gamma_1+\gamma_2\right)\right) s_{\varepsilon}+\left(\gamma_1 \gamma_2 /\left(\gamma_1+\gamma_2\right)\right) \zeta, &-\gamma_2 \zeta\leq s_{\varepsilon}\leq c \\
w_1, & c \leq s_{\varepsilon}\end{cases} \\
& T_2\left(s_{\varepsilon}\right)=\mathcal{T}_2\left(\Lambda\left(s_{\varepsilon}\right)\right)= \begin{cases}s_{\varepsilon}, & 0\leq s_{\varepsilon}\leq -\gamma_2 \zeta \\
\left(\gamma_2 /\left(\gamma_1+\gamma_2\right)\right) s_{\varepsilon}-\left(\gamma_1 \gamma_2 /\left(\gamma_1+\gamma_2\right)\right) \zeta, & -\gamma_2 \zeta\leq s_{\varepsilon}\leq c \\
s_{\varepsilon}-w_1, & c\leq s_{\varepsilon}\end{cases}
\end{aligned}
$$
\item \textbf{Step 4}: Find the unique AFPO taxation rule by adding the fairness constraint.\bigskip
$$
\begin{aligned}
\mathbb{E}\left[T_1(S_\varepsilon ; \zeta)\right]= & (1-p_0) \left(\int_{-\gamma_2 \zeta}^c\frac{1}{w_1 + w_2}\left(\left(\frac{\gamma_1}{\left(\gamma_1+\gamma_2\right)}\right) s_{\varepsilon}+ \left(\frac{\gamma_1 \gamma_2}{\left(\gamma_1+\gamma_2\right)} \right) \zeta\right) d s_\varepsilon  + \int_c^{w_1+w_2} \frac{w_1}{w_1+w_2}  d s_\varepsilon \right) \\
= & (1-p_0) \left(\frac{\gamma_1}{\gamma_1+\gamma_2} \frac{(c+\gamma_2 \zeta)^2}{2(w_1+w_2)}+w_1-\frac{w_1}{w_1+w_2}c\right).
\end{aligned}
$$
It follows from the equation $\mathbb{E}\left[T_1(S_\varepsilon ; \zeta)\right]=\mu_1$ that
\begin{equation*}
  \zeta=\frac{1}{\gamma_2} \left(\left(1+\gamma\right)\frac{w_1}{2} + \frac{2(\mu_1+\mu_2)}{1-p_0} \left(\frac{1}{1-p_0}\frac{\mu_1}{w_1}-1\right)\right).
  \label{fixpointv3}
\end{equation*}
The set of these parameters that leads to this solution $\zeta$ with $ \frac{w_1}{\gamma_1}- \frac{w_2}{\gamma_2}\leq\zeta \leq 0$ is given by: 
$$A_2:=\Bigl\{\,(\mu_i,\gamma_i,w_i)_{i=1,2} \; / \; M \in \left[\left(\gamma + 1\right)  \left(\frac{w_1}{\gamma_1}\right)^2, \left(1-\gamma \right)  \left(\frac{w_1}{\gamma_1}\right)^2 + 2\gamma\frac{w_1}{\gamma_1}\frac{w_2}{\gamma_2} \right] \Bigr\}.$$

\end{itemize}

\paragraph{Case 3: $-\frac{w_2}{\gamma_2} \leq \zeta \leq \frac{w_1}{\gamma_1}-\frac{w_2}{\gamma_2}$}

This case is equivalent to the case $1$ where the region $1$ is exchanged to the region $2$ that is:

\[
\zeta  = -\frac{w_2}{\gamma_2}
  + \sqrt{\frac{M}{\gamma}- \frac{1}{\gamma}\bigl(\frac{w_1}{\gamma_1}\bigr)^2},
\quad
T_1(s_\varepsilon)=
\begin{cases}
0, & 0 \leq s_\varepsilon\leq-\gamma_2\zeta ,\\
\frac{\gamma_1}{\gamma_1+\gamma_2}\,s_\varepsilon
 -\frac{\gamma_1\gamma_2}{\gamma_1+\gamma_2}\,\zeta ,
 & -\gamma_2\zeta  \leq s_\varepsilon\leq\frac{(\gamma+1)(w_1+w_2)-c}{\gamma},\\
s_\varepsilon-w_2, & \frac{(\gamma+1)(w_1+w_2)-c}{\gamma} \leq s_\varepsilon\leq w_1+w_2.
\end{cases}
\]
The set of the input parameters that leads to this solution $\zeta$ with $-\frac{w_2}{\gamma_2} \leq \zeta \leq \frac{w_1}{\gamma_1}-\frac{w_2}{\gamma_2}$ is given by: 

$$A_3:=\Bigl\{\,(\mu_i,\gamma_i,w_i)_{i=1,2} \; / \; M \in \left[ \left(\frac{w_1}{\gamma_1}\right)^2  , \left(\gamma + 1\right)  \left(\frac{w_1}{\gamma_1}\right)^2\right] \Bigr\}.$$

\section{Additional illustrations for Cases 2 and 3 in Section \ref{sec:subsec:sensitivity} \label{addsen}}
\begin{figure}[h!]
    \centering
    \includegraphics[width=.8\linewidth]{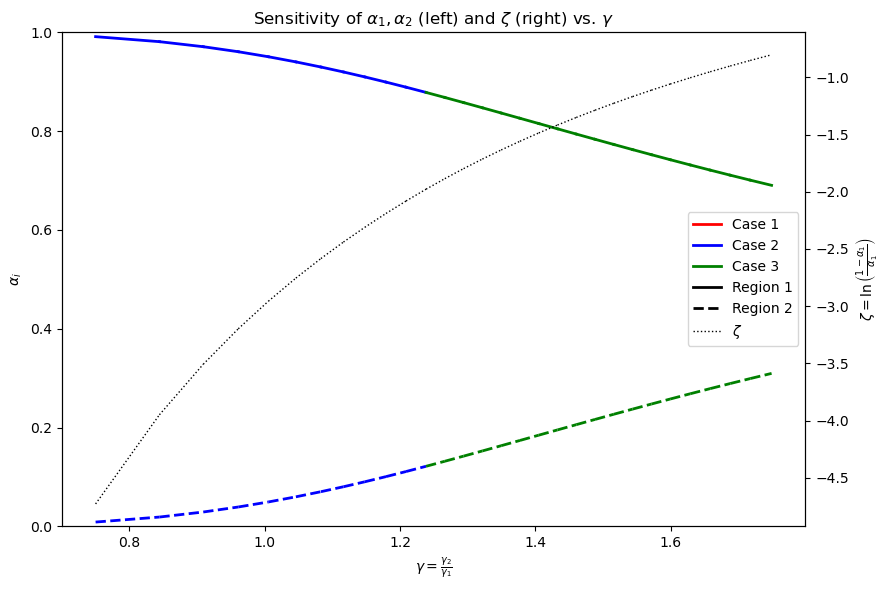}
    \caption{Sensitivity of $\zeta  = \ln \left(\frac{1-\alpha_1}{\alpha_1}\right)$ (dotted) and the optimal weights \(\alpha_1\) (solid) and \(\alpha_2 = 1 - \alpha_1\) (dashed), with respect to the relative risk tolerance ratio $\gamma=\frac{\gamma_2}{\gamma_1}$.  
    Curve segments are colored by the active case: \textcolor{blue}{Case 2}, and \textcolor{ForestGreen}{Case 3}, determined by the parametric regions.  
    Parameters are fixed to: $w_1 = 4.5$, $w_2 = 8$, $\mu_2=4, \mu_1=(1-p_0)\frac{w_1+w_2}{2}-\mu_2$, $\gamma_1 = 1$, and $p_0 = 0.1$.}
    \label{fig:sensitivity_analysisV2}
\end{figure}

For completeness, we include further sensitivity analysis with respect to risk tolerance for Cases 2 and 3 in complement of Cases 1 and 2 in Section \ref{sec:subsec:sensitivity} by changing the parameter $w_2$ from $10$ to $8$. Figure \ref{fig:sensitivity_analysisV2} highlights the behavior of $\zeta  = \ln \left(\frac{1-\alpha_1}{\alpha_1}\right)$ (dotted) and the optimal weights \(\alpha_1\) (solid) and \(\alpha_2 = 1 - \alpha_1\) (dashed), with respect to the relative risk tolerance ratio $\gamma=\frac{\gamma_2}{\gamma_1}$.

\end{appendices}

\pagebreak
\listoffigures


\end{document}